\documentclass[11pt]{article}
\pdfoutput = 1

\usepackage[utf8]{inputenc}
\usepackage{color,graphicx}
\usepackage{epsfig}
\usepackage{amsmath}
\usepackage{verbatim}
\usepackage{amssymb}
\usepackage{mathabx}
\usepackage[mathcal]{eucal}
\usepackage{stmaryrd}
\usepackage{esint}
\usepackage{braket}
\usepackage{physics}
\usepackage{amsfonts}
\usepackage{cite}
\usepackage{array}
\usepackage{setspace}
\usepackage{braket}
\usepackage{float}
\usepackage{url}
\usepackage{mathtools}
\usepackage{bbold}
\usepackage{slashed}
\usepackage{tikz}
\usepackage{graphicx}
\usepackage{epstopdf}
\usepackage{subfigure}
\usepackage[labelfont=bf,font={sf}]{caption}
\usepackage{pgfplots}
\usepackage{mathrsfs}
\usepackage{fancybox}
\usepackage{eurosym}
\usepackage{tcolorbox}
\usepackage{tensor}
\allowdisplaybreaks

\usepackage[margin = 2.2cm]{geometry}
\usepackage[ragged]{footmisc}
\usepackage[bookmarks=true,bookmarksnumbered=false,
hyperindex=true,bookmarksopen=true,hyperfigures=true,
colorlinks=true,linkcolor=webblue,citecolor=webgreen,urlcolor=webblue,breaklinks]{hyperref}
\definecolor{webgreen}{rgb}{0, 0.5, 0}
\definecolor{webblue}{rgb}{0, 0, 0.5}
\definecolor{webred}{rgb}{0.5, 0, 0}
\definecolor{darkgreen}{rgb}{0,0.5,0}

\renewcommand{\d}{\mathrm{d}}
\renewcommand{\i}{\mathrm{i}}

\newcommand{\average}[1]{\left\langle #1 \right\rangle}

\newcommand{\Vol}{\text{Vol}}

\def\ben{\begin{equation}}
\def\een{\end{equation}}

\let\a=\alpha \let\b=\beta  \let\d=\delta 
   
\let\l=\lambda \let\m=\mu \let\n=\nu  \let\p=\phi \let\r=v
\let\s=\sigma

\let\w=\omega

\def\be{\begin{equation}}
\def\ee{\end{equation}}
\def\ba{\begin{array}}
\def\ea{\end{array}}

\def\dalemb#1#2{{\vbox{\hrule height .#2pt
       \hbox{\vrule width.#2pt height#1pt \kern#1pt
               \vrule width.#2pt}
       \hrule height.#2pt}}}

\newcommand{\bea}{\begin{eqnarray}}
\newcommand{\eea}{\end{eqnarray}}

\let\tilde=\widetilde

\def\cone{\raisebox{0.3em}{\tikz[baseline=0.1ex,scale=0.8]{
\filldraw[fill=blue!15!white!90] (1,2) .. controls (0,1.5) .. (-3,0) .. controls (0,-1.5) .. (1,-2) .. controls (0.6,-0.8) and (0.6,0.8) .. (1,2);
\filldraw[fill=gray!30!white!90] (1.05,0) ellipse (0.4 and 2);
\draw[red,thick] (-3.1,-0.1) -- (-2.9,0.1);
\draw[red,thick] (-3.1,0.1) -- (-2.9,-0.1);
\draw [green!40!black!60,thick] plot [smooth cycle] coordinates {(-3,0) (-1.3,0.4) (-0.9,0.2) (0,0.7) (-0.5,-0.1) (-0.6,-0.6) (-1.5,-0.2)};}
}}

\renewcommand{\d}{\mathrm{d}}
\renewcommand{\i}{\mathrm{i}}

\numberwithin{equation}{section}

\title{}

\pgfplotsset{compat=1.18} 
\begin{document}

\thispagestyle{empty}
\begin{center}
    ~\vspace{5mm}

     {\LARGE \bf 
   Gravitons on the edge
   }
    
   \vspace{0.4in}
    
    {\bf Andreas Blommaert$^1$, Sean Colin-Ellerin$^2$ }

    \vspace{0.4in}
    {$^1$SISSA and INFN, Via Bonomea 265, 34127 Trieste, Italy\\
    $^2$ Center for Theoretical Physics and Department of Physics,\\
University of California, Berkeley, California 94720, U.S.A.\\}
    \vspace{0.1in}
    
    {\tt ablommae@sissa.it, scolinellerin@berkeley.edu }
\end{center}

\vspace{0.4in}

\begin{abstract}
\noindent We study free graviton entanglement between Rindler wedges in the Minkowski vacuum state via the Euclidean path integral. We follow Kabat's method for computing the conical entropy, using the heat kernel on the cone with the tip removed, whose resulting von Neumann entropy for photons correctly predicted electromagnetic edge modes. We find that, in addition to the bulk graviton contributions, the conical entropy has a contact term that can be attributed to a vector field anchored to the $(d-2)$-dimensional (Euclidean) Rindler horizon whose contribution equals $d-2$ times Kabat's contact term for photons. We suggest that graviton edge modes are hence the $d-2$ large diffeomorphisms which act internally within the Rindler horizon. Along the way, we address several known issues regarding graviton entanglement. We furthermore sketch how our results may be used to study edge modes in closed bosonic string theory.
\end{abstract}

\pagebreak
\setcounter{page}{1}
\tableofcontents

\section{Introduction}
\label{sec:intro}
We would like to understand the degrees of freedom associated with subregions in quantum gravity. An important benchmark in this sense is entanglement entropy between bulk subregions. This is a daunting challenge in full, non-perturbative quantum gravity which would require understanding entanglement in a UV-complete theory of gravity, such as string theory. One reason why one might be interested in this is to get a more complete picture of the interplay between the area and entanglement entropy terms in the generalized entropy of black holes, when gravity is fully quantized.

It is reasonable to start with a less ambitious challenge. One could ask the simpler question of what are the degrees of freedom of subregions in \emph{perturbative} gravity. Even in the \emph{free} graviton approximation, there are yet many open questions. The goal of this work is to (partially) address those. One phenomena which is expected to play a major role in understanding degrees of freedom associated with subregions in gravity is the concept of `edge modes' \cite{Donnelly:2011hn}, which we shall now introduce.

To understand entanglement entropy between a subregion $A$ and its complement on a Cauchy slice, or even simply the degrees of freedom associated with $A$, one would like to factorize the Hilbert space $\mathcal{H} \overset{?}{=} \mathcal{H}_{A} \otimes \mathcal{H}_{A^{c}}$.\footnote{This is technically speaking not possible in any quantum field theory due to entanglement between UV modes. Typically one introduces a cutoff in order to separate $A$ from $A^{c}$ by a gap to deal with this, for example, by putting the theory on the lattice. However, this does not resolve the lack of factorization for gauge theories we discuss here. They are orthogonal problems.} In a theory with constraints, for instance a gauge theory, the Hilbert space does not naively factorize as the constraints relate degrees of freedom in $\mathcal{H}_{A}$ to those in $\mathcal{H}_{A^{c}}$. This is solved by introducing an `extended' Hilbert space, in which the constraints are relaxed; and then embedding the constrained Hilbert space $\mathcal{H}$ inside this extended Hilbert space \cite{Casini:2013rba}. Schematically, the result is that $\mathcal{H}$ decomposes as a direct product
\begin{equation}
    \mathcal{H}=\underset{\ket{\psi}}{\oplus}\,(\ket{\psi}_\text{edge}\otimes \mathcal{H}_{A\,\text{naive}})\otimes (\ket{\psi}_\text{edge}\otimes \mathcal{H}_{A^c\,\text{naive}})\,. 
\end{equation}
Here it is understood that the Hilbert space associated with a subregion $A$
\begin{equation}
    \mathcal{H}_A=\underset{\ket{\psi}}{\oplus}\,\ket{\psi}_\text{edge}\otimes \mathcal{H}_{A\,\text{naive}}\,,\label{1.2 introintro}
\end{equation}
contains `extra' degrees of freedom $\ket{\psi}_\text{edge}$ associated with the boundary of $A$, known as edge modes. These edge modes contribute non-trivially to entanglement entropy.

Edge modes are rather well-understood in topological gauge theories, such as Chern-Simons theory \cite{Dong:2008ft,Qi_2012,Das:2015oha,Wen:2016snr,Wong:2017pdm,Fliss:2020cos} or BF theory \cite{Geiller:2019bti,Fliss:2023dze,Fliss:2023uiv}. For Maxwell theory, the formalism of edge modes is also well-developed \cite{Buividovich:2008gq,Donnelly:2011hn,Casini:2013rba,Radicevic:2014kqa,Donnelly:2014fua,Donnelly:2015hxa,Huang:2014pfa,Ghosh:2015iwa,Soni:2015yga,Soni:2016ogt,Blommaert:2018rsf,Blommaert:2018oue,Ball:2024hqe,Harlow:2015lma}. In gauge theories, one can understand these edge modes as the large gauge transformations which do not fall off at the boundary. The lesson is that large `gauge' transformations are not actually redundant, rather they are physical degrees of freedom which contribute to the entropy between subregions.  

Despite this progress in gauge theories, gravitational entanglement has proven to remain a delicate matter even in the free graviton approximation. The redundancy of general relativity is diffeomorphism symmetry. Thus, based on other gauge theories, one expects that gravitational edge modes are related with large diffeomorphisms. However, several conflicting proposals exist for graviton edge modes. The proposals differ in determining which large diffeomorphisms should be included as physical, and hence which contribute to the entanglement entropy \cite{Donnelly:2016auv,Geiller:2017xad,Geiller:2017whh,Freidel:2019ees,Freidel:2020xyx,Donnelly:2020xgu,Ciambelli:2021vnn,Ciambelli:2021nmv,Ciambelli:2022cfr,Donnelly:2022kfs,David:2022jfd}. Which proposal is correct? What are the edge degrees of freedom of gravity associated with a subregion (in the free graviton approximation)?

Historically we have often struggled in correctly counting degrees of freedom and computing entropy in a Hilbert space language. In cases where we struggle, it has proven very fruitful to first compute the entropy using the Euclidean path integral. Famous examples are the Bekenstein-Hawking entropy \cite{bekenstein1973black} and the Page curve \cite{Page:1993df} of evaporating black holes. More relevant to our work are the aforementioned gauge theories. In those cases, precise matches between Euclidean path integral calculations and Hilbert space calculations of entanglement entropy were found \cite{Dong:2008ft,Qi_2012,Das:2015oha,Wen:2016snr,Wong:2017pdm,Fliss:2020cos,Fliss:2023dze,Fliss:2023uiv,Kabat:1995eq,Donnelly:2015hxa,Huang:2014pfa,Blommaert:2018rsf,Soni:2016ogt}. Notably, the Euclidean calculation only matches a Hilbert space calculation \emph{if} the correct edge modes are included in the latter. In this sense, the Euclidean path integral \emph{predicts} edge modes. This gives us a test to select the correct proposals for gravitational edge modes. Our main goal in this work is to obtain a Euclidean prediction for free graviton entanglement in the Minkowski vacuum, following Kabat's work for photons \cite{Kabat:1995eq}. The Euclidean calculation in question is to compute the `conical entropy' \cite{Carlip:1993sa,Susskind:1994vu}. 

\begin{figure}[h]
\begin{center}
\begin{tikzpicture}[scale=0.8]

\filldraw[fill=blue!15!white!90] (1,2) .. controls (0,1.5) .. (-3,0) .. controls (0,-1.5) .. (1,-2) .. controls (0.6,-0.8) and (0.6,0.8) .. (1,2);
\filldraw[fill=gray!30!white!90] (1.05,0) ellipse (0.4 and 2);

\draw[<-] (1.5,1.8) .. controls (1.9,0.8) and (1.9,-0.8) .. (1.5,-1.8);
\node at (3,0) {$\tau \sim \tau+\beta$};

\draw[red,thick] (-3.1,-0.1) -- (-2.9,0.1);
\draw[red,thick] (-3.1,0.1) -- (-2.9,-0.1);
\node at (-3,-0.65) {\color{red}{$r=0$}};

\end{tikzpicture}
\end{center}
\caption{Rindler cone whose partition function $Z(\beta)$ is used to compute conical entropy. The angular time direction $\tau$ has length $\beta$ and the tip of the cone ($r=0$) is removed.}
\label{fig:1}
\end{figure}
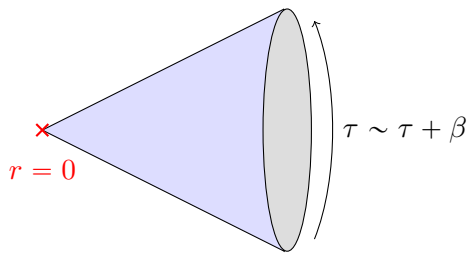

Let us recall the logic behind this conical entropy. A right Rindler wedge of $d$-dimensional Minkowski space has the metric
\begin{equation}\label{eqn:Rindlermetric}
    \d s^{2} = -r^{2}\d t^{2}+\d r^{2}+\sum_{i=1}^{d-2}\d x_i^2\,.
\end{equation}
The standard statement is that the fiducial observer sees quantum fields at Unruh temperature $\beta=2\pi$. Therefore the reduced density matrix for the Minkowski vacuum state restricted to a half space \eqref{eqn:Rindlermetric} is
\begin{equation}
    \rho=e^{-2\pi H}\,,\label{1.3 rho}
\end{equation}
where $H$ is the boost generating Rindler time $t$ evolution. Equation \eqref{1.3 rho} is simply the statement that evolving for a Euclidean Rindler time $\tau=2\pi$ produces Euclidean Minkowski. What is the entanglement entropy between two half-spaces in the Minkowski vacuum state? For the reduced density matrix \eqref{1.3 rho}, one finds the von Neumann entropy
\begin{equation}\label{eqn:entropy}
    S=-\Tr \rho \log \rho=(1-\beta \partial_\beta)\log Z(\beta)\rvert_{\beta=2\pi}\,,\quad Z(\beta)=\Tr e^{-\beta H}\,.
\end{equation}
Here $Z(\beta)$ is computed by evolving for a Euclidean Rindler time $\beta\neq 2\pi$ prior to gluing the initial and final states. In a path integral language, this corresponds to considering quantum fields on Rindler spacetime with a different Euclidean time periodicity, resulting in a cone-shaped spacetime \cite{Carlip:1993sa,Susskind:1994vu}
\begin{equation}\label{eqn:Rindlercone}
    \d s^{2} = r^{2}\d \tau^{2}+\d r^{2}+\sum_{i=1}^{d-2}\d x_i^2\,,\quad \tau\sim \tau+\beta\,,\quad r>0\,.
\end{equation}
This is shown in figure \ref{fig:1}.

Our main technical goal is to compute the path integral $Z(\beta)_\text{graviton}$ of free gravitons on the Rindler cone \eqref{eqn:Rindlercone}, and interpret the associated entanglement entropy \eqref{eqn:entropy} in terms of gravitational edge modes.

\subsection{Summary and structure}
 We now summarize our main results and present the outline of this work.
 
In \textbf{section \ref{sec:makingsense}} we deal with two conceptual puzzles that arise when one wants to study free graviton entanglement using conical entropy. In particular, we explain how to define subregions in (perturbative) gravity, and we point out that one should consider free gravitons on the Rindler cone \eqref{eqn:Rindlercone} with the tip deleted. More explanation on why this seems to be a well defined procedure can be found in the discussion \textbf{section \ref{sect:disc}}.

We emphasize that our choice to deal with the gravitational path integral by ``removing the tip of the cone'' following Kabat, is indeed a \emph{choice}. We do not claim this is the only reasonable approach one could take. However, it is the only reasonable one that we are currently aware of, for reasons explained in more detail in the discussion \textbf{section \ref{sect:disc}}. Removing the tip of the cone makes QFT on the cone well defined as long as one assumes regularity of the fields at the tip (in a way made more precise in section \ref{sect:disc}).\footnote{In other words, after removing the tip, QFT on the cone is well defined and no additional arbitrary choices (asides from the first choice, namely to remove the tip) need to be made.}

In \textbf{section \ref{sect:photons}} we revisit Kabat's calculation of the Maxwell path integral $Z(\beta)_\text{photon}$ function on the Rindler cone \cite{Kabat:1995eq}, in order to streamline calculations in subsequent sections. One obtains \eqref{3.39 Kphotcont}
\begin{equation}
    \log Z_\text{photon}(\beta)=(d-2) \int_{\epsilon^{2}}^\infty \frac{\d s}{2s}\int_\text{cone} \d^d x \sqrt{g}\,K_\text{scal}(x|s)-\int_{\epsilon^{2}}^\infty \d s\int_\text{horizon} \d^{d-1} x \,K_\text{scal}(x|s)\,.
\end{equation}
Aside from the naive bulk contributions one recognizes the contact term
\begin{equation}
    \log Z^\text{contact}_\text{photon}= -\int_{\epsilon^{2}}^\infty \d s\int_\text{horizon} \d^{d-1} x \,K_\text{scal}(x|s)\,.
\end{equation}
As we recall in \textbf{section \ref{sect:interpret}}, this can be rewritten as the following path integral over edge modes \cite{Donnelly:2014fua,Donnelly:2015hxa,Blommaert:2018rsf,Blommaert:2018oue}
\begin{equation}
    Z_\text{photon}^\text{contact}=\int \mathcal{D}q\,\exp\bigg(-\frac{1}{2}\int_\text{horizon}\d^{d-2}x\,q(x)\int_\text{horizon}\d^{d-2}y\,q(y)\,G_\text{scal horizon}(x|y)\bigg)\,.\label{1.8 photoncontact}
\end{equation}
Here the edge modes $q$ physically represent electromagnetic charges on the entangling surface
\begin{equation}
    q= \frac{\delta S_\text{on-shell}}{\delta A_\tau}\Big\rvert_\text{bdy} =\sqrt{h}\,n_\mu F^{\mu\tau}\rvert_\text{bdy}\,.
\end{equation}
The charges $q$ are canonically conjugate to (or generate) large gauge transformations $A_\mu\to A_\mu+\nabla_\mu \phi$, which were indeed referred to as `edge modes' earlier \cite{Donnelly:2014fua,Donnelly:2015hxa,Blommaert:2018rsf,Blommaert:2018oue}.

In \textbf{section \ref{sect:kabatgraviton}} we compute the free graviton partition function $Z(\beta)_\text{graviton}$ on the Rindler cone and obtain
\begin{align}
    \log Z_\text{graviton}(\beta) = \frac{d(d-3)}{2} \int_{\epsilon^{2}}^\infty \frac{\d s}{2s}\int_\text{cone} \d^d x \sqrt{G}\,K_\text{scal}(x|s)-\int_{\epsilon^{2}}^\infty \d s\int_\text{horizon} \d^{d-1} x \,K_\text{vec}(x|s)\,.\label{4.23 zgravcontINTRO}
\end{align}
The first term represents the usual graviton polarizations. The second `contact' term, given by a vector propagator anchored on the Rindler horizon, evaluates to $d-2$ times the photon contact term \eqref{1.8 photoncontact}
\begin{equation}
   \log Z^\text{contact}_{\text{graviton}}=-\int_{\epsilon^{2}}^\infty \d s\int_\text{horizon} \d^{d-1} x \,K_\text{vec}(x|s)=(d-2)\log Z^\text{contact}_\text{photon}\,.
\end{equation}
This results in the following conical entropy
\begin{equation}
    S_\text{graviton}=2\pi A \int_{\epsilon^{2}}^\infty \frac{\d s}{(4\pi s)^{d/2}}\bigg(\frac{1}{6}\frac{d(d-3)}{2}-(d-2)\bigg)\,.
\end{equation}
We conclude that, according to this Euclidean path integral, \textbf{there are $\bf{d-2}$ graviton edge modes}. This claim was proposed as one possible answer in \cite{Freidel:2020xyx} based on earlier phase space methods introduced in \cite{Donnelly:2016auv}. The other possibilities raised in \cite{Freidel:2020xyx} have more than $d-2$ gravitational edge modes. This includes in particular a ``maximal'' option corresponding with the edge mode counting proposed in the seminal work of Donnelly and Freidel \cite{Donnelly:2016auv}. As explained in \cite{Freidel:2020xyx}, it is not the case that this ``maximal'' option is the only mathematically or physically consistent one. These different possibilities amount to a choice of boundary terms in the gravitational action. So, our work suggests, that rather than this ``maximal'' edge mode count of \cite{Donnelly:2016auv}, we should rather only include the ``minimal'' consistent amount of $d-2$ edge modes.

In \textbf{section \ref{sect:interpret}} we observe that the graviton contact term can be rewritten in terms of edge charges as
\begin{equation}
        Z_\text{graviton}^\text{contact}=\int \mathcal{D}q^i_\text{ADM}\exp\bigg(-\frac{1}{2}\int_\text{horizon}\d^{d-2}x\,q^i_\text{ADM}(x)\int_\text{horizon}\d^{d-2}y\,q^j_\text{ADM}(y)\,G_\text{vec horizon}(x|y)_{i\,j}\bigg)\,.
\end{equation}
We speculate that the edge charges are the ADM charges which act internally in the $(d-2)$-dimensional Rindler horizon as suggested (among other options) in \cite{Freidel:2020xyx}
\begin{equation}
    q^i_\text{ADM}=\frac{\delta S_\text{on-shell}}{\delta h_{i\tau}}\rvert_\text{bdy}\,.
\end{equation}
These generate large diffeomorphisms \eqref{eqn:quantumdiffsymkappa0} \emph{within} the Rindler horizon.\footnote{These large diffeomorphisms tangent to the entangling surface are often referred to as corner symmetries \cite{Ciambelli:2021nmv,Ciambelli:2021vnn,Ciambelli:2022cfr,Freidel:2020xyx}.} More details about this edge mode interpretation will be presented in future work \cite{Blommaert:2024}. We furthermore return to one of the original questions about edge modes: entanglement of strings in Rindler \cite{Susskind:1994vu,Susskind:1993ws,Susskind:1994sm,Kabat:1995jq,Balasubramanian:2018axm,Ahmadain:2022eso,Donnelly:2020teo,Geiller:2019bti,Witten:2018xfj,He:2014gva,Dabholkar:1994ai,Dabholkar:1994gg,Mertens:2016tqv,Mertens:2015adr}. By combining our graviton calculation with the Kalb-Ramond partition function computed in \textbf{appendix \ref{sec:KBfield}}, we match the contact term of the massless fields in the bosonic string to that coming from two photons living on the $(d-1)$-dimensional boundary
\begin{equation}
    \log Z_\text{closed string massless}^\text{contact}=2(d-3)\log Z^\text{contact}_\text{photon}\,.
\end{equation}
Two photons fields are indeed the gauge degrees of freedom of the closed bosonic string at the massless level. We conjecture that such an edge mode counting continues to hold for all massive string levels.

Finally, in \textbf{section \ref{sect:disc}} we briefly compare with earlier attempts at computing graviton entanglement using the Euclidean path integral in various contexts \cite{Fursaev:1996uz,Solodukhin:2015hma,He:2014gva,Anninos:2020hfj}, and we discuss why removing the tip of the Rindler cone in \eqref{eqn:Rindlercone} is the mathematically correct procedure.

\section{Free graviton entanglement makes sense}
\label{sec:makingsense}
It may not be obvious that the question of graviton entanglement is even well-defined. Aside from the non-factorization of Hilbert spaces due to constraints in gauge theories which we already discussed in section \ref{sec:intro}, here we point out two conceptual subtleties and explain how they are avoided in our setup. In particular, in \textbf{section \ref{sec:subregion}}, we explain how to define subregions for free gravitons. Then, in \textbf{section \ref{sec:onshell}}, we emphasize the need for on-shell backgrounds, which forces us to consider the Rindler cone \eqref{eqn:Rindlercone} with the tip $r=0$ \emph{deleted}. We will later return to this in section \ref{sect:disc}.

\subsection{Quantum diffeomorphisms and subregions}
\label{sec:subregion}

A standard formulation of entanglement in quantum field theory is to foliate the spacetime by Cauchy surfaces on which the Hilbert space of the theory is defined and then consider a codimension-$1$ subregion $A$ of the Cauchy slice. Given any density matrix in the theory defined on the Cauchy slice, the reduced density matrix obtained by tracing out the complement subregion $A^{c}$ diagnoses their entanglement.\footnote{One can formulate entanglement in terms of algebras. This may resolve some of the issues we discuss here \cite{Witten:2023xze}, but this is not be the subject of this work.} 

However, this is subtle when we quantize gravity. How does one define subregions in a path integral over metrics? This is unclear at a non-perturbative level, as the subregion itself can fluctuate. However, perturbatively in Newton's constant $G_\text{N}$, one can proceed using the background field expansion \cite{Bastianelli:2013tsa,Veltman:1975vx,Grisaru:1975ei} familiar from Yang-Mills theory \cite{Schwartz:2014sze,Grisaru:1975ei}. The metric $g_{\mu\nu}$ is decomposed as the sum of a fixed classical background metric $G_{\mu\nu}$, and a perturbative graviton field $h_{\mu\nu}$ which is path integrated over, viz.,
\begin{equation}\label{eqn:metricexp}
    g_{\mu\nu} = G_{\mu\nu}+\kappa h_{\mu\nu}\,, \qquad \kappa = \sqrt{16\pi G_{\text{N}}}\,.
\end{equation}
In our case, this background metric $G_{\mu\nu}$ is the Rindler cone \eqref{eqn:Rindlercone}
\begin{equation}
    \d s^{2} =G_{\mu\nu}\d x^\mu\d x^\nu= r^{2}\d \tau^{2}+\d r^{2}+\sum_{i=1}^{d-2}\d x_i^2\,,\quad \tau\sim \tau+\beta\,,\quad r>0\,.\label{2.2 G}
\end{equation}
This metric $G_{\mu\nu}$ is \emph{not} path integrated over so perturbatively in $G_{\text{N}}$ one can define subregions $A$ simply by using coordinates $x^\mu$ with this background metric $G_{\mu\nu}$. So, in this formalism, there are no quantum fluctuation in our definition of $A$. 

What about diffeomorphisms? We want to understand what the gauge symmetries are in the path integral over $h_{\mu\nu}$. For gravity, the diffeomorphism gauge symmetry acts on the full metric as\footnote{We use the subscript $\dots^{(g)}$ to label objects like covariant derivatives which are defined with respect to the full dynamical metric $g_{\mu\nu}$. Most indices and covariant derivatives in this work refer to the background Rindler cone metric $G_{\mu\nu}$.}
\begin{equation}\label{eqn:fullmetric_diffs}
g_{\mu\nu} \to g_{\mu\nu}+\pounds^{(g)}_{\kappa\epsilon}g_{\mu\nu} = g_{\mu\nu} + \kappa\nabla_{\mu}^{(g)}\epsilon_{\nu} + \kappa\nabla_{\nu}^{(g)}\epsilon_{\mu}\,.
\end{equation}
In the background field expansion \eqref{eqn:metricexp}, one \emph{defines} $G_{\mu\nu}$ as the part of $g_{\mu\nu}$ that scales as $\kappa^0$. With this definition of the dynamical quantum field $h_{\mu\nu}$, the redundancies \eqref{eqn:fullmetric_diffs} are simply
\begin{equation}\label{eqn:quantumdiffsymkappa0}
    \boxed{G_{\mu\nu} \to G_{\mu\nu}, \qquad h_{\mu\nu} \to h_{\mu\nu}+\nabla_{\mu}\epsilon_{\nu}+\nabla_{\nu}\epsilon_{\mu}+\mathcal{O}(\kappa^1)}\,.
\end{equation}
In the free graviton approximation, where one restricts to the quadratic action \eqref{eqn:EHactionexp_2ndorder} below, all positive powers of $\kappa$ in \eqref{eqn:quantumdiffsymkappa0} may be ignored. The transformations \eqref{eqn:quantumdiffsymkappa0} were called `quantum diffeomorphisms' in \cite{Bastianelli:2013tsa,Grisaru:1975ei}, and must be modded out in the free graviton path integral, which we will study throughout this work. The key point is that the transformations \eqref{eqn:quantumdiffsymkappa0} leave the background metric $G_{\mu\nu}$ unchanged, therefore the definition of a region $A$ that use coordinates and metric $G_{\mu\nu}$ in \eqref{2.2 G} is \emph{manifestly} invariant under the gauge symmetry \eqref{eqn:quantumdiffsymkappa0}.

We can contrast these `quantum diffeomorphisms' \eqref{eqn:quantumdiffsymkappa0} with the `classical diffeomorphisms' \cite{Bastianelli:2013tsa,Veltman:1975vx}, where one considers $\kappa\epsilon\sim \epsilon_\text{class}\sim \kappa^0$. Then, since $G_{\mu\nu}$ is the order $\kappa^0$ part of the metric $g_{\mu\nu}$, one finds instead the transformation
    \begin{align}\label{eqn:classicaldiffsymkappa0}
        G_{\mu\nu} &\to G_{\mu\nu}+\pounds_{\epsilon_\text{class}}^{(g)}G_{\mu\nu} = G_{\mu\nu}+\nabla_{\mu}\epsilon_{\text{class}\,\nu}+\nabla_{\nu}\epsilon_{\text{class}\,\mu} + \mathcal{O}(\kappa)\nonumber\\  h_{\mu\nu} &\to h_{\mu\nu}+\pounds_{\epsilon_\text{class}}^{(g)}h_{\mu\nu} = h_{\mu\nu}+\epsilon^{\rho}_\text{class}\nabla_{\rho}h_{\mu\nu}+h_{\rho\nu}\nabla_{\mu}\epsilon^{\rho}_\text{class}+h_{\mu\rho}\nabla_{\nu}\epsilon^{\rho}_\text{class} + \mathcal{O}(\kappa)\,.
    \end{align}
    The claim of \cite{Bastianelli:2013tsa,Veltman:1975vx} is that these coordinate transformations should not be gauged in the graviton path integral, just like coordinate choices in any covariant quantum field theory are not gauged.
    \footnote{Note also that \eqref{eqn:quantumdiffsymkappa0} already results in the correct number of graviton polarizations, further clarifying that there is not some huge redundancy being missed.}

The `background field expansion' \eqref{eqn:metricexp} is nothing but what one would do in quantum field theory, we expand around a classical saddle $G_{\mu\nu}$ and path integrate over fluctuations $h_{\mu\nu}$ around that saddle. Most confusions about defining subregions arise non-perturbatively in $\kappa$, which we do not study here.

We only consider entanglement entropy at $O(\kappa^{0})$. Because of this, we do not need to worry about fluctuations in the position of the entangling surface. At higher orders in $\kappa$, one can find an extremal surface in the perturbed geometry, and use this as the new entangling surface. However, it is always possible to go to a gauge for the graviton where the entangling surface is not perturbed, as discussed in detail in \cite{Colin-Ellerin:2025dgq}.

Next, we address a second related subtlety that arises in defining conical entropy \eqref{eqn:entropy} for gravitons.

\subsection{Going on-shell via tip-less cones}
\label{sec:onshell}
The graviton action is obtained by expanding the Einstein Hilbert action in powers of $\kappa$ using \eqref{eqn:metricexp}\footnote{Throughout this section we will ignore boundary terms for presentation purposes.}
\begin{equation}\label{eqn:EHaction}
    S_\text{EH} = -\frac{1}{\kappa^{2}}\int \d^{d}x\sqrt{g}\,R^{(g)}=\frac{1}{\kappa^{2}}S^{(0)}+\frac{1}{\kappa}S^{(1)}+S^{(2)}+\mathcal{O}(\kappa)\,.
\end{equation}
The higher order terms in $\kappa$ give interaction vertices for $h_{\mu\nu}$. In this work, as a first step, we study free graviton entanglement, thus ignoring such terms. The free theory does not suffer from the renormalization issues in quantizing general relativity.\footnote{The theory is indeed renormalizable in Minkowski space with the explicit counterterms computed by 't Hooft and Veltman \cite{t1974one}.} One obtains the following linear graviton action \cite{t1974one,Christensen:1979iy}
\begin{equation}\label{eqn:EHactionexp_1storder}
    S^{(1)} = \int \d^{d}x \sqrt{G}\, h^{\mu\nu}E_{\mu\nu}\,,\quad E_{\mu\nu} = R_{\mu\nu} - \frac{1}{2}G_{\mu\nu}R\,.
\end{equation}
This linear piece only vanishes when the background $G_{\mu\nu}$ is on-shell
\begin{equation}
    E_{\mu\nu}=0\,.\label{2.8 E}
\end{equation}
In other words, the graviton expansion \eqref{eqn:metricexp} only makes sense around background metrics $G_{\mu\nu}$ that solve the classical Einstein equations, as always when we intend to do perturbative quantum field theory. Other inconsistencies also arise if one would insist on trying to expand around off-shell `backgrounds', for instance the actions $S^{(1)}$ and $S^{(2)}$ would not be invariant under the gauge redundancy \eqref{eqn:quantumdiffsymkappa0}.

This raises an immediate problem with trying to compute graviton entanglement via conical entropy. The conical spacetime \eqref{eqn:Rindlercone} has a delta function curvature singularity at the tip of the cone $r=0$ \cite{Louko:1995jw}, and therefore solves the Einstein equations everywhere \emph{except} at $r=0$. To deal with this we will follow Kabat's method \cite{Kabat:1995eq}, where we instead study the graviton path integral on the Rindler cone with the tip \emph{deleted}. In other words, we consider $G_{\mu\nu}$ to be the metric for the spacetime \eqref{eqn:Rindlercone} with strictly $r>0$.\footnote{Smoothing out this conical singularity (as was done for photons in \cite{Nelson:1994na,Fursaev:1995ef,Frolov:1995xe,Donnelly:2014fua,Donnelly:2015hxa}) would not resolve this issue, because the smooth cone still does not solve the Einstein equations.} This tip-less cone \eqref{eqn:Rindlercone} is flat, and one obtains the simple quadratic, free graviton action \cite{t1974one,Christensen:1979iy}
\begin{equation}\label{eqn:EHactionexp_2ndorder}
    S^{(2)} = -\frac{1}{2}\int_\text{cone} \d^{d}x \, \sqrt{G}\left(\frac{1}{2}h^{\mu\nu}\Box h_{\mu\nu}-\frac{1}{2}h\Box h-\nabla_{\mu}h^{\mu\nu}\nabla_{\nu}h+\nabla_{\nu}h^{\nu\alpha}\nabla^{\mu}h_{\mu\alpha}\right)\,,
\end{equation}
where we defined the trace of the graviton as
\begin{equation}
    h=G^{\mu\nu}h_{\mu\nu}\,.
\end{equation}
This theory \eqref{eqn:EHactionexp_2ndorder} is a massless symmetric spin-$2$ quantum field propagating on a fixed background $G_{\mu\nu}$. It is the simplest action with quantum diffeomorphism symmetry \eqref{eqn:quantumdiffsymkappa0}. This action \eqref{eqn:EHactionexp_2ndorder} is sufficiently simple that we will be able to explicitly compute the partition function
\begin{align}
    Z_\text{graviton}(\beta) = \int \frac{\mathcal{D}h}{\text{Vol}(G_\text{diff})}e^{-S^{(2)}}\label{2.11 zgrav}
\end{align}
in section \ref{sect:kabatgraviton}, and extract from this the conical entropy \eqref{eqn:entropy}.

Before proceeding, let us make some comments about this procedure of deleting the tip of the cone. One would think that on the cone \emph{without} the tip, one may need to impose boundary conditions at the tip for the theory to be well-defined. We only demand that the graviton modes are regular at $r=0$ \cite{KayStuder}. The fact that this leads to a mathematically well-defined theory is diagnosed via the inner product on modes, and as we show in appendix \ref{app:modesortho}, smoothness is sufficient for graviton mode orthogonality so long as $\beta<2\pi$. Morally speaking, we do not restrict graviton fluctuations at the tip in any significant way, therefore we expect to capture all degrees of freedom.\footnote{Smoothness would indeed also be a necessary condition on the modes in Minkowski at $r=0$.}

Instead of using the conical entropy, one might have thought one could compute the entanglement entropy via the replica trick \cite{Callan:1994py,Larsen:1995ax} where one considers the multiple cover geometry $\beta=2\pi n$ in \eqref{eqn:entropy} and then analytically continues in $n$ to $n=1$. This replica geometry has a conical excess at $r=0$ so the spacetime still suffers from a curvature singularity at the tip of the cone.
For free graviton entanglement one would want to again delete the tip, however, this would not be consistent as such spacetimes do not satisfy $\beta<2\pi$. We conclude that one cannot study free graviton entropy using the replica trick.

In the photon case, other mathematical options are available besides removing the tip of the cone, such as smoothing out the singularity \cite{Nelson:1994na,Fursaev:1995ef,Frolov:1995xe,Donnelly:2014fua,Donnelly:2015hxa}. In that case, all methods agree on the Rindler entropy. That answer has an elegant Lorentzian interpretation in terms of edge modes and large gauge transformations \cite{Donnelly:2014fua,Donnelly:2015hxa,Blommaert:2018rsf,Blommaert:2018oue,Ball:2024hqe}. As we discuss in section \ref{sect:interpret}, our results for the graviton path integral on the tip-less cone also suggest a natural interpretation in terms of edge modes. We view the fact that our answer \eqref{eqn:gravcontact} has a `sensible' physical interpretation as a posteriori motivation for this procedure.

These statements remain true if one wants to compute the conical entropy for non-flat backgrounds. One must still delete the tip for the expansion around $G_{\mu\nu}$ to make sense. What changes is the quadratic action \eqref{eqn:EHactionexp_2ndorder} which becomes more complicated.\footnote{Explicitly, one obtains the following additional quadratic action for non-flat metrics
\begin{align}\label{eqn:EHactionexp_2ndorder_gen}
\begin{split}
    S^{(2)}_\text{generic} &=S^{(2)} -\frac{1}{2}\int \d^{d}x \, \sqrt{G}\bigg(\frac{1}{4}h^{2}R - h h^{\mu\rho}R_{\mu\rho}+\frac{1}{2}h_{\alpha\beta}h^{\alpha\beta}R+h^{\nu\rho}{h^{\beta}}_{\omega}{R_{\nu\beta\rho}}^{\omega}-{h_{\rho}}^{\alpha}h^{\rho\nu}R_{\nu\alpha}\bigg)\,.
\end{split}
\end{align}}

\section{Photons revisited}\label{sect:photons}
We now review the essential ingredients of Kabat's calculation of the conical entropy for photons \cite{Kabat:1995eq}, paving the way for the graviton calculation in section \ref{sect:kabatgraviton}. First, in \textbf{section \ref{sect:scalar}}, we review the calculation of the massless scalar partition function on the Rindler cone, as this will feature prominently in both the photon and in the graviton partition function. In \textbf{section \ref{sect:photon}}, we discuss the decomposition of the photon path integral on the Rindler cone into a ratio of determinants of a massless vector and of two massless scalars. Finally, in \textbf{section \ref{sect:photonthermo}}, we review Kabat's computation of the photon heat kernel and the resulting contact term contribution to the photon entanglement entropy.

\subsection{Scalar heat kernel}\label{sect:scalar}
Consider the partition function of a scalar field on the Rindler cone \eqref{eqn:Rindlercone}
\begin{align}
    Z_\text{scal}(\beta)&=\int\mathcal{D} \phi\, \exp\bigg(- \frac{1}{2}\int_{\text{cone}}\d^{d}x\,\sqrt{g}\left(\nabla^{\mu}\phi\nabla_{\mu}\phi+m^{2}\phi^{2}\right)\bigg)\nonumber\\&=\int\mathcal{D} \phi\, \exp\bigg(- \frac{1}{2}\int_{\text{cone}}\d^{d}x\,\sqrt{g}\,\phi\left(-\Box+m^{2}\right)\phi\bigg)\,,\label{3.1zzzz}
\end{align}
where the d'Alembertian is given by 
\begin{equation}
\Box=\partial_\mathbf{x}^2+\frac{1}{r^2}\partial_\tau^2+\frac{1}{r}\partial_r\, r\partial_r\,.
\end{equation}
In going from the first line to the second line of \eqref{3.1zzzz} one has to be careful about boundary terms. The boundary terms may be ignored if we assume regularity of the eigenfunctions at the co-dimension 2 tip of the cone, taking into account the fact that we remove the tip of the cone. Indeed, regularity is required for orthonormality in the absence of any particular boundary conditions (coming from boundary terms in the action), as we detail more in appendix \ref{app:modesortho}.

This path integral is computed by diagonalization of the operator $-\Box+m^2$
\begin{equation}
    \phi=\sum_\lambda c_\lambda \phi_\lambda\,,\quad \left(-\Box+m^2\right)\phi_\lambda=\lambda\, \phi_\lambda \,,\quad \int_\text{cone}\d^d x\sqrt{g}\,\phi^*_{\lambda_i}\phi_{\lambda_2}=\delta_{\lambda_1\,\lambda_2}\,,
\end{equation}
leading to the path integral measure
\begin{equation}
    \int\mathcal{D} \phi=\prod_\lambda \int_{-\infty}^{+\infty}\d c_\lambda\,.
\end{equation}
This gives the standard result
\begin{equation}
    Z_\text{scal}(\beta)=\prod_\lambda \lambda^{-1/2}=\det(-\Box+m^2)^{-1/2}\,.\label{2.3 zscal}
\end{equation}
On a technical level, it is easier to compute traces than determinants of operators, making the following rewriting useful (from hereon we will specify to massless scalars $m^2=0$)\footnote{The integral representation of the logarithm works only for positive eigenvalues so one should be careful with exact zero modes of the d'Alembertian. Such zero modes will not play an important role in our work, and we will implicitly ignore them in all ensuing equations.}
\begin{equation}
    \log Z_\text{scal}(\beta)=-\frac{1}{2}\Tr \log(-\Box)=\int_0^\infty \frac{\d s}{2s}\Tr e^{s\,\Box}=\int_0^\infty \frac{\d s}{2s}\int_\text{cone} \d^d x \sqrt{g}\,K_\text{scal}(x|s)\,.\label{2.4 zscal}
\end{equation}
This leads one to compute the scalar heat kernel $K_\text{scal}(x|s)$
\begin{equation}
    K_\text{scal}(x|s)=\sum_\lambda e^{-s\lambda}\phi_\lambda(x)\phi_\lambda(x)^*\,,
\end{equation}
the trace of which matches the partition function of a quantum particle on the Rindler cone\footnote{Indeed, going to phase space variables, the Schrodinger equation for this quantum mechanics reduces to $-\Box\, \phi_E=E\phi_E$.}
\begin{equation}
    \int_\text{cone} \d^d x \sqrt{g}\,K^\text{scal}(x|s)=\int \mathcal{D}x\exp\bigg(-\frac{1}{4}\int_0^s\d \tau\,g_{\mu\nu}\frac{\d x^\mu}{\d\tau}\frac{\d x^\nu}{\d\tau}\bigg)\,,\label{2.6 first quantized}
\end{equation}
whereas the heat kernel $K_\text{scal}(x|s)$ is a propagator from $x$ to $x$ in this language. Hence \eqref{2.4 zscal} is the usual Feynman diagram rule that the partition function of a free field is an exponential of particle loops. For a comprehensive review of heat kernel methods, see \cite{Vassilevich:2003xt}.

The orthonormal eigenfunctions of $-\Box$ on the Rindler cone \eqref{eqn:Rindlercone} are identified as Bessel functions\footnote{Bessel functions satisfy $J_{\nu}(e^{i m\pi}z) = e^{i m\pi\nu}J_{\nu}(z)$ so $k<0$ eigenfunctions are not linearly independent from $k>0$ ones.}
\begin{align}\label{eqn:scalarLap_eigfns}
    -\Box\,\phi_{k,\ell,\mathbf{k}}(x)=(k^2+\mathbf{k}^2)\,\phi_{k,\ell,\mathbf{k}}(x)\,,\quad \phi_{k,\ell,\mathbf{k}}(x) = \sqrt{\frac{k}{(2\pi)^{d-2}\beta}}e^{\frac{2\pi i \ell}{\beta}\tau}e^{i \mathbf{k}\cdot \mathbf{x}}\,J_{\frac{2\pi |\ell|}{\beta}}(kr)\,,\quad k>0\,.
\end{align}
We choose boundary conditions such that the eigenfunctions are regular at the origin,\footnote{For clarity, regularity at a point (in this case the cone) is defined in the same way it would be for functions on say the complex plane. There are no poles in the Taylor series. In this case it means the function goes as $r^\alpha$ with $\alpha\geq0$ for $r\to 0$.} which is known as the Friedrichs extension on the cone \cite{KayStuder}. Using these modes, one computes the propagator
\begin{equation}
    K_\text{scal}(x\rvert s)=\sum_{l=-\infty}^{\infty}\int_{0}^{\infty}\d k\,\int_{-\infty}^{+\infty} \d\mathbf{k}\,e^{-s(k^{2}+\mathbf{k}^2)}|\phi_{k,\ell,\mathbf{k}}(x)|^{2}=\frac{2\pi}{\beta} \frac{1}{(4\pi s)^{d/2}}\,e^{-\frac{r^{2}}{2s}}\sum_{l=-\infty}^{\infty}I_{\frac{2\pi |\ell|}{\beta}}\left(\frac{r^{2}}{2s}\right)\,,\label{2.8 K}
\end{equation}
where the integral over $k$ is evaluated using formula 6.633.2 in \cite{gradshteyn2014table}. A detailed evaluation of the sum over $\ell$ is presented in appendix \ref{app:scalarkernel}. One obtains
\begin{align}
    K_\text{scal}(x\rvert s)=\,&\frac{1}{(4\pi s)^{d/2}}\sum_{\abs{n}\beta<\pi}e^{-\frac{r^2}{s}\sin(\frac{n\beta}{2})^2}\nonumber\\&\qquad-\frac{1}{(4\pi s)^{d/2}}\,\frac{1}{\beta}\sin(\frac{2\pi^2}{\beta})\int_{-\infty}^{+\infty}\d y\,\frac{e^{-\frac{r^2}{s} \cosh(\frac{y}{2})^2}}{\cosh(\frac{2\pi y}{\beta})-\cos(\frac{2\pi^2}{\beta})}\,.\label{2.9 Kres}
\end{align}
Despite appearances, this function is analytic for $0<\beta<2\pi$, with the sum over $n$ accounting for poles in the integral that cross the $y$ integration contour at $\beta=\pi/\abs{m}$. The relevant properties of this heat kernel for our work are its trace
\begin{equation}
    \int_\text{cone} \d^d x \sqrt{g}\,K_\text{scal}(x|s)=\frac{1}{(4\pi s)^{d/2}}\,\int_\text{cone} \d^d x \sqrt{g}\,+\frac{A}{(4\pi s)^{d/2}}\bigg(\frac{2\pi^2}{3}\frac{s}{\beta}-\frac{s\beta}{6}\bigg)\,,\label{2.10 traceK}
\end{equation}
and its first two Taylor expansion coefficients around the Rindler horizon (or the tip of the cone) $r=0$\footnote{Around $r=\infty$ the only term is the constant, which can be computed for instance by replacing the sum over $\ell$ by an integral and using the Bessel asymptotics in \eqref{2.8 K}.}
\begin{equation}
    K_\text{scal}(0|s)=\frac{1}{(4\pi s)^{d/2}}\,\frac{2\pi}{\beta}\,,\quad K_\text{scal}(\infty|s)=\frac{1}{(4\pi s)^{d/2}}\,,\quad \frac{1}{r}\partial_r K_\text{scal}(0|s)=-\frac{1}{(4\pi s)^{d/2}}\,\frac{2\pi}{\beta}\frac{1}{s}\,.\label{2.11 series}
\end{equation}
Here $A$ is the area of the $(d-2)$-dimensional Rindler horizon. The derivations of \eqref{2.10 traceK} and \eqref{2.11 series} can be found in appendix \ref{app:scalarkernel}. From \eqref{2.10 traceK}, one then finds the scalar partition function on the cone \eqref{2.4 zscal}\footnote{Following conventions we introduced a UV cutoff $\epsilon$ to make this integral finite, see \cite{Donnelly:2014fua,Donnelly:2015hxa} for more details.}
\begin{equation}
    \log Z_\text{scal}(\beta)=\frac{1}{\beta}\frac{\pi^2}{3}\, \bigg(1-\left(\frac{\beta}{2\pi}\right)^{2}\bigg) A\int_{\epsilon^{2}}^\infty \frac{\d s}{(4\pi s)^{d/2}} \,.\label{2.12 zcal}
\end{equation}
The terms linear in $\beta$ represent a (non-physical) shift in the ground state energy and do not contribute to the conical entropy \eqref{eqn:entropy}. The term proportional to $1/\beta$ gives the known scalar entanglement entropy between the left and right Rindler wedges in the Minkowski vacuum
\begin{equation}
    S_\text{scal}=2\pi A\int_{\epsilon^{2}}^\infty \frac{\d s}{(4\pi s)^{d/2}}\,\frac{1}{6}\,.\label{2.13 sscal}
\end{equation}

\subsection{Path integral decomposition}\label{sect:photon}
Next, we compute the path integral of Maxwell theory on the Rindler cone \eqref{eqn:Rindlercone}. The partition function of Maxwell theory is defined as
\begin{equation}
    Z_\text{photon}(\beta)=\int \frac{\mathcal{D}A}{\text{Vol}(G)}\exp\bigg( -\frac{1}{4}\int_{\text{cone}}\d^{d}x\sqrt{g}\,F_{\mu\nu}F^{\mu\nu}\bigg)\,,\label{2.14 lagr}
\end{equation}
where $F_{\mu\nu} = \nabla_{\mu}A_{\nu}-\nabla_{\nu}A_{\mu}$. The $U(1)$ gauge symmetry consists of a scalar field $\a$ with $A_\m\sim A_\m+\nabla_\m \a$. 

We will fix this gauge symmetry using the standard $R_{\xi}$-gauge-fixing procedure for the path integral, see for instance \cite{Donnelly:2013tia}. The Faddeev-Popov determinant $\Delta^{\mathrm{FP}} = \det(-\Box)$ satisfies the following identity\footnote{We will ignore zero modes $k=\mathbf{k}=0$ throughout.}
\begin{equation}\label{eqn:photonFPdetidentity}
    \det(-\Box)\int \mathcal{D}\beta\prod_{\lambda}\delta((\nabla^\m A_\m-\Box \b)_\lambda+\omega_\lambda)=1\,.
\end{equation}
Here, $\omega$ is an auxiliary scalar field. Indeed, for each of the modes $\beta_\lambda$ the integral of the delta function gives a factor $1/\lambda$ which cancels with the factor $\det(-\Box)$ in front. One may then perform a Gaussian integral over the expansion coefficients of the auxiliary scalar field $\omega$\footnote{The Gaussian integral over $\omega$ gives a constant proportional to the volume of spacetime and therefore does not contribute to entropy \cite{Donnelly:2013tia}.\label{fn:constant}}
\begin{align}
    1&=\det(-\Box)\int \mathcal{D}\beta\int \mathcal{D}\omega\,\exp\bigg(-\frac{1}{2\xi}\int_\text{cone} \d^{d}x \sqrt{g}\,\w^2\bigg)\prod_{\lambda}\delta((\nabla^\m A_\m-\Box \b)_\lambda+\omega_\lambda)\nonumber\\
    &=\det(-\Box)\int \mathcal{D}\b\, \exp\bigg(-\frac{1}{2\xi}\int_\text{cone} \d^d x \sqrt{g}\,(\nabla^\m(A_\m-\nabla_\m \b ))^2\bigg).
\end{align}

We can insert this identity into the partition function \eqref{2.14 lagr} to obtain the gauge-fixed path integral
\begin{equation}
    Z_\text{photon}(\beta)=\det(-\Box)\int \mathcal{D}A\,\exp\bigg( -\int_{\text{cone}}\d^{d}x\,\sqrt{g}\bigg(\frac{1}{4}F_{\mu\nu}F^{\mu\nu}+\frac{1}{2\xi}\nabla^\mu A_\mu\nabla^\nu A_\nu\bigg)  \bigg)\int\frac{\mathcal{D\beta}}{\text{Vol}(G)}\,.\label{2.17}
\end{equation}
Here, we have simply shifted $A_\m\to A_\m+\nabla_\m \b$ within the $A$ integral, which is a symmetry of the original Maxwell Lagrangian \eqref{2.14 lagr} and removes all $\beta$ dependence from the action. We can then use the formal identity
\begin{equation}
    \int\frac{\mathcal{D\beta}}{\text{Vol}(G)}=1\,,
\end{equation}
since the action of the $U(1)$ gauge symmetries corresponds to the path integral over a scalar field $\alpha$. By construction, the path integral is independent of $\xi$. This discussion reduces to the usual Faddeev-Popov procedure associated with gauge-fixing $\nabla^\mu A_\mu = 0$ for $\xi=0$ where the path integral \eqref{2.17} localizes on this modulus. However, for our purposes it is more convenient to choose $\xi=1$. After doing integration by parts, one arrives at
\begin{align}
     Z_\text{photon}(\beta)&=\det(-\Box)\int \mathcal{D}A\,\exp\bigg(-\frac{1}{2}\int_\text{cone} \d^d x \sqrt{g}\,A^\mu (-\Box\, \tensor{\delta}{_\mu^\nu}+\tensor{R}{_\mu^\nu})A_\nu \bigg)\nonumber\\
     &=\det(-\Box)\det(-\Box\, \delta^\mu_\nu )^{-1/2}\,.\label{2.aa zphot}
\end{align}
Here we used the fact that we compute $Z(\beta)$ on a cone \eqref{eqn:entropy} with the tip removed. Away from the tip of the cone, the Rindler cone is flat such that $R_{\mu\nu}=0$. See also section \ref{sec:makingsense}. We conclude from \eqref{2.aa zphot} that
\begin{equation}
    \log Z_\text{photon}(\beta)=\log Z_\text{vec}(\beta)-2\log Z_\text{scal}(\beta)\,,\label{2.24 zphot}
\end{equation}
where $Z_\text{scal}(\beta)$ represents, according to \eqref{2.3 zscal}, the scalar determinant in \eqref{2.aa zphot}. The vector path integral representing the vector determinant in \eqref{2.aa zphot} is
\begin{equation}
    Z_\text{vec}(\beta)=\int \mathcal{D}A \, \exp\bigg(-\frac{1}{2}\int_\text{cone}\d ^d x\,\sqrt{g}\, \nabla^\nu A^\mu \nabla_\nu A_\mu\bigg)=\det(-\Box\, \delta^\mu_\mu)^{-1/2}\,.
\end{equation}
Following the same steps that led to \eqref{2.4 zscal}, one rewrites this path integral as
\begin{equation}
    \log Z_\text{vec}(\beta)=\sum_\lambda \int_0^\infty \frac{\d s}{2s}\,e^{-s\lambda}=\int_0^\infty \frac{\d s}{2s}\int_\text{cone}\d^d x\sqrt{g}\,K_\text{vec}(x|s)\,,\label{2.26 logzvector}
\end{equation}
where we have introduced the vector heat kernel defined as
\begin{equation}
    K_\text{vec}(x|s)=\sum_\lambda e^{-s\lambda}g^{\mu\nu}\phi_{\lambda\,\mu}(x)^*\phi_{\lambda\,\nu}(x)\,,
\end{equation}
which features the complete set of orthonormal eigenfunctions of the vector d'Alembertian $\Box$
\begin{equation}
    -\Box\, \phi_{\lambda\,\mu}=\lambda\, \phi_{\lambda\,\mu}\,,\quad \int_\text{cone}\d^d x\sqrt{g}\,g^{\mu\nu}\phi_{\lambda_1\,\mu}^*\phi_{\lambda_2\,\nu}=\delta_{\lambda_1\,\lambda_2}\,.
\end{equation}
The vector heat kernel has an interpretation akin to \eqref{2.6 first quantized} where one adds a polarization for the particle, see for instance \cite{Bastianelli:2013tsa}.

\subsection{Heat kernel and contact terms}\label{sect:photonthermo}
We now proceed to construct the vector eigenfunctions required to evaluate the vector heat kernel. We claim that for generic tensor fields $T_{\mu_1\mu_2\dots}$ a basis of eigenfunctions of the d'Alembertian $\Box$
\begin{equation}\label{eqn:tensoreigenfneqn}
    -\Box\, \phi_{\lambda\,\m_1\m_2\dots}=\lambda \,\phi_{\lambda\,\m_1\m_2\dots}
\end{equation}
consists of the following tensor modes (with $\alpha_i=0,1,a$ indices taking $d$ values):
\begin{equation}
    \phi_{\lambda\,\m_1\m_2\dots}^{(\a_1\a_2\dots)}=e^{(\a_1)}_{\m_1} e^{(\a_2)}_{\m_2}\dots \phi_\lambda\,.\label{2.30 modesgen}
\end{equation}
Here, $\phi_\lambda$ are the eigenmodes of the scalar d'Alembertian \eqref{eqn:scalarLap_eigfns} with identical eigenvalues $\lambda=k^2+\mathbf{k}^2$, and we introduced the following `operators' $e_\mu^{(\a)}$ (which one should think of as orthonormal vectors):
\begin{align}\label{eqn:basis}
    e^{(a)}_\m =(0,0,n^a)\,,\quad e^{(0)}_\m=\frac{1}{k}\nabla_\mu^{\text{2d}}\,,\quad e^{(1)}_\m=\frac{1}{k}\epsilon_{\m\n}^{\text{2d}}\nabla^{\n\,\text{2d}}\,.
\end{align}
The vectors $\{n^a\}$ form a basis in the $(d-2)$-dimensional space spanned by the coordinates $x_i$. We also introduced the superscript $\dots^{\text{2d}}$ for objects referring to the remaining 2d piece of the Rindler cone
\begin{equation}
    \d s^{2\,\text{2d}}=r^2\d \tau^2+\d r^2\,,\quad \tau\sim \tau+\beta\,.
\end{equation}
Furthermore, $\epsilon_{\m\n}^{\text{2d}}$ is the 2d Levi-Cevita tensor with $\epsilon_{\tau r}^\text{2d}=\sqrt{g}$. This simple decomposition \eqref{2.30 modesgen} relies on the fact that the $d-2$ transverse directions $x_i$ in the Rindler cone decouple, in that the only non-zero Christoffel symbols have $\tau$ and $r$ indices. It is a mild modification of modes discussed in section 5 and 7 of \cite{Blommaert:2018rsf}. The proof goes as follows. Firstly, the modes \eqref{2.30 modesgen} are all solutions because $\phi_\lambda$ are scalar eigenmodes of $\Box$ and this operator commutes with the Levi-Cevita tensor $\varepsilon_{\m\n}^{\text{2d}}$ and with $\nabla_\mu^{\text{2d}}$. Secondly, we show in appendix \ref{app:modesortho} that the modes are orthonormal. Thirdly, they are complete since the number of polarizations is correct.

The vector heat kernel with $\Box$ spectrum $\phi^{(\a)}_{\lambda\,\mu}$ involves therefore a sum over polarizations $\alpha$
\begin{equation}
    K_\text{vec}(x\rvert s)=\sum_{\alpha}\sum_{\ell=-\infty}^{\infty}\int_{0}^{\infty}\d k\,\int_{-\infty}^{+\infty} \d\mathbf{k}\,e^{-s(k^{2}+\mathbf{k}^2)}g^{\mu\nu}\phi^{(\alpha)}_{\mu\,k,\ell,\mathbf{k}}(x)^*\phi^{(\alpha)}_{\nu\,k,\ell,\mathbf{k}}(x)\,.\label{3.29 vecKdef}
\end{equation}
The contribution from each of the $d-2$ polarizations $\alpha=a$ immediately reduces to a scalar heat kernel \eqref{2.8 K}. The polarizations $\alpha=0,1$ require more work. One obtains
\begin{align}
    K_\text{vec}(x\rvert s) &= (d-2)K_\text{scal}(x\rvert s)+2\int \d^{d}x\,\sqrt{g}\sum_{\ell=-\infty}^{\infty}\int_{0}^{\infty}\frac{\d k}{k^{2}}\,\int \d\mathbf{k}\,e^{-s(k^{2}+\mathbf{k}^2)}\nabla^{\mu\,\text{2d}}\phi_{k,\ell,\mathbf{k}}(x)^*\nabla_{\mu}^\text{2d}\phi_{k, \ell, \mathbf{k}}(x)\nonumber\\&=(d-2)K_\text{scal}(x\rvert s)+\frac{1}{(4\pi s)^{d/2-1}}K_\text{vec}^\text{2d}(x\rvert s)\,.\label{eqn:vectorheatkernel}
\end{align}
In order to streamline calculations, we have introduced the vector propagator in 2d denoted $K_\text{vec}^\text{2d}(x\rvert s)$. As the heat kernel in fact only depends on the $r$ coordinate we can henceforth write $K_\text{vec}^\text{2d}(r\rvert s)$. It can be evaluated explicitly and written in terms of the scalar heat kernel as follows:
\begin{align}\label{eqn:Kvec}
    K^{\text{2d}}_\text{vec}(r|s)&=2\sum_{\ell=-\infty}^{+\infty}\int_0^\infty \d k\, e^{-s k^2}\,\frac{1}{k^2} \nabla^{\mu\,\text{2d}}\phi^\text{2d}_{k, \ell}(x)^*\nabla_\mu^\text{2d} \phi^{\text{2d}}_{k, \ell}(x)\nonumber\\
    &=2\sum_{\ell=-\infty}^{+\infty}\int_0^\infty \d k\, e^{-s k^2}\phi_{k, \ell}^\text{2d}(x)^*\phi^\text{2d}_{k, \ell}(x)+\Box^{\text{2d}}\sum_{\ell=-\infty}^{+\infty}\int_0^\infty \d k\, e^{-s k^2}\frac{1}{k^2} \phi_{k, \ell}^\text{2d}(x)^*\phi_{k, \ell}^\text{2d}(x) \nonumber\\
    &=2K^{\text{2d}}_\text{scal}(r\rvert s)+\int_s^\infty \d p\,\Box^\text{2d}\,K^\text{2d}_\text{scal}(r\rvert p) \nonumber\\
    &=2K^\text{2d}_\text{scal}(r\rvert s)+\frac{2s}{r}\partial_{r}K^\text{2d}_\text{scal}(r\rvert s)\,.
\end{align}
To obtain the second equality, one can write out the action of the scalar $\Box^{\text{2d}}$ on the product of wavefunctions $\phi_{k, \ell}^\text{2d}(x)$ and their conjugates and use the fact that each individually is an eigenfunction of $\Box^{\text{2d}}$ with eigenvalue $-k^2$. For the third equality, we wrote $e^{-sk^2}/k^2$ as the result of the $p$ integral of $e^{-p k^2}$, and did the $k$ integral first. The last equality uses the fact that the explicit expression for $K^\text{2d}_\text{scal}(r\rvert s)$ in \eqref{2.8 K} satisfies
\begin{equation}\label{eqn:rtos}
    \Box^\text{2d}\,K^\text{2d}_\text{scal}(r\rvert s)=\frac{1}{r}\partial_r\,r\partial_r \,K^\text{2d}_\text{scal}(r\rvert s)\,,\quad r\partial_r \,K^\text{2d}_\text{scal}(r\rvert s)=-2\partial_s\,s K^\text{2d}_\text{scal}(r\rvert s)\,.
\end{equation}
The second equation above allows one to carry out the $p$ integral using integration by parts.\footnote{A corollary of the expression \eqref{eqn:Kvec} for the heat kernel is that by inserting the radial Taylor series of the scalar heat kernel \eqref{2.11 series} around $r=0$ one finds
\begin{equation}
    K^{\text{2d}}_\text{vec}(0|s)=0\,.\label{3.33 property}
\end{equation}
This will be useful in section \ref{sect:kabatgraviton}.} 

Inserting the 2d expression \eqref{eqn:Kvec} into  \eqref{eqn:vectorheatkernel} results in the vector heat kernel
\begin{equation}
    K_\text{vec}(r|s)=d\, K_\text{scal}(r|s)+\frac{2s}{r}\partial_r\,K_\text{scal}(r|s)\,.\label{3.33 Kvec}
\end{equation}
To compute the vector path integral \eqref{2.26 logzvector} one needs the trace of this heat kernel\footnote{The Rindler horizon is technically $(d-2)$-dimensional, we will be somewhat loose with what we call horizon throughout. In this case, due to the flat measure, the integral over the time direction is responsible for a factor of $\beta$.}
\begin{equation}
    \int_\text{cone} \d^d x \sqrt{g}\,K_\text{vec}(x|s)=d \int_\text{cone} \d^d x \sqrt{g}\,K_\text{scal}(x|s)-2s \int_\text{horizon} \d^{d-1} x \,K_\text{scal}(x|s)\,, \label{2.38 kvectrace}
\end{equation}
resulting in
\begin{equation}
    \log Z_\text{vec}(\beta)=d \int_{\epsilon^{2}}^\infty \frac{\d s}{2s}\int_\text{cone} \d^d x \sqrt{g}\,K_\text{scal}(x|s)-\int_{\epsilon^{2}}^\infty \d s\int_\text{horizon} \d^{d-1} x \,K_\text{scal}(x|s)\,.
\end{equation}
Accounting for the ghost correction in \eqref{2.24 zphot} gives the final result for the photon path integral on the Rindler cone
\begin{align}
    \boxed{\log Z_\text{photon}(\beta)=(d-2) \int_{\epsilon^{2}}^\infty \frac{\d s}{2s}\int_\text{cone} \d^d x \sqrt{g}\,K_\text{scal}(x|s)-\int_{\epsilon^{2}}^\infty \d s\int_\text{horizon} \d^{d-1} x \,K_\text{scal}(x|s)}\,.\label{3.39 Kphotcont}
\end{align}
As stressed by Kabat \cite{Kabat:1995eq}, this is the partition function of $d-2$ scalar fields, which one would associate with the $d-2$ physical bulk polarizations of the photon, but with an additional (negative) contact term contribution\footnote{Technically, there is also a contribution coming from $r=\infty$ when we integrate by parts in \eqref{2.38 kvectrace} which has an additional factor $-\beta/2\pi$. Such terms proportional to $\beta$ are not physical and do not contribute to entropy. However, these terms have the property that at the smooth point $Z(\beta=2\pi)$ is the expected answer without any contact term.}
\begin{equation}
    \log Z_\text{photon}(\beta)=(d-2)\log Z_\text{scal}(\beta)+\log Z^\text{contact}_\text{photon}\,,
\end{equation}
where the interesting contact term is explicitly
\begin{equation}\label{eqn:Zcontact}
    \log Z^\text{contact}_\text{photon}= -\int_{\epsilon^{2}}^\infty \d s\int_\text{horizon} \d^{d-1} x \,K_\text{scal}(x|s)=-2\pi A\int_{\epsilon^{2}}^\infty \frac{\d s}{(4\pi s)^{d/2}}\,.
\end{equation}
As this is independent of $\beta$, it corresponds to zero energy excitations and contributes an extra overall (negative) degeneracy to the density of states. Combined with \eqref{2.13 sscal}, one finds the photon entanglement entropy in the Minkowski vacuum
\begin{equation}
    \boxed{S_\text{photon}=2\pi A \int_{\epsilon^{2}}^\infty \frac{\d s}{(4\pi s)^{d/2}}\bigg(\frac{1}{6}(d-2)-1\bigg)}\,.
\end{equation}
We will discuss various interpretations of this contact term in section \ref{sect:interpret} after first presenting a similar technical analysis for graviton entanglement in the Minkowski vacuum.

\section{Gravitons}\label{sect:kabatgraviton}
We now repeat this analysis for gravitons, following the preliminary work of section \ref{sec:makingsense}. In particular, we will compute the free graviton path integral \eqref{2.11 zgrav} on the Rindler cone \eqref{eqn:Rindlercone}. First, in \textbf{section \ref{sect:4.1 gravPI}}, we rewrite this path integral as a simple product of determinants. The new ingredient is the determinant of a massless symmetric tensor field on the Rindler cone. In \textbf{section \ref{sect:4.2 gravK}}, we compute this determinant via a direct evaluation of the associated heat kernel. Finally, in \textbf{section \ref{sect:4.3 gravContact}}, we obtain the resulting entanglement entropy in the right Rindler wedge, and identify the graviton contact term. The physical interpretation of this new contact term is discussed separately in section \ref{sect:interpret}.

\subsection{Path integral decomposition}\label{sect:4.1 gravPI}

Following the discussion of section \ref{sec:makingsense}, we seek to compute the free graviton path integral on the Rindler cone \eqref{eqn:Rindlercone}, i.e., the quadratic part of the perturbative expansion of the Einstein-Hilbert action
\begin{align}
    &Z_\text{graviton}(\beta) = \int \frac{\mathcal{D}h}{\text{Vol}(G_{\mathrm{diff}})} \exp\left(-S^{(2)}\right) \nonumber
    \\&= \int \frac{\mathcal{D}h}{\text{Vol}(G_{\mathrm{diff}})}\exp\bigg(\frac{1}{2}\int_{\text{cone}} \d^{d}x \, \sqrt{G}\left(\frac{1}{2}h^{\nu\alpha}\nabla^{2}h_{\nu\alpha}-\frac{1}{2}h\nabla^{2}h-\nabla_{\nu}h^{\nu\alpha}\nabla_{\alpha}h+\nabla_{\nu}h^{\nu\alpha}\nabla^{\mu}h_{\mu\alpha}\right)\bigg)\,.\label{eqn:gravitonpartfn}
\end{align}
As discussed in section \ref{sec:makingsense}, the gauge symmetry $G_{\mathrm{diff}}$ in question is the `quantum diffeomorphisms' which consist of a vector field $\varepsilon_\mu$ of degrees of freedom that transform the graviton $h_{\mu\nu}$ as \eqref{eqn:quantumdiffsymkappa0}
\begin{equation}
    h_{\mu\nu}\to  h_{\mu\nu}+\nabla_{\mu}\epsilon_{\nu}+\nabla_{\nu}\epsilon_{\mu}\,.
\end{equation}
As for Maxwell theory in section \ref{sect:photon}, we deal with the gauge symmetry in the path integral using the $R_{\xi}$-gauge fixing procedure. Our choice of gauge-fixing is related to the de Donder gauge
\begin{equation}\label{eqn:deDondergauge}
    \nabla^{\mu}\left(h_{\mu\nu}-\frac{1}{2}G_{\mu\nu}h\right) = 0\,.
\end{equation}
The Faddeev-Popov determinant is  $\Delta^{\text{FP}}[h] = \det(-\Box\, \delta^\sigma_\n-R_\n^\s)$. We proceed by inserting the following identity
\begin{equation}\label{eqn:Rxigaugeidentity}
    \det(-\Box\, \delta^\sigma_\n-R_\n^\s)\int \mathcal{D}\eta\, \delta(\nabla^\mu (h_{\m\n}-\frac{1}{2} G_{\m\n}h)-(\Box \,\delta^\sigma_\n+R_\n^\s)\eta_\s+\w_\n)=1
\end{equation}
into the graviton path integral and shift $h_{\mu\nu} \to h_{\mu\nu}+\nabla_{\mu}\eta_{\nu}+\nabla_{\nu}\eta_{\mu}$. As this is a symmetry of the original action \eqref{eqn:gravitonpartfn} this removes any $\eta_\mu$ dependence from the integrand. The path integral over the vector $\eta_{\mu}$ then cancels the $\text{Vol}(G_{\mathrm{diff}})$, which represented a path integral over a similar vector $\varepsilon_\mu$. Finally, we may perform the following Gaussian integral over $\omega_{\mu}$\footnote{This is modulo irrelevant normalization factors, see footnote \ref{fn:constant}.}
\begin{equation}\label{eqn:omegaGaussian}
    \int \mathcal{D}\omega\,\exp\bigg(-\frac{1}{2\xi}\int_\text{cone} \d^{d}x \sqrt{G}\,G^{\mu\nu}\omega_{\mu}\omega_{\nu}\bigg) = 1\,,
\end{equation}
resulting in the following expression for the graviton path integral:
\begin{align}\label{eqn:gravitygaugefixedPI}
   Z_\text{graviton}(\beta) &= \det(-\Box \delta^\sigma_\n)\int \mathcal{D}h\,\exp\bigg(-\frac{1}{2\xi}\int_\text{cone} \d^{d}x\, \sqrt{G}\, \nabla_{\mu}\left(h^{\mu\nu}-\frac{1}{2}G^{\mu\nu}h\right)\nabla^{\alpha}\left(h_{\nu\alpha}-\frac{1}{2}G_{\nu\alpha}h\right)\bigg)  \nonumber
   \\ & \qquad \times \exp\bigg(\frac{1}{2}\int_{\text{cone}} \d^{d}x \, \sqrt{G}\left(\frac{1}{2}h^{\nu\alpha}\nabla^{2}h_{\nu\alpha}-\frac{1}{2}h\nabla^{2}h-\nabla_{\nu}h^{\nu\alpha}\nabla_{\alpha}h+\nabla_{\nu}h^{\nu\alpha}\nabla^{\mu}h_{\mu\alpha}\right)\bigg) \,.
\end{align}
For $\xi= 0$, this path integral localizes on the de Donder gauge \eqref{eqn:deDondergauge}. Instead, it is more convenient for us to choose $\xi=1$, which results in the following `gauge-fixed' path integral
\begin{align}\label{eqn:gravitonpartfn_gaugefixed_simpl}
    Z_\text{graviton}(\beta) &= \det(-\Box\, \delta^\sigma_\n)\int \mathcal{D}h\,\exp{-\frac{1}{4}\int_\text{cone} \d^{d}x\, \sqrt{G}\left(-h_{\m\n}\Box h^{\m\n}+\frac{1}{2}h\Box h\right)}\nonumber\\ &= \det(-\Box\, \delta^\mu_\nu)\det(-\Box\,\delta^\mu_\sigma \delta^\nu_\kappa+1/2\, G^{\mu\nu} G_{\sigma\kappa}\Box)^{-1/2}\,.
\end{align}
The first determinant also appeared in the photon partition function \eqref{2.aa zphot} and can thus be computed using the vector heat kernel \eqref{3.29 vecKdef}.\footnote{As in the photon case, this determinant (written in general form in \eqref{eqn:Rxigaugeidentity}) only makes sense for the tip-less cone.} In a Lagrangian formulation, this Faddeev-Popov determinant would be written as a path integral over vector ghosts, so we will refer to it as the ghost contribution. The second determinant may look as if it is more tricky to evaluate, due to the extra term in the action for configurations with non-zero trace. We show in appendix \ref{app:tensorpathintegral} that, despite this, the final answer is rather simple:
\begin{align}
    \log Z_\text{sym tensor}(\beta)&=\log\left[ \int \mathcal{D}h\,\exp{-\frac{1}{4}\int_\text{cone} \d^{d}x\sqrt{G}\,\left(-h_{\m\n}\Box h^{\m\n}+\frac{1}{2}h\Box h\right)}
    \right] \nonumber
    \\&=\int_0^\infty \frac{\d s}{2s}\int_\text{cone}\d^d x\sqrt{G}\,K_\text{sym tensor}(x|s)\,.\label{4.8 tensor determinant}
\end{align}
Here, the symmetric tensor heat kernel is given by
\begin{equation}
    K_\text{sym tensor}(x\rvert s)=\sum_{\alpha\leq \beta}\sum_{\ell=-\infty}^{\infty}\int_{0}^{\infty}\d k\,\int_{-\infty}^{+\infty} \d\mathbf{k}\,e^{-s(k^{2}+\mathbf{k}^2)}G^{\mu\sigma}G^{\nu\kappa}\phi^{(\alpha\beta)}_{\mu\nu\,k,\ell,\mathbf{k}}(x)^*\phi^{(\alpha\beta)}_{\sigma\kappa\,k,\ell,\mathbf{k}}(x)\,.\label{4.9 Ktensdef}
\end{equation}
The eigenmodes appearing here are the symmetric $2$-tensor eigenfunctions of the d'Alembertian $\Box$ which will be constructed momentarily. According to \eqref{eqn:gravitonpartfn_gaugefixed_simpl}, the graviton path integral is then computed as
\begin{equation}
    \log Z_\text{graviton}(\beta)=\log Z_\text{sym tensor}(\beta)-2\log Z_\text{vec}(\beta)=\int_0^\infty \frac{\d s}{2s}\int_\text{cone}\d^d x\sqrt{G}\,K_\text{graviton}(x|s)\,.\label{4.11 logZgrav}
\end{equation}
We will evaluate the symmetric tensor heat kernel \eqref{4.9 Ktensdef} on the cone in section \ref{sect:4.2 gravK} below and therefore, in combination with the vector heat kernel \eqref{3.33 Kvec}, we obtain the graviton heat kernel.

\subsection{Heat kernel}\label{sect:4.2 gravK}

The remaining task is to compute the symmetric tensor heat kernel \eqref{4.9 Ktensdef}, which is quite tedious. The reader who is not interested in the details may skip directly to the answer \eqref{eqn:gravprop}. 

For the symmetric tensor heat kernel \eqref{4.9 Ktensdef}, one can use the basis of modes of the d'Alembertian $\Box$ constructed in \eqref{eqn:tensoreigenfneqn}, \eqref{2.30 modesgen}, and \eqref{eqn:basis}. The normalized modes for a symmetric $2$-tensor eigenfunction are
\begin{align}\label{eqn:sym2eigenfns}
        \p^{(\a\b)}_{\lambda\,\m\n}=\frac{1}{\sqrt{2}}(e^{(\a)}_\m e^{(\b)}_\n+e^{(\b)}_\m e^{(\a)}_\n)\,\p_\lambda\,, \quad \a \neq \b\,\quad\text{and}\quad \p^{(\a\a)}_{\lambda\,\m\n}=e^{(\a)}_\m e^{(\a)}_\n\p_\lambda\,,\quad \a=\b\,.
\end{align}
The modes $\phi^{(ab)}$ for $a \leq b$ contribute one scalar heat kernel each. There are $(d-1)(d-2)/2$ such modes so
\begin{equation}
    K_\text{sym tensor}(r\rvert s)\supset \,\frac{(d-1)(d-2)}{2} K_\text{scal}(r|s)\,\quad \text{from all modes $\phi^{(ab)}$ with $a \leq b$.}\label{4.12 Kab}
\end{equation}
The contributions from the $\phi^{(0a)}$ and $\phi^{(1a)}$ modes combine nicely using
\begin{equation}
    G^{\mu\sigma}G^{\nu\gamma}\phi^{(0 a)}_{\lambda\,\mu\nu}(x)^*\phi^{(0a)}_{\lambda\,\sigma\gamma}(x)+G^{\mu\sigma}G^{\nu\gamma}\phi^{(1 a)}_{\lambda\,\mu\nu}(x)^*\phi^{(1a)}_{\lambda\,\sigma\gamma}(x)=\frac{2}{k^2}\nabla^{\mu\,\text{2d}}\phi_{\lambda}(x)^*\nabla_{\mu}^\text{2d}\phi_{\lambda}(x)\,.
\end{equation}
We recognize a contribution to the heat kernel that we previously encountered in the vector calculation \eqref{eqn:vectorheatkernel} for each of the $d-2$ choices $a$. Therefore, we obtain a total contribution
\begin{equation}
    K_\text{sym tensor}(x\rvert s)\supset 2 (d-2)\, K_\text{scal}(r|s)+(d-2)\,\frac{2s}{r}\partial_r\,K_\text{scal}(r|s)\,\quad \text{from all combination of $\phi^{(0a)}$ and $\phi^{(0a)}$.}\label{4.14 Koa1a}
\end{equation}
More care is required for the polarizations $\phi^{(01)}$, $\phi^{(00)}$ and $\phi^{(11)}$ constructed out of vectors along the 2d cone. For notational comfort, we isolate the pure 2d kernel due to these polarizations
\begin{equation}
    K_\text{sym tensor}(x\rvert s)\supset\frac{1}{(4\pi s)^{d/2-1}}K^{\text{2d}\,(01)}(x\rvert s)+\frac{1}{(4\pi s)^{d/2-1}}K^{\text{2d}\,(00)}(x\rvert s)+\frac{1}{(4\pi s)^{d/2-1}}K^{\text{2d}\,(11)}(x\rvert s)\,.\label{4.15 Kcontr}
\end{equation}
The 2d kernel contribution from the $\phi^{(00)}$ and $\phi^{(11)}$ modes is computed as follows
\begin{align}\label{eqn:Kdiag}
    &\nonumber K^{\text{2d}\,(00)}(r\rvert s)+K^{\text{2d}\,(11)}(r\rvert s) = \sum_{\ell=-\infty}^{\infty}\int_0^\infty \,\d k e^{-s k^2}\frac{2}{k^4} \nabla^{\mu\,\text{2d}} \nabla^{\nu\,\text{2d}}\phi_{k,\ell}(x)^*\nabla_\mu^{\text{2d}} \nabla_\nu^\text{2d} \phi_{k,\ell}(x)
    \\  
    &\nonumber = \sum_{\ell=-\infty}^{\infty}\int_0^\infty \,\d k e^{-s k^2}\frac{2}{k^2} \phi_{k,\ell}(x)^*\phi_{k,\ell}(x)+ \Box^{\text{2d}}\sum_{\ell=-\infty}^{\infty}\int_0^\infty \,\d k e^{-s k^2}\frac{1}{k^4} \nabla^{\mu\,\text{2d}} \phi_{k,\ell}(x)^*\nabla_\mu^{\text{2d}} \phi_{k,\ell}(x)\\&\qquad \qquad\qquad\qquad\qquad\qquad\qquad\qquad\qquad\qquad+\Box^\text{2d}\sum_{\ell=-\infty}^{\infty}\int_0^\infty \,\d k e^{-s k^2}\frac{1}{k^2} \phi_{k,\ell}(x)^*\phi_{k,\ell}(x)
    \\
    &\nonumber = 2K_\text{scal}^{\text{2d}}(r\rvert s)+\int_{s}^{\infty}\d p\,  \Box^{2d}\left(\frac{1}{2}K^{\text{2d}}_\text{vec}(r|p)+K_\text{scal}^{\text{2d}}(x\rvert p)\right)= 2K_\text{scal}^{\text{2d}}(r\rvert s)+\frac{s}{r}\partial_{r}K_\text{vec}^{2d}(r\rvert s)+\frac{2s}{r}\partial_{r}K_\text{scal}^{\text{2d}}(r\rvert s)\,.
\end{align}
In the second equality one uses the explicit action of the $\Box^{\text{2d}}$, in the third equality we recognized the scalar \eqref{2.8 K} and vector \eqref{eqn:Kvec} heat kernels in 2d, and for the last equality we used the fact that \eqref{eqn:rtos} holds for both $K_\text{scal}^{\text{2d}}(r\rvert s)$ and $K_\text{vec}^{2d}(r\rvert s)$ and evaluated the $p$ integral by integration by parts.\footnote{The fact $K_\text{vec}^{2d}(r\rvert s)$ also satisfies the second equality in \eqref{eqn:rtos} can be checked from the expression for $K_\text{vec}^{2d}(r\rvert s)$ as a function of $K_\text{scal}^{2d}(r\rvert s)$ in \eqref{eqn:Kvec}, from which one sees that both are of the form $(1/s) f(r^2/2s)$ with different $f$. This functional form suffices to show that the differential equation \eqref{eqn:rtos} holds.} The first and third terms are the kinds of terms we found for the photon propagator \eqref{eqn:Kvec}. However, there is also a genuinely new type of term appearing for the graviton propagator, namely a radial derivative of the vector heat kernel. Finally, the 2d kernel from the modes $\phi^{(01)}$ evaluates to
\begin{align}\label{eqn:K01}
    K^{\text{2d}\,(01)}(r\rvert s) &= \sum_{\ell=-\infty}^{\infty}\int_0^\infty \,\d k e^{-s k^2}\frac{1}{k^4} \nabla^{\mu\,\text{2d}} \nabla^{\nu\,\text{2d}}\phi_{k,\ell}(x)^*\nabla_\mu^{\text{2d}} \nabla_\nu^\text{2d} \phi_{k,\ell}(x)\nonumber\\&\qquad\qquad\qquad\qquad\qquad+\sum_{\ell=-\infty}^{\infty}\int_{s}^{\infty}\d p\,\int_0^\infty \d k\, e^{-p k^2}\frac{1}{k^2} \epsilon^{\nu\alpha\,\text{2d}}\epsilon_{\rho\beta}^\text{2d}\nabla^{\rho\,\text{2d}} \nabla_\alpha^\text{2d}\phi_{k, l}(x)^*\nabla_\nu^\text{2d} \nabla^{\beta\,\text{2d}} \phi_{k, l}(x)
    \\&= K_\text{scal}^{\text{2d}}(r\rvert s)+\int_{s}^{\infty}\d p\,  \Box^{2d}\left(\frac{1}{2}K^{\text{2d}}_\text{vec}(r|p)+K_\text{scal}^{\text{2d}}(x\rvert p)\right) \nonumber
    \\&= K_\text{scal}^{\text{2d}}(r\rvert s)+\frac{s}{r}\partial_{r}K_\text{vec}^{2d}(r\rvert s)+\frac{2s}{r}\partial_{r}K_\text{scal}^{\text{2d}}(r\rvert s)\,,\nonumber
\end{align}
where the first term in the second equality used that this term is equal to $1/2$ times the expression in \eqref{eqn:Kdiag}. For the second term, we demonstrate in appendix \ref{app:c2} that
\begin{align}
    \nonumber\int_{s}^{\infty}\d p\,\int_0^\infty \d k\, e^{-p k^2}\frac{1}{k^2}\sum_\ell \epsilon^{\nu\alpha\,\text{2d}}\epsilon_{\rho\beta}^\text{2d}\nabla^{\rho\,\text{2d}} \nabla_\alpha^\text{2d}&\phi_{k, l}(x)^*\nabla_\nu^\text{2d} \nabla^{\beta\,\text{2d}} \phi_{k, l}(x)\\&=\frac{1}{2}\int_{s}^{\infty}\d p\,  \Box^{2d}\left(\frac{1}{2}K^{\text{2d}}_\text{vec}(r|s)+K_\text{scal}^{\text{2d}}(x\rvert s)\right)\,.\label{c2 eq}
\end{align}
The final equality in \eqref{eqn:K01} is similar to the final equality in the calculation in \eqref{eqn:Kdiag}. Combining \eqref{eqn:K01} and \eqref{eqn:Kdiag}, one finds the following contribution to the symmetric $2$-tensor propagator
\begin{align}\label{4.19 Kspecial}
    &K_\text{sym tensor}(r|s) \\&\quad\supset 3\,K_\text{scal}(r\rvert s)+2\,\frac{2s}{r}\partial_{r}K_\text{scal}(r\rvert s)+\frac{2s}{r}\partial_{r}\,\frac{1}{(4\pi s)^{(d-2)/2}}K_\text{vec}^{\text{2d}}(r\rvert s)\,\quad \text{from modes $\phi^{(00)}$, $\phi^{(11)}$ and $\phi^{(01)}$\,.}\nonumber
\end{align}
Therefore, by combining all the contributions \eqref{4.12 Kab}, \eqref{4.14 Koa1a} and \eqref{4.19 Kspecial}, we arrive at the full propagator for the symmetric $2$-tensor
\begin{equation}\label{eqn:tensprop}
    K_\text{sym tensor}(r\rvert s) = \frac{d(d+1)}{2}K_\text{scal}(r\rvert s)+d\,\frac{2s}{r}\partial_{r}K_\text{scal}(r\rvert s)+\frac{2s}{r}\partial_{r}\,\frac{1}{(4\pi s)^{(d-2)/2}}K_\text{vec}^{\text{2d}}(r\rvert s)\,.
\end{equation}
The contributions from each of the modes to this propagator are summarized in table \ref{tab:entropycount}.

To compute the graviton propagator \eqref{4.11 logZgrav}, we must include the contribution from the vector ghost fields whose heat kernel equals that of (minus) two vector fields \eqref{2.38 kvectrace}. We emphasize (as also captured in table \ref{tab:entropycount}) that this includes the ``standard'' contact term contribution that arises in the evaluation of such vector determinants. The free graviton propagator on the Rindler cone is thus
\begin{align}\label{eqn:gravprop}
\begin{split}
  K_\text{graviton}(r\rvert s) &= K_\text{sym tensor}(r\rvert s) - 2\,K_\text{vec}(r\rvert s) 
  \\    &= \frac{d(d-3)}{2}K_\text{scal}(r\rvert s)+(d-2)\frac{2s}{r}\partial_{r}K_\text{scal}(r\rvert s)+\frac{2s}{r}\partial_{r}\,\frac{1}{(4\pi s)^{(d-2)/2}}K_\text{vec}^{\text{2d}}(r\rvert s)
  \\    &= \frac{d(d-3)}{2}K_\text{scal}(r\rvert s)+\frac{2s}{r}\partial_{r}K_\text{vec}(r\rvert s)\,.
\end{split}
\end{align}
In the last equality, we recognize a radial derivative of the full vector heat kernel \eqref{eqn:vectorheatkernel}.

\begin{table}[ht]
\caption{Summary of graviton entropy contributions}
\centering
\begin{tabular}{|c|c|c|c|}
\hline
Modes & Scalar term & Photon contact term & New term \\ \hline
$\p^{(ab)}$ & $(d-1)(d-2)/2$ & $0$ & $0$ \\ \hline
$\p^{(0a)}$ and $\p^{(1a)}$ & $2(d-2)$ & $d-2$ & $0$ \\ \hline
$\p^{(00)}$ and $\p^{(11)}$ & $2$ & $1$ & $1$\\ \hline
$\p^{(01)}$ & $1$ & $1$ & $1$ \\ \hline
vector ghosts & $-2d$ & $-2$ & 0 \\ \hline
Total & $d(d-3)/2$ & $d-2$ & $2$\\ \hline
\end{tabular}
\label{tab:entropycount}
\end{table}
\subsection{Entropy and graviton contact terms}\label{sect:4.3 gravContact}

We now have all the pieces that we need to assemble the graviton partition function. The trace of the graviton heat kernel \eqref{eqn:gravprop} is
\begin{equation}
    \int_\text{cone} \d^d x \sqrt{G}\,K_\text{grav}(x|s) = \frac{d(d-3)}{2} \int_\text{cone} \d^d x \sqrt{G}\,K_\text{scal}(x|s)-2s \int_\text{horizon} \d^{d-1} x \,K_\text{vec}(x|s)\,. \label{2.38 kgravtrace}
\end{equation}
Inserting this into \eqref{4.11 logZgrav} gives the final answer for the free graviton path integral on the Rindler cone:
\begin{align}
    \boxed{\log Z_\text{graviton}(\beta) = \frac{d(d-3)}{2} \int_{\epsilon^{2}}^\infty \frac{\d s}{2s}\int_\text{cone} \d^d x \sqrt{G}\,K_\text{scal}(x|s)-\int_{\epsilon^{2}}^\infty \d s\int_\text{horizon} \d^{d-1} x \,K_\text{vec}(x|s)}\,.\label{4.23 zgravcont}
\end{align}
We see that the graviton partition function gives $\frac{d(d-3)}{2}$ times the partition function for a scalar field, as expected for the number of physical bulk polarizations, along with a (negative) contribution contact term given by a \emph{vector} propagator anchored on the horizon. We will explore the physical interpretation of this more in section \ref{sect:interpret}. We can write this partition function as
\begin{equation}\label{eqn:gravF}
    \log Z_\text{graviton}(\beta) = \frac{d(d-3)}{2}\log Z_\text{scal}(\beta)+\log Z^\text{contact}_{\text{graviton}}\,.
\end{equation}
As seen from the second line of \eqref{eqn:gravprop}, the graviton contact term $\log Z^\text{contact}_{\text{graviton}}$ equals $(d-2)$ times the photon contact term $Z^\text{contact}_\text{photon}$ \eqref{eqn:Zcontact} \emph{plus} a new type of term. This new term actually gives a vanishing contribution to the path integral because, according to equation \eqref{3.33 property}, we have $K_\text{vec}^{\text{2d}}(0|s)=0$, and so
\begin{equation}\label{eqn:gravcontact_new}
   \log Z^\text{contact}_{\text{graviton}} \supset \beta A\int_{\epsilon^{2}}^\infty \d s\,\frac{1}{(4\pi s)^{(d-2)/2}} K_\text{vec}^{\text{2d}}(0|s) =0\,.
\end{equation}
The conclusion is that the full graviton contact term reduces to simply $d-2$ times the photon contact term \eqref{eqn:Zcontact}. This is our main technical result:
\begin{equation}\label{eqn:gravcontact}
   \boxed{\log Z^\text{contact}_{\text{graviton}}=(d-2)\log Z^\text{contact}_\text{photon}}\,.
\end{equation}
Armed with \eqref{eqn:gravcontact}, we can immediately read off the graviton entropy from \eqref{eqn:gravF}, \eqref{2.13 sscal}, and \eqref{eqn:Zcontact} to obtain
\begin{equation}\label{eqn:graventropy}
    \boxed{S_\text{graviton}=2\pi A \int_{\epsilon^{2}}^\infty \frac{\d s}{(4\pi s)^{d/2}}\bigg(\frac{1}{6}\frac{d(d-3)}{2}-(d-2)\bigg)}\,.
\end{equation}
Thus, the entropy is equal to that of $\frac{d(d-3)}{2}$ scalars corresponding to the physical polarizations, with a negative contribution from $d-2$ contact terms, indicating that there are $d-2$ edge modes for gravitons.

\section{Interpretation}\label{sect:interpret}
In this section, we seek the physical interpretation of the contact term contribution to the free graviton partition function on the Rindler cone \eqref{4.23 zgravcont}:
\begin{align}
    \log Z_\text{graviton}^\text{contact} =-\int_{\epsilon^{2}}^\infty \d s\int_\text{horizon} \d^{d-1} x \,K_\text{vec}(x|s)=-\int_\text{horizon}\d^{d-1}x\,G_\text{vect}(x\rvert x)\,.\label{5.1 zgrav}
\end{align}
Here $G_\text{vect}(x\rvert x)$ is the horizon-to-horizon propagator of a massless vector field through the $d$-dimensional bulk. From the point of view of our calculation, it seems a quite non-trivial conspiracy that the graviton heat kernel features \emph{precisely} such a vector propagator between two points on the horizon, as shown in equation \eqref{eqn:gravprop}. This generalizes Kabat's interpretation \cite{Kabat:1995eq} for the photon contact term \eqref{3.39 Kphotcont}:
\begin{align}
    \log Z_\text{photon}^\text{contact} =-\int_{\epsilon^{2}}^\infty \d s\int_\text{horizon} \d^{d-1} x \,K_\text{scal}(x|s)=-\int_\text{horizon}\d^{d-1}x\,G_\text{scal}(x\rvert x)\,.\label{5.2 phot}
\end{align}
For the graviton contact term, one might thus be lead to a picture of massless vector particle loops that begin and end on the horizon, and contribute non-trivially to the graviton partition function
\begin{equation}\label{eqn:Zgravcontact_paths}
    \log Z_\text{graviton}^\text{contact}\overset{?}{=}-\int \mathcal{D}\text{\color{green!25!black!75}{paths}}\quad \cone \,\,.
\end{equation}
With this being said, the physical value of such an interpretation is currently \emph{not} clear to us. Even in the photon case \eqref{5.2 phot}, this picture has never been made precise.

In the spirit of equation \eqref{1.2 introintro} one would like to interpret the contact term as counting `new' degrees of freedom that live on the horizon. However, physical configurations do not contribute to the (log of the) path integral with a \emph{negative} sign. Indeed, due to the negative sign in front of \eqref{5.1 zgrav}, the truth is
\begin{equation}
    Z_\text{graviton}^\text{contact} =\frac{1}{\exp\bigg(\int_\text{horizon}\d^{d-1}x\,G_\text{vect}(x\rvert x)\bigg)}\,.\label{5.4 less}
\end{equation}
From this point of view, it seems as if the bulk polarizations over-counted degrees of freedom and that the contact term is correcting for this by dividing out by this overcounting. In other words, the contact term would appear to not be counting `extra' degrees of freedom on the horizon, but \emph{removing} them.

For photons, this confusion \cite{Barvinsky:1995dp,DeNardo:1996kp,Iellici:1996jx,Cognola:1997xp,Kabat:2012ns,Donnelly:2012st,Solodukhin:2012jh} was largely resolved by \cite{Donnelly:2014fua,Donnelly:2015hxa,Blommaert:2018rsf,Blommaert:2018oue} who pointed out that the photon contact term \eqref{5.2 phot} \emph{does} have an interpretation as counting `extra' degrees of freedom on the horizon. Recalling some details of this `edge mode' construction, in \textbf{section \ref{sect:edge}} we will sketch an interpretation of the graviton contact term \eqref{5.4 less} in terms of `graviton edge modes'. We claim that the extra horizon degrees of freedom are the $d-2$ ADM charges $q^i_\text{ADM}$ which generate large diffeomorphisms $\epsilon_i$ confined within the $(d-2)$-dimensional entangling surface.

In \textbf{section \ref{sect:string}}, we briefly return to what is arguably the roots of edge modes, namely entanglement of strings in Rindler space \cite{Susskind:1994vu,Susskind:1993ws,Susskind:1994sm,Kabat:1995jq,Balasubramanian:2018axm,Ahmadain:2022eso,Donnelly:2020teo,Geiller:2019bti,Witten:2018xfj,He:2014gva,Dabholkar:1994ai,Dabholkar:1994gg,Mertens:2016tqv,Mertens:2015adr}. Combining our graviton calculation of section \ref{sect:kabatgraviton} with the Kalb-Ramond partition function on the Rindler cone whose calculation is presented in appendix \ref{sec:KBfield}, we match the contact term of the massless fields in the bosonic string to that coming from two photon fields $A^\alpha$ and $\bar{A}^\alpha$ living on the $(d-1)$-dimensional boundary (including the infinitesimal time direction). These are the gauge degrees of freedom of the closed bosonic string at the massless level and we speculate that this relation between edge modes and null states continues to hold at all massive string levels.

Both section \ref{sect:edge} and section \ref{sect:string} are written compactly and more details will be presented elsewhere.

\subsection{Towards graviton edge modes}\label{sect:edge}
For photons, the edge mode construction may be summarized as follows \cite{Donnelly:2014fua,Donnelly:2015hxa,Blommaert:2018rsf,Blommaert:2018oue}.\footnote{In order to present an intuitive overview, we gloss over significant subtleties, such as the distinction between Neumann and Dirichlet boundary conditions on the bulk modes \cite{Donnelly:2014fua,Donnelly:2015hxa,Blommaert:2018rsf,Blommaert:2018oue,Ball:2024hqe}. All equations in this section are morally correct but should not be viewed as precise. More precise equations can be found in the original references \cite{Donnelly:2014fua,Donnelly:2015hxa,Blommaert:2018rsf,Blommaert:2018oue,Ball:2024hqe}.} The relevant horizon degrees of freedom are electric charges $q$ living on the $(d-2)$-dimensional entangling surface. These charges source electric fields in the bulk through the boundary condition
\begin{equation}
    q= \frac{\delta S_\text{on-shell}}{\delta A_\tau}\Big\rvert_\text{bdy} =\sqrt{h}\,n_\mu F^{\mu\tau}\rvert_\text{bdy}\,.
\end{equation}
The bulk polarizations satisfy homogeneous boundary conditions $q=0$. The action for these boundary charges $q$ is computed by sourcing the theory, and evaluating the Maxwell action on-shell. This gives rise to a bilocal boundary action
\begin{equation}
    Z(q)=\exp\bigg(-\frac{1}{2}\int_\text{horizon}\d^{d-2}x\,q(x)\int_\text{horizon}\d^{d-2}y\,q(y)\,G_\text{scal horizon}(x|y)\bigg)\,.\label{5.6 zq}
\end{equation}
The propagator of a scalar field in the $(d-2)$-dimensional flat boundary is given by
\begin{equation}
    G_\text{scal horizon}(x|y)=\bra{x}\frac{1}{-\Box_\text{horizon}}\ket{y}=\sum_\lambda \frac{1}{\lambda}\,\phi_{\lambda\,\text{horizon}}(x)^*\,\phi_{\lambda\,\text{horizon}}(y)\,.
\end{equation}
In the momentum basis, this becomes simply equation (6.22) in \cite{Blommaert:2018rsf}, or similarly equation (7) in \cite{Donnelly:2014fua}:
\begin{equation}
    Z(q)=\exp\bigg(-\frac{1}{2}\int_{-\infty}^{+\infty}\d\mathbf{k}\,\frac{1}{\mathbf{k}^2}\,q(\mathbf{k})q(-\mathbf{k}) \bigg)\,.\label{5.8 zq}
\end{equation}
The edge mode or contact term partition function is obtained by path integrating over these boundary charges $q$, as follows:
\begin{equation}
    Z_\text{photon}^\text{contact}=\int \mathcal{D}q\,Z(q)=\int \mathcal{D}q\,\exp\bigg(-\frac{1}{2}\int_\text{horizon}\d^{d-2}x\,q(x)\int_\text{horizon}\d^{d-2}y\,q(y)\,G_\text{scal horizon}(x|y)\bigg)\,.
\end{equation}
This path integral is formally evaluated using \eqref{5.8 zq} as
\begin{equation}
    \log Z_\text{photon}^\text{contact}=-\frac{1}{2}\Tr \log \bigg(\frac{1}{-\Box_\text{horizon}}\bigg)\,.
\end{equation}
To appreciate that this is identical to the photon contact term in \eqref{3.39 Kphotcont} or \eqref{5.2 phot}, one can perform the following manipulations
\begin{align}\label{5.11 edgetocontact}
    -\frac{1}{2}\Tr \log \bigg(\frac{1}{-\Box_\text{horizon}}\bigg)=\frac{1}{2}\Tr \log (-\Box_\text{horizon})&=-\int_{\epsilon^{2}}^\infty \frac{\d s}{2s}\int_\text{horizon} \d^{d-2} x \,K_\text{scal\,\text{horizon}}(x|s)\\&=-\int_{\epsilon^{2}}^\infty \d s\int_\text{horizon} \d^{d-1} x \,K_\text{scal}(x|s)\,.\nonumber
\end{align}
In the final equality, one uses the expression for the bulk heat kernel $K_\text{scal}(x|s)$ evaluated at a horizon point $x$, which can be found in \eqref{eqn:Zcontact}. One indeed recognizes the heat kernel of a scalar field that lives purely in the $d-2$ dimensional boundary. The first equality shows how edge modes resolve the negative sign confusion discussed around \eqref{5.4 less}. Instead of \emph{modding out} by a scalar field with Laplacian $\Box_\text{horizon}$, one \emph{adds} a nonlocal scalar field $q$ with `Laplacian' $1/\Box_\text{horizon}$. As discussed in the Introduction, there is ample evidence that this edge mode interpretation, where one includes boundary charges $q$ in the extended Hilbert space and counts their contribution to the entanglement entropy, is physically correct.

For gravitons, starting from \eqref{5.1 zgrav} and using \eqref{eqn:gravcontact}, we similarly have the observation that
\begin{align}
    \log Z_\text{graviton}^\text{contact} =-\int_{\epsilon^{2}}^\infty \d s\int_\text{horizon} \d^{d-1} x \,K_\text{vec}(x|s)=-\frac{(d-2)}{2}\Tr \log \bigg(\frac{1}{-\Box_\text{horizon}}\bigg)\,.
\end{align}
This can then be rewritten as arising form the path integral of a non-local vector field $q^i_\text{ADM}$ constrained to the $(d-2)$-dimensional Rindler horizon as follows:
\begin{equation}
        \boxed{Z_\text{graviton}^\text{contact}=\int \mathcal{D}q^i_\text{ADM}\exp\bigg(-\frac{1}{2}\int_\text{horizon}\d^{d-2}x\,q^i_\text{ADM}(x)\int_\text{horizon}\d^{d-2}y\,q^j_\text{ADM}(y)\,G_\text{vec horizon}(x|y)_{i\,j}\bigg)}\,.\label{5.13 gravedge}
\end{equation}
But what is the physical interpretation of these boundary charges $q^i_\text{ADM}$? What are the extra degrees of freedom associated with subregions in gravity? We believe \cite{Blommaert:2024} that they are essentially the $d-2$ ADM charges of gravity that act internally in the $(d-2)$-dimensional horizon, i.e., the entangling surface
\begin{equation}
    \boxed{q^i_\text{ADM}=\frac{\delta S_\text{on-shell}}{\delta h_{i\tau}}\rvert_\text{bdy}}\,.\label{5.14 adm}
\end{equation}
These generate large (would-be) gauge transformations $\epsilon^i$ \eqref{eqn:quantumdiffsymkappa0}, with indices $i$ in the Rindler horizon. The idea is that those gauge transformations would be physical, in the sense that they would show up in the physical phase space, i.e., the symplectic form; whereas other large gauge transformations would remain redundant.

Earlier work on gravitational edge modes includes \cite{Freidel:2019ees,Donnelly:2016auv,Geiller:2017xad,Geiller:2017whh,Lin:2018xkj,Freidel:2020xyx,Jafferis:2019wkd,Donnelly:2020xgu,Ciambelli:2021vnn,David:2022jfd,Ciambelli:2021nmv,Ciambelli:2022cfr,Donnelly:2022kfs,Mertens:2022ujr,Chen:2023tvj,Takayanagi:2019tvn,Anninos:2020hfj,Blommaert:2018iqz}, and in particular the $d-2$ large would-be gauge transformations $\epsilon_i$ were discussed as one option (out of many) for the correct gravitational edge modes in \cite{Freidel:2020xyx}. To the best of our knowledge, we are the first to attempt to match a precise Euclidean path integral calculation to validate such an edge mode proposal.

Let us stress the main point. The confusing minus sign in \eqref{5.1 zgrav} arises not because we are \emph{subtracting} degrees of freedom, but because we are counting degrees of freedom, namely graviton edge modes $q^i_\text{ADM}$, with a \emph{non-local} action.

In the remainder of this section, we present a preliminary attempt to embed our work in the context of entanglement of bosonic strings.

\subsection{String contact terms and horizon physics?}\label{sect:string}
Closed string path integrals are notoriously difficult to interpret and compute due to the mapping class group. As a small step to understanding entanglement in string theory, one could instead simply study the path integrals of all fields in the string spectrum in the free approximation.\footnote{In the free approximation the only difference with actual closed string theory is whether one integrates the moduli space of the torus over Teichmuller space or the fundamental domain of the modular group. So one obtains $\infty$ times the string free energy.} Here we discuss the contact term for the massless fields of the closed bosonic string: the graviton, a scalar dilaton, and a two-form Kalb-Ramond field (whose path integral on the Rindler cone is computed in appendix \ref{sec:KRfield_PI}).

We \emph{define} the combined partition function as $Z_\text{closed string massless}$. By combining the graviton path integral \eqref{eqn:gravF} with the Kalb-Ramond answer in \eqref{a.22 KRz}, we obtain
\begin{align}\label{eqn:closedstringmasslessZ}
    \log Z_\text{closed string massless}(\beta)&=\log Z_\text{graviton}(\beta)+\log Z_\text{Kalb-Ramond}(\beta)+\log Z_\text{scal}(\beta) \nonumber\\
    &=(d-2)^2 \log Z_\text{scal}(\beta)+2(d-3) \log Z^\text{contact}_\text{photon}\,.
\end{align}
The first term reproduces the correct number of physical polarizations, i.e., that of a 2-tensor in $d-2$ dimensions.\footnote{The fact that two dimensions are lost can be seen for instance in lightcone quantization.} More interesting for our purposes is the last term which states that the contact term for closed string field theory (up to mapping class group caveats) is $2(d-3)$ times the photon contact term
\begin{equation}
    \boxed{\log Z_\text{closed string massless}^\text{contact}=2(d-3)\log Z^\text{contact}_\text{photon}}\,.\label{5.16 string}
\end{equation}
It would be interesting to compare this with string theory calculations \cite{Balasubramanian:2018axm,Witten:2018xfj,He:2014gva,Mertens:2016tqv,Mertens:2015adr,Dabholkar:1994ai}. A preliminary analysis suggests a match.

Let us \emph{speculate} about the potential interpretation of this $2(d-3)$ counting of edge mode polarizations. For this, recall the structure of the closed string Hilbert space in covariant quantization (see \cite{zwiebach2004first}). Physical polarizations are due to worldsheet states that satisfy the Virasoro constraints and the on-shell condition. These constraints for the open string are given by
\begin{equation}
(L_n-\delta_{n,0})\ket{\psi}=0\,,\label{osvirasoro}
\end{equation}
with $n\geq 0$, while for the closed string one has
\begin{gather}
\begin{aligned}
(L_n+\bar{L}_n-2\delta_{n,0})\ket{\psi}&=0\,,\quad (L_n-\bar{L}_n)\ket{\psi}&=0\,.
\end{aligned}\label{csvirasoro}
\end{gather}
We are interested in the physical states \emph{and} the null states. 

For the open string, the massless level consists of a target space Maxwell field in Lorenz gauge with the usual scalar gauge mode corresponding to a null state. The closed string massless fields only consist of a target space 2-tensor field $Q_{\mu\nu}$. The Virasoro constraints on the worldsheet implement a version of Lorenz gauge in the target space:
\begin{align}\label{eqn:Qconstr}
\nabla^\mu\nabla_\mu Q_{\nu\sigma}&=0\,,\quad \nabla^\nu Q_{\nu\sigma}=\nabla^\sigma Q_{\nu\sigma}=0\,.
\end{align}
In a string field theory language, the null states correspond to the gauge redundancy
\begin{equation}\label{eqn:Qred}
Q_{\nu\sigma}\to Q_{\nu\sigma}+\nabla_\nu A_\sigma+\nabla_\nu\bar{A}_\sigma\,,
\end{equation}
where the one-form fields $A_{\sigma}$ and $\bar{A}_{\sigma}$ themselves satisfy Lorenz gauge
\begin{equation}
\nabla^\mu A_\mu=\nabla^\mu\bar{A}_\mu=0\,.\label{1nullconstr}
\end{equation}
These null fields carry a redundancy $A_{\mu} \to A_{\mu} + \nabla_{\mu}\phi$ and thus in $d$ dimensions would have $d-2$ polarizations.\footnote{We see therefore from the constraints of the second equation in \eqref{eqn:Qconstr} and the redundancy \eqref{eqn:Qred} that the total number of physical bulk polarizations for the field $Q_{\mu\nu}$ is equal to $d^2-2d-2(d-2)=(d-2)^{2}$ in agreement with the number appearing in \eqref{eqn:closedstringmasslessZ}.}  It is this structure which suggests a natural interpretation for \eqref{5.16 string}.

To this end, we first reconsider the photon contact term and rewrite \eqref{5.6 zq} as follows\footnote{This is purely a path integral for the boundary theory. We are also being careless with the $\epsilon$ factors required to go from the $(d-1)$-dimensional boundary to the $(d-2)$-dimensional horizon.}
\begin{align}
    Z(q )&= \exp\bigg(-\frac{1}{2}\int_\text{horizon}\d^{d-2}x\,q(x)\int_\text{horizon}\d^{d-2}y\,q(y)\,G_\text{scal horizon}(x|y)\bigg)\nonumber\\&=\bigg\langle \exp\bigg(\i\int_\text{horizon} \d^{d-1} x\,q(x)\,\phi(x)\bigg)\bigg\rangle\nonumber
    \\ &= \frac{\int \mathcal{D}\phi\,\exp\bigg(-\frac{1}{2}\int_\text{horizon}\d^{d-1} x\sqrt{h}\,\nabla^\mu \phi\nabla_\mu\phi+\i\int_\text{horizon} \d^{d-1} x\,q\,\phi\bigg)}{\int \mathcal{D}\phi\,\exp\bigg(-\frac{1}{2}\int_\text{horizon}\d^{d-1} x\sqrt{h}\,\nabla^\mu \phi\nabla_\mu\phi\bigg)}\,.\label{5.22 zq}
\end{align}
These steps are related to the usual logic that one uses in holography \cite{Witten:1998qj}. We put boundary conditions on a bulk theory, where the boundary conditions are labeled by $q(x)$, and then we evaluated the bulk theory on-shell resulting in the first line. As in AdS/CFT, this results in a bilocal action for $q(x)$, and the kernel appearing in that action is the propagator of a scalar field which lives in the boundary. For example, for a massless scalar we have
\begin{equation}
    \average{\phi(x)\phi(y)}_\text{bdy}=G_\text{scal horizon}(x|y)\,.
\end{equation}
Here, we put the scalar field $\phi$ on the boundary with metric
\begin{equation}
    \d s^2_\text{horizon}=\lim_{\epsilon\to 0} \epsilon^2 \d \tau^2+\sum_{i=1}^{d-2}\d x_i^2\,,\label{5.24 ds2}
\end{equation}
and do the Kaluza-Klein (KK) reduction with $\epsilon\to 0$ so that temporal modes are not excited. One can interpret $Z(q)$ as turning on a source in the boundary for a scalar field $\phi$ living on the boundary, just like one would do in AdS/CFT, as indicated in the second and third line of \eqref{5.22 zq}.

Let us pause to emphasize that we are not suggesting that this \emph{is} holography, rather we are merely using the analogy to usefully manipulate the boundary part of the theory. We have the dynamical bulk theory \emph{and} the edge mode boundary theory together.

Unlike in the AdS/CFT context, we are also instructed to path integrate over boundary sources $q$, and keep track of the normalization of the partition function, which gives
\begin{equation}
    Z^\text{contact}_\text{open string massless}=\int \mathcal{D}q\,Z(q)\,.
\end{equation}
Inserting \eqref{5.22 zq}, the path integral over $q$ localizes in the numerator of the last line of \eqref{5.22 zq} to $\phi=0$, which leaves only the normalization factor, i.e., the denominator
\begin{equation}
    Z^\text{contact}_\text{open string massless} = \frac{1}{\int \mathcal{D}\phi\,\exp\bigg(-\frac{1}{2}\int_\text{horizon}\d^{d-1} x\sqrt{h}\,\nabla^\mu \phi\nabla_\mu\phi\bigg)}\,.\label{5.26 phot}
\end{equation}
This reproduces again the last term on the top line of equation \eqref{5.11 edgetocontact}

We claim that a very similar interpretation exists for the massless closed string contact term \eqref{5.16 string}:
\begin{equation}
    Z^\text{contact}_\text{closed string massless}=\int \mathcal{D} J^\alpha \mathcal{D}\bar{J}^\alpha\,Z(J^\alpha)Z(\bar{J}^\alpha)\,.\label{5.27}
\end{equation}
In this expression, $Z(J^\alpha)$ corresponds to turning on a current source $J^\alpha$ in a Maxwell theory living in the boundary
\begin{align}
    Z(J^\alpha)&=\bigg\langle \exp\bigg(\i\int_\text{horizon} \d^{d-1} x\,J^\alpha(x)\,A_\alpha(x)\bigg)\bigg\rangle\nonumber
    \\ &=\frac{\int \frac{\mathcal{D}A}{\text{Vol}(G)}\exp\bigg( -\frac{1}{4}\int_{\text{horizon}}\d^{d-1}x\,\sqrt{g}F_{\mu\nu}F^{\mu\nu}+\i\int_\text{horizon} \d^{d-1} x\,J^\alpha\,A_\alpha \bigg)}{\int \frac{\mathcal{D}A}{\text{Vol}(G)}\exp\bigg( -\frac{1}{4}\int_{\text{horizon}}\d^{d-1}x\,\sqrt{g}F_{\mu\nu}F^{\mu\nu}\bigg)}\,.\label{5.28 zj}
\end{align}
The partition function $Z(J^\alpha)Z(\bar{J}^\alpha)$ should correspond to evaluating the graviton and Kalb-Ramond actions on-shell with suitable boundary sources as in \eqref{5.13 gravedge}. Here, we limit ourselves to simply showing that the path integral \eqref{5.27} is correct. Path integrating over the currents results in localization in the numerator, and therefore
\begin{align}
    &Z^\text{contact}_\text{closed string massless}\nonumber\\&=\frac{1}{\int \frac{\mathcal{D}A}{\text{Vol}(G)}\exp\bigg( -\frac{1}{4}\int_{\text{horizon}}\d^{d-1}x\,\sqrt{g}F_{\mu\nu}F^{\mu\nu}\bigg)\int \frac{\mathcal{D}\bar{A}}{\text{Vol}(G)}\exp\bigg( -\frac{1}{4}\int_{\text{horizon}}\d^{d-1}x\,\sqrt{g}\bar{F}_{\mu\nu}\bar{F}^{\mu\nu}\bigg)}\,.
\end{align}
At this point we can already see that this reproduces \eqref{5.16 string} since a gauge-fixed photon in $d-1$ dimensions has indeed $d-3$ polarizations. To be more precise, we can do the KK reduction by sending $\epsilon\to 0$ in the metric \eqref{5.24 ds2}. This leaves two photons and two scalars in $(d-2)$-dimensional flat space. Each photon has $d-4$ physical polarizations, so the total partition function reproduces the $2(d-4)+2=2(d-3)$ scalars on the horizon, in accordance with \eqref{5.16 string}.

The potentially intriguing part of this interpretation of the closed string massless contact term as due to two photons $A$ and $\bar{A}$ living on the horizon is that they are exactly the pure gauge fields for the massless closed string level, as shown in \eqref{eqn:Qred}. This resonates well with the gauge theory intuition that large gauge degrees of freedom are physical. Combined with the massless open string case \eqref{5.26 phot}, this suggest the following \textbf{conjecture}:
\begin{center}
\emph{The field theory content at each level in string theory has contact terms which may be accounted for by promoting the pure gauge fields (coming from the null states on the worldsheet) to physical edge modes.}
\end{center}
It would be interesting to check this conjecture with string theory calculations \cite{Balasubramanian:2018axm,Witten:2018xfj,He:2014gva,Mertens:2016tqv,Mertens:2015adr,Dabholkar:1994ai}. If this turns out to be correct, it would be tempting to search for a D-brane type interpretation of equations such as \eqref{5.22 zq} and \eqref{5.28 zj}, perhaps making more concrete some of the ideas in \cite{Susskind:1993ws,Susskind:1994vu,Susskind:1994sm}. Of course this revitalizes the minus sign confusion discussed around \eqref{eqn:Zgravcontact_paths}, and it is not obvious to us how to interpret these theories on the horizon, especially the normalization (the denominator on the last line of \eqref{5.28 zj}).

\section{Concluding remarks}\label{sect:disc}
In this work, we computed the conical entropy for gravitons on the Rindler cone \eqref{eqn:Rindlercone} and found that the entropy received the expected contribution coming from bulk physical polarizations, along with a contribution from $d-2$ contact terms. We interpreted these contact terms as due to edge modes in the putative extended Hilbert space and gave a suggestion of how they may arise in closed bosonic string theory. 

To conclude we briefly comment again on the procedure we discussed in section \ref{sec:onshell} of removing the tip of the cone, and compare our results to previous (different) calculations in the literature \cite{Fursaev:1996uz,Solodukhin:2015hma,He:2014gva,Anninos:2020hfj}.

\vspace{4mm}

\noindent \textbf{Removing the tip of the cone}

\noindent We saw in section \ref{sec:onshell} that in order to make sense of the conical entropy for gravitons, one must delete the tip of the Rindler cone \eqref{eqn:Rindlercone}. This serves two purposes:
\begin{enumerate}
    \item The background metric $G_{\mu\nu}$ becomes on-shell, so perturbation theory of gravitons is well-defined (see section \ref{sec:onshell}).
    \item It removes the singular (delta-function) curvature at the tip, leading to simpler determinants of smooth operators in the evaluation of the graviton partition function (see sections \ref{sect:photon} and \ref{sect:4.1 gravPI}).
\end{enumerate}

Removing the tip makes mathematical sense as long as the symmetric $2$-tensor eigenmodes of the d'Alembertian are orthonormal \emph{without} imposing boundary conditions at the tip. We show in appendix \ref{app:modesortho} that this only holds for $\beta < 2\pi$ so it would be wrong to trust the results of the heat kernel calculation for $\beta>2\pi$ to compute the conical entropy. Indeed, the integrals needed to compute the graviton heat kernel \eqref{eqn:STTcontactterm_explicit} diverge for $\beta > 2\pi$. So the whole procedure is ill-defined in that range of $\beta$. On the other hand, we computed the conical entropy for $\beta < 2\pi$ and analytically continued the result to the physical temperature $\beta=2\pi$, which is well-defined. Therefore, the final answer \eqref{eqn:graventropy} is a rigorous heat kernel calculation of the graviton entropy for Rindler with the horizon removed.\footnote{For higher spin fields we expect demanding orthonormality restricts to $\beta<2\pi/n$, and one should analytically continue from this trustworthy region to calculate the conical entropy. Amusingly, this is similar to the analytic continuation used ad-hoc in \cite{He:2014gva}.}

Nevertheless, it would be reassuring to have an alternate calculation of the entropy for gravitons. For the photon one can instead smooth out the tip of the cone and reproduce the tip-less cone result \cite{Donnelly:2014fua,Donnelly:2015hxa}. That does not work for the graviton as the background would be off-shell. One possibility is to consider a Schwarzschild black hole spacetime whose thermal entropy is obtained from the partition function on the smooth Euclidean cigar geometry. There is a smooth Euclidean cigar for any temperature $\beta$, if we allow the background metric $G_{\mu\nu}$ to depend on $\beta$ by tuning the mass $M$ (or horizon radius) to the appropriate value. So, in this setup one allows for backreaction of $G_{\mu\nu}$ to smooth out the tip, as opposed to keeping $M$ fixed whilst varying $\beta$ which leads to a cigar with a conical tip. It is not obvious that these setups lead to the same entropy, however it has been argued that for Schwarzschild they do \cite{Jacobson:2012ek}. To compute the partition function for gravitons in Schwarzschild, one needs to use the graviton action \eqref{eqn:EHactionexp_2ndorder_gen}, which includes various curvature terms, so the evaluation of the heat kernel seems intractable at finite mass, but perhaps at leading order in inverse powers of the mass one could manage.\footnote{The infinite mass limit of the Schwarzschild black hole spacetime is the Rindler spacetime \cite{Susskind:1994sm} so this limit is the natural one to take to compare to our results.} It would be interesting to check if the result obtained from the smooth geometry agrees with the one from the tip-less cone.

\vspace{4mm}

\noindent \textbf{Comparison to literature}

\noindent Finally we comment on previous attempts to compute the entanglement entropy of gravitons. 

By working in the first quantization formalism and ignoring the off-shell issues discussed above, the authors of \cite{He:2014gva} obtained the partition function for a massless spin-$2$ particle on orbifolds of Euclidean Rindler and found a contribution to entropy given by $(d-2)$ times the photon contact term contribution \eqref{eqn:Zcontact}, so their result agrees with our \eqref{eqn:graventropy}. Regardless of their ignorance of the off-shell issues, their analysis only holds for orbifolds $\beta=2\pi/N$, whereas our calculation holds for \emph{all} $\beta<2\pi$. One might be worried that `analytic continuation' from a discrete set of orbifold points gives a wrong result. Indeed, this is a famous critique on calculations of string entanglement via orbifolds \cite{Susskind:1994vu,Susskind:1993ws,Susskind:1994sm,Kabat:1995jq,Balasubramanian:2018axm,Ahmadain:2022eso,Donnelly:2020teo,Witten:2018xfj,He:2014gva,Dabholkar:1994ai,Dabholkar:1994gg,Mertens:2016tqv,Mertens:2015adr}. For the graviton case, our calculation removes this issue. In this sense one could view our analysis as indirect motivation (given our match with \cite{He:2014gva}) to take the orbifold analytic continuation seriously.

In the second quantization formalism, the authors of \cite{Fursaev:1996uz,Solodukhin:2015hma} computed the partition function of the graviton on compact manifolds with conical singularities, but importantly \emph{without} removing the tip. We do not think this makes sense as a starting point because of the problems with off-shell backgrounds that we discuss in section \ref{sec:onshell}. This is reflected in the fact that the result of \cite{Fursaev:1996uz} depends on whether one considers the sharp cone or one chooses to smooth out the cone. With this being said, for the sharp cone, in \cite{Fursaev:1996uz} they also find $d-2$ photon contact terms. Different results were found in equation (5.28) in \cite{Solodukhin:2015hma}, which we however are not trusting for the aforementioned reasons.

In their study of de Sitter entropy, the authors of \cite{Anninos:2020hfj} compute the partition function for gravitons on a sphere using the characters of massless spin-$2$ representations of the isometry group. These characters split into a `bulk' piece minus an `edge' piece, where the edge piece is equal to the characters of $d+2$ scalars in $d-2$ dimensions. The fact that the edge theory lives on the $(d-2)$-dimensional horizon and contributes with a minus sign to the entropy is consistent with what we obtained in this work, but the fact that there are $d+2$ scalars in their analysis is different. It would be interesting to understand the discrepancy.

Besides these path integral calculations, there are many claims about (partial) Hilbert space analysis of gravitational edge modes \cite{Donnelly:2016auv,Geiller:2017xad,Geiller:2017whh,Freidel:2019ees,Freidel:2020xyx,Donnelly:2020xgu,Ciambelli:2021vnn,Ciambelli:2021nmv,Ciambelli:2022cfr,Donnelly:2022kfs,David:2022jfd}. We will not comment on these aside from saying that we did not find comparisons in those papers with any of the aforementioned Euclidean path integral calculations. Therefore, it is not clear which calculation `validates' the (inequivalent) claims in \cite{Donnelly:2016auv,Geiller:2017xad,Geiller:2017whh,Freidel:2019ees,Freidel:2020xyx,Donnelly:2020xgu,Ciambelli:2021vnn,Ciambelli:2021nmv,Ciambelli:2022cfr,Donnelly:2022kfs,David:2022jfd}. We intend to fill this gap in the future \cite{Blommaert:2024}.
\section*{Acknowledgments}
We thank Thomas G. Mertens for useful discussions. AB was supported by the ERC-COG Grant NP-QFT No. 864583 and by INFN Iniziativa Specifica GAST, and thanks Berkeley, Princeton and CERN for hospitality during various stages of this work. SC-E was supported by the National Science Foundation under Award Number 2112880 and by the Department of Energy, Office of Science, Office of High Energy Physics under QuantISED Award DE-SC0019380. SC-E would like to thank University of Milano-Bicocca and SISSA for hospitality during the final stages of this work, and would like to thank KITP for hospitality during the workshop ``What is String Theory? Weaving Perspectives Together'' where this research was supported in part by grant NSF PHY-2309135 to the Kavli Institute for Theoretical Physics (KITP).

\appendix

\section{Kalb-Ramond field}
\label{sec:KBfield}
In this appendix, we evaluate the partition function of the Kalb-Ramond field on the Rindler cone \eqref{eqn:Rindlercone} using similar analysis to the graviton case in section \ref{sect:kabatgraviton}. We extract the contact terms and compute the entropy.

\subsection{Path integral decomposition}
\label{sec:KRfield_PI}

Consider an antisymmetric $2$-tensor field $B_{\mu\nu}$ with field strength $H_{\mu\nu\sigma}=\nabla_{[\mu}B_{\nu\sigma]}$ and action
\begin{equation}
    S = \frac{1}{48} \int\d^{d}x\sqrt{g}\,H^{\mu\nu\sigma}H_{\mu\nu\sigma}\,.
\end{equation}
The partition function of the Kalb-Ramond theory is defined by
\begin{equation}\label{eqn:KBaction}
    Z_\text{Kalb-Ramond}(\beta)=\int \frac{\mathcal{D}B}{\text{Vol}(G_{B})}\exp\left(-\frac{1}{48} \int\d^{d}x\sqrt{g}\,H^{\mu\nu\sigma}H_{\mu\nu\sigma}\right)\,.
\end{equation}
Here the gauge symmetry $G_B$ in question is
\begin{equation}\label{eqn:KRgaugesym}
    B_{\mu\nu} \sim B_{\mu\nu} + \nabla_{\mu}\epsilon_{\nu}-\nabla_{\nu}\epsilon_{\mu}\,,\qquad \epsilon_\mu\sim \epsilon_\mu+\nabla_\mu \zeta\,.
\end{equation}
Such transformations leave invariant $H_{\mu\nu\sigma}$ and hence also the action. The redundancies $\epsilon_\mu=\nabla_\mu \zeta$ do not actually transform $B_{\mu\nu}$, and one only wants to quotient out the path integral by non-trivial gauge transformations. As in section \ref{sect:photon} and \ref{sect:4.1 gravPI}, we will use the $R_{\xi}$-gauge fixing procedure related to the natural generalization of Lorenz gauge for Maxwell theory
\begin{equation}
    \nabla^{\mu}B_{\mu\nu} = 0\,.
\end{equation}
Implementing this $R_{\xi}$-gauge-fixing is more subtle for the Kalb-Ramond field than one might naively think \cite{Townsend:1979hd,Siegel:1980jj,Schwarz:1984wk,Buchbinder:1988tj}. The subtleties come from two issues, both of which stem from the fact that we want an identity like \eqref{eqn:photonFPdetidentity} or \eqref{eqn:Rxigaugeidentity} where the Faddeev-Popov determinant times an integral of a delta-function is 1. Firstly, we have to deal with the zero modes $\epsilon_{\s}=\nabla_{\s}\zeta$ which are excluded from $G_B$. We address this later. Secondly, we want to consider deltas of $\nabla^{\mu}B_{\mu\nu}$ which has the property that $\nabla^{\nu}\nabla^{\mu}B_{\mu\nu}=0$ owing to the anti-symmetry of $B_{\mu\nu}$. The standard definition of a delta-function in quantum field theory for a vector field $v_{\mu}$
\begin{equation}
    \delta(v_{\mu}) = \int \mathcal{D}\gamma \exp\left(\i\int \d^{d}x \sqrt{g}\,v_{\mu}\gamma^{\mu}\right)\,,
\end{equation}
is not correct for fields $v_{\mu}$ which are manifestly divergence-free $\nabla^{\mu}v_{\mu}=0$ since $\gamma_{\mu}$ has a gauge symmetry $\gamma_{\mu} \to \gamma_{\mu}+\nabla_{\mu}\alpha$. We need to mod out by this redundancy to obtain the correct delta function $\delta_\text{div-free}(v_{\mu})$ on manifestly divergence-free vectors $v_\mu$\footnote{Here $G$ is the gauge group corresponding to the scalar field $\alpha$.}
\begin{equation}\label{eqn:deltahat}
    \delta_\text{div-free}(v_{\mu}) = \int \frac{\mathcal{D}\gamma}{\Vol(G)} \exp\left(\i\int \d^{d}x \sqrt{g}\,v_{\mu}\gamma^{\mu}\right)\,.
\end{equation}
For this path integral we can proceed as in section \ref{sect:photon} with Lorenz gauge $\nabla^\mu\gamma_\mu=0$ by inserting into this equation the following identity:
\begin{equation}
    \det(-\Box)\int \mathcal{D}\alpha\, \delta(\nabla_{\mu}\gamma^{\mu}+\Box\alpha) = 1\,,
\end{equation}
then shifting $\gamma_{\mu} \to \gamma_{\mu} - \nabla_{\mu}\alpha$ and performing the path integral over $\alpha$ to cancel the $\Vol(G)$. This results in
\begin{align}
    \delta_\text{div-free}(v_{\mu}) &= \det(-\Box)\int \mathcal{D}\gamma \, \delta(\nabla_{\mu}\gamma^{\mu})\exp\left(\i\int \d^{d}x \sqrt{g}\,v_{\mu}\gamma^{\mu}\right)\label{eqn:deltahat_gauge-fix2}
    \\  &= \det(-\Box)\int \mathcal{D}\gamma \,\mathcal{D}\varphi\,\exp\left(\i\int \d^{d}x \sqrt{g}\,\left(v_{\mu}+\nabla_{\mu}\varphi\right)\gamma^{\mu}\right) = \det(-\Box)\int\mathcal{D}\varphi\, \delta(v_{\mu}+\nabla_{\mu}\varphi)\,.\nonumber
\end{align}

Next, we must deal with the zero-modes. We want to gauge-fix $B_{\mu\nu}$ by considering a path integral over $\epsilon_{\mu}$ of $\delta_\text{div-free}\left(\nabla^{\mu}B_{\mu\nu}+\nabla^{\mu}\left(\nabla_{\mu}\epsilon_{\nu}-\nabla_{\nu}\epsilon_{\mu}\right)\right)$, but $\epsilon_{\mu}$ has the gauge symmetry $\epsilon_{\mu} \to \epsilon_{\mu} + \nabla_{\mu}\zeta$ which leaves this $\delta_\text{div-free}$ invariant. In appendix \ref{sec:identityproof}, we prove the following identity
\begin{equation}\label{eqn:KRid1}
    \det(-\Box)^{-2}\det(-\Box\delta_{\mu}^{\nu})\int \frac{\mathcal{D}\epsilon}{\Vol(G)}\,\delta_\text{div-free}\Big(\nabla^{\mu}B_{\mu\nu}+\nabla^{\mu}\big(\nabla_{\mu}\epsilon_{\nu}-\nabla_{\nu}\epsilon_{\mu}\big)+\theta_{\nu}\Big) = 1\,,
\end{equation}
where $\theta_{\mu}$ is a divergence-free vector.\footnote{The vector determinants can be understood as vector ghosts and the scalar determinants are the well-known ghosts of ghosts for a Kalb-Ramond field. The reason these determinants can be understood as ghosts of ghosts is the following \cite{Townsend:1979hd}. Naively, one would write the Faddeev-Popov determinant using two fermionic vectors as
\begin{equation}
\det(\Box \delta^\s_\n-\nabla_\n\nabla^\s) = \int \mathcal{D}C\,\mathcal{D}C^\ast\, \exp\left(\int\d^{d}x\sqrt{G}\,C_{\m}^{\ast}\left(\Box \delta^{\m\n}-\nabla^\m\nabla^\n\right)C_{\n}\right)\,.
\end{equation}
However, this has redundancies $C_{\m} \to C_{\m}+\nabla_{\m} a$ and $C_{\m}^{\ast} \to C_{\m}^{\ast}+\nabla_{\m} \tilde{a}$ where $a$ and $\tilde{a}$ are fermionic scalars. Therefore, the correct determinant is
\begin{equation}\label{eqn:FPdetB_nozeromode}
\int \frac{\mathcal{D}C}{\Vol(G_{C})}\,\frac{\mathcal{D}C^\ast}{\Vol(G_{C^{\ast}})}\, \exp\left(\int\d^{d}x\sqrt{G}\,C_{\m}^{\ast}\left(\Box \delta^{\m\n}-\nabla^\m\nabla^\n\right)C_{\n}\right)\,.
\end{equation}
These redundancies in $C_{\mu}$, $C_{\mu}^{\ast}$ can then again be dealt with by introducing another set of ghosts, hence ghosts for ghosts.} Inserting this identity into the Kalb-Ramond path integral \eqref{eqn:KBaction} and using \eqref{eqn:deltahat_gauge-fix2} gives
\begin{align}
    &Z_\text{Kalb-Ramond}(\beta)\\ &= \det(-\Box)^{-1}\det(-\Box\delta_{\mu}^{\nu})\int \frac{\mathcal{D}B}{\text{Vol}(G_{B})} \frac{\mathcal{D}\epsilon}{\Vol(G)}\,\mathcal{D}\varphi\,\exp\left(-\frac{1}{48} \int\d^{d}x\sqrt{-G}\,H^{\mu\nu\sigma}H_{\mu\nu\sigma}\right) \nonumber
    \\  &\qquad\qquad\qquad\qquad\qquad\qquad\qquad\qquad \qquad\qquad\qquad\qquad \delta\Big(\nabla^{\mu}B_{\mu\nu}+\nabla^{\mu}\big(\nabla_{\mu}\epsilon_{\nu}-\nabla_{\nu}\epsilon_{\mu}\big)+\theta_{\nu}+\nabla_{\nu}\varphi\Big)  \nonumber
    \\  &= \det(-\Box)^{-1}\det(-\Box\delta_{\mu}^{\nu})\int \mathcal{D}B\,\mathcal{D}\varphi\, \exp\left(-\frac{1}{48} \int\d^{d}x\sqrt{-G}\,H^{\mu\nu\sigma}H_{\mu\nu\sigma}\right)\,\delta\Big(\nabla^{\mu}B_{\mu\nu}+\theta_{\nu}+\nabla_{\nu}\varphi\Big)  \nonumber
    \\  &= \det(-\Box)^{-1}\det(-\Box\delta_{\mu}^{\nu})\int \mathcal{D}B\,\mathcal{D}\varphi\,\exp\bigg\{- \int\d^{d}x\sqrt{-G}\,\bigg(\frac{1}{48}H^{\mu\nu\sigma}H_{\mu\nu\sigma}\nonumber\\&\qquad\qquad\qquad\qquad\qquad\qquad\qquad\qquad\qquad\qquad\qquad\qquad\qquad+\frac{1}{2\xi}\nabla^{\mu}B_{\mu\nu}\nabla_{\gamma}B^{\gamma\nu}+\frac{1}{2\xi}\nabla_{\nu}\varphi\nabla^{\nu}\varphi\bigg)\bigg)\nonumber\,.
\end{align}
In the second equality, we shifted $B_{\mu\nu} \to B_{\mu\nu}-\nabla_{\mu}\epsilon_{\nu}+\nabla_{\nu}\epsilon_{\mu}$ such that we can indeed cancel the volume
\begin{equation}\label{eqn:KRgaugevolume}
    \frac{1}{\text{Vol}(G_{B})}\int \frac{\mathcal{D}\epsilon}{\Vol(G)} = 1\,.
\end{equation}
For the third equality we performed a Gaussian integral over $\theta_{\mu}$\footnote{One uses again with appropriate normalization\begin{equation}
    \int \mathcal{D}\theta \exp\left(-\frac{1}{2\xi}\int\d^{d}x\sqrt{-G}\,\theta_{\mu}\theta^{\mu}\right) = 1\,.
\end{equation}}
and used the fact that $\nabla^{\mu}B_{\mu\nu}$ is divergence-free to drop cross-terms containing $B_{\mu\nu}$ and $\varphi$. The most convenient choice is $\xi=1$ which gives
\begin{align}\label{eqn:KBpartfn_gaugefixed_simpl}
    Z_\text{Kalb-Ramond}(\beta) &= \det(-\Box)^{-3/2}\det(-\Box\, \delta^\sigma_\n)\int \mathcal{D}B\,\exp{-\frac{1}{4}\int_\text{cone} \d^{d}x\, \sqrt{G}\left(-B_{\m\n}\Box B^{\m\n}\right)}\nonumber
    \\ &= \det(-\Box)^{-3/2}\det(-\Box\, \delta^\sigma_\n)\det(-\Box\,\delta^\mu_\sigma \delta^\nu_\kappa)_\text{anti-sym}^{-1/2}\,.
\end{align}
Here we stressed that the final determinant is taken in the space of antisymmetric $2$-tensors as opposed to the symmetric $2$-tensor determinant arising for the graviton in equation \eqref{eqn:gravitonpartfn_gaugefixed_simpl}. 

\subsection{Entropy and contact terms}
\label{sec:KR_entropy}

We can evaluate \eqref{eqn:KBpartfn_gaugefixed_simpl} using heat kernel methods like we did in section \ref{sect:kabatgraviton}. A great deal of this analysis carries over from the graviton calculations so we will be brief. The final determinant in \eqref{eqn:KBpartfn_gaugefixed_simpl} gives
\begin{equation}
    \log Z_\text{anti-sym tensor}(\beta)=\int_0^\infty \frac{\d s}{2s}\int_\text{cone}\d^d x\sqrt{G}\,K_\text{anti-sym tensor}(x|s)\,,\label{4.8 anti-sym tensor determinant}
\end{equation}
with the anti-symmetric tensor heat kernel given by
\begin{equation}
    K_\text{anti-sym tensor}(x\rvert s)=\sum_{\alpha < \beta}\sum_{\ell=-\infty}^{\infty}\int_{0}^{\infty}\d k\,\int_{-\infty}^{+\infty} \d\mathbf{k}\,e^{-s(k^{2}+\mathbf{k}^2)}G^{\mu\sigma}G^{\nu\kappa}\phi^{[\alpha\beta]}_{\mu\nu\,k,\ell,\mathbf{k}}(x)^*\phi^{[\alpha\beta]}_{\sigma\kappa\,k,\ell,\mathbf{k}}(x)\,.\label{4.9 Kanti-symtensdef}
\end{equation}
Here, the normalized anti-symmetric $2$-tensor eigenfunctions are
\begin{equation}\label{eqn:antisym2eigenfns}
        \p^{[\a\b]}_{\lambda\,\m\n}=\frac{1}{\sqrt{2}}(e^{(\a)}_\m e^{(\b)}_\n-e^{(\b)}_\m e^{(\a)}_\n)\,\p_\lambda\,.
\end{equation}
The $\phi^{[ab]}$, $\phi^{[0a]}$, and $\phi^{[1a]}$ polarizations all give the same contributions as in the graviton analysis. The only polarization that behaves differently is $\phi^{[01]}$. As usual $K^{[01]}(r\rvert s) = (4\pi s)^{1-d/2}K^{\text{2d}\,[01]}(r\rvert s)$ and for this 2d kernel one obtains
\begin{align}
    K^{\text{2d}\,[01]}(r\rvert s) &= \sum_{\ell=-\infty}^{\infty}\int_0^\infty \,\d k e^{-s k^2}\frac{1}{k^4} \nabla^{\mu\,\text{2d}} \nabla^{\nu\,\text{2d}}\phi_{k,\ell}(x)^*\nabla_\mu^{\text{2d}} \nabla_\nu^\text{2d} \phi_{k,\ell}(x)\nonumber\\&\qquad\qquad -\sum_{\ell=-\infty}^{\infty}\int_{s}^{\infty}\d p\,\int_0^\infty \d k\, e^{-p k^2}\frac{1}{k^2} \epsilon^{\nu\alpha\,\text{2d}}\epsilon_{\rho\beta}^\text{2d}\nabla^{\rho\,\text{2d}} \nabla_\alpha^\text{2d}\phi_{k, l}(x)^*\nabla_\nu^\text{2d} \nabla^{\beta\,\text{2d}} \phi_{k, l}(x)\nonumber
    \\&= K_\text{scal}^{\text{2d}}(r\rvert s)\,,\label{eqn:K01KR}
\end{align}
where we used $1/2$ times equation \eqref{eqn:Kdiag} and equation \eqref{c2 eq}. We see that the radial derivatives of the scalar or vector heat kernel cancel between the two terms in this equation and so we only have a scalar heat kernel contribution, i.e, there is no contact term from this mode, unlike in the graviton case \eqref{eqn:K01}. 

Summing up all these contributions, the heat kernel for the anti-symmetric tensor is thus
\begin{equation}\label{eqn:tenspropKR}
    K_\text{anti-sym tensor}(r\rvert s) = \frac{d(d-1)}{2}K_\text{scal}(r\rvert s)+(d-2)\,\frac{2s}{r}\partial_{r}K_\text{scal}(r\rvert s)\,.
\end{equation}
The contributions of the different modes are summarized in table \ref{tab:entropycount_KR}. By including the ghosts and ghosts of ghosts, we find the heat kernel for the Kalb-Ramond field
\begin{align}\label{eqn:KRprop}
\begin{split}
  K_\text{Kalb-Ramond}(r\rvert s) &= K_\text{anti-sym tensor}(r\rvert s) - 2\,K_\text{vec}(r\rvert s) + 3K_\text{scal}(r\rvert s)
  \\    &= \frac{(d-2)(d-3)}{2}K_\text{scal}(r\rvert s)+(d-4)\frac{2s}{r}\partial_{r}K_\text{scal}(r\rvert s)\,.
\end{split}
\end{align}
Therefore, the partition function is
\begin{equation}
    \log Z_\text{Kalb-Ramond}(\beta) = \frac{(d-2)(d-3)}{2} \log Z_\text{scal}(\beta)+\log Z_\text{Kalb-Ramond}^\text{contact}\,,\label{a.22 KRz}
\end{equation}
with contact term
\begin{equation}
    \boxed{\log Z_\text{Kalb-Ramond}^\text{contact}=(d-4) \log Z_\text{photon}^\text{contact}}\,.
\end{equation}
The resulting conical entropy \eqref{eqn:entropy} is
\begin{equation}
    \boxed{S_\text{Kalb-Ramond}=2\pi A \int_{\epsilon^{2}}^\infty \frac{\d s}{(4\pi s)^{d/2}}\bigg(\frac{1}{6}\frac{(d-2)(d-3)}{2}-(d-4)\bigg)}\,.
\end{equation}
The authors of \cite{Moitra:2018lxn,Mukherjee:2023ihb} argued that there are $d-2$ edge modes for the Kalb-Ramond field. However, they did not address the zero modes in the gauge transformations $\epsilon_{\mu}=\nabla_{\mu}\zeta$. Perhaps a more careful treatment of those zero modes would result in $d-4$ edge modes, as predicted by our current calculation. It would be interesting to reconcile these calculations in more detail. The calculations of \cite{He:2014gva} seem to predict $d-4$ but only hold for orbifolds $\beta=2\pi/N$. Nevertheless, this is reassuring.

\begin{table}[ht]
\caption{Summary of Kalb-Ramond entropy contributions}
\centering
\begin{tabular}{|c|c|c|c|}
\hline
Modes & Scalar term & Photon contact term & New term \\ \hline
$\p^{[ab]}$ & $(d-2)(d-3)/2$ & $0$ & $0$ \\ \hline
$\p^{[0a]}$ and $\p^{[1a]}$ & $2(d-2)$ & $d-2$ & $0$ \\ \hline
$\p^{[01]}$ & $1$ & $0$ & $0$ \\ \hline
vector ghosts & $-2d$ & $-2$ & 0 \\ \hline
ghosts of ghosts & $3$ & $0$ & 0 \\ \hline
Total & $(d-2)(d-3)/2$ & $d-4$ & $0$\\ \hline
\end{tabular}
\label{tab:entropycount_KR}
\end{table}

\subsection{Useful identity}
\label{sec:identityproof}

We now present the proof of the identity \eqref{eqn:KRid1}:
\begin{equation}\label{eqn:KRid2}
    \det(-\Box)^{-2}\det(-\Box\delta_{\mu}^{\nu})\int \frac{\mathcal{D}\epsilon}{\Vol(G)}\,\delta_\text{div-free}\Big(\nabla^{\mu}B_{\mu\nu}+\nabla^{\mu}\big(\nabla_{\mu}\epsilon_{\nu}-\nabla_{\nu}\epsilon_{\mu}\big)+\theta_{\nu}\Big) = 1\,.
\end{equation}
Firstly, one gauge-fixes the redundancies of $\epsilon_{\mu}$ using the gauge $\nabla_{\mu}\epsilon^{\mu}=\chi$ for an arbitrary scalar field $\chi$
\begin{align}\label{eqn:gauge-fixingthegauge-fixing}
   \int \frac{\mathcal{D}\epsilon}{\Vol(G)}\,\delta_\text{div-free}\Big(\nabla^{\mu}B_{\mu\nu}+&\nabla^{\mu}\big(\nabla_{\mu}\epsilon_{\nu}-\nabla_{\nu}\epsilon_{\mu}\big)+\theta_{\mu}\Big)
   \\   &= \det(-\Box)\int \mathcal{D}\epsilon\,\delta(\nabla_{\mu}\epsilon^{\mu}-\chi)\,\delta_\text{div-free}\left(\nabla^{\mu}B_{\mu\nu}+\nabla^{\mu}\left(\nabla_{\mu}\epsilon_{\nu}-\nabla_{\nu}\epsilon_{\mu}\right)+\theta_{\mu}\right).\nonumber
\end{align}
Furthermore, we will use a version of \eqref{eqn:deltahat_gauge-fix2} for an arbitrary scalar field $\rho$
\begin{equation}\label{eqn:deltahat_gauge-fix3}
    \delta_\text{div-free}(v_{\mu}) = \det(-\Box)\int \mathcal{D}\gamma \, \delta(\nabla_{\mu}\gamma^{\mu}-\rho)\exp\left(\i\int \d^{d}x \sqrt{g}\,v_{\mu}\gamma^{\mu}\right)\,.
\end{equation}
Inserting \eqref{eqn:gauge-fixingthegauge-fixing} and subsequently using \eqref{eqn:deltahat_gauge-fix3} leads to the identity
\begin{align}\label{eqn:KRidentity}
    &\det(-\Box)^{-2}\int \frac{\mathcal{D}\epsilon}{\Vol(G)}\,\delta_\text{div-free}\Big(\nabla^{\mu}B_{\mu\nu}+\nabla^{\mu}\big(\nabla_{\mu}\epsilon_{\nu}-\nabla_{\nu}\epsilon_{\mu}\big)+\theta_{\nu}\Big) 
    \\  &\qquad = \int \mathcal{D}\epsilon\,\mathcal{D}\gamma \,\delta(\nabla_{\mu}\epsilon^{\mu}-\chi)\delta(\nabla_{\mu}\gamma^{\mu}-\rho)\exp\left(\i\int \d^{d}x \sqrt{g}\,\left(\nabla^{\mu}B_{\mu\nu}+\nabla^{\mu}\left(\nabla_{\mu}\epsilon_{\nu}-\nabla_{\nu}\epsilon_{\mu}\right)+\theta_{\nu}\right)\gamma^{\nu}\right).\nonumber
\end{align}
We now path integrate both sides with exponential weighting
\begin{equation}
    \int \mathcal{D}\chi\,\mathcal{D}\rho\,\exp\left(-\i\int \d^{d}x \sqrt{G}\,\chi\rho\right)=1\,.
\end{equation}
This removes the delta-functions on the right side of \eqref{eqn:KRidentity} and results in the claimed identity \eqref{eqn:KRid2}
\begin{align}\label{eqn:KRidentity_final}
    \det(-\Box)^{-2}\det(-\Box\delta_{\mu}^{\nu})&\int \frac{\mathcal{D}\epsilon}{\Vol(G)}\,\delta_\text{div-free}\Big(\nabla^{\mu}B_{\mu\nu}+\nabla^{\mu}\big(\nabla_{\mu}\epsilon_{\nu}-\nabla_{\nu}\epsilon_{\mu}\big)+\theta_{\nu}\Big)
    \\  &= \det(-\Box\delta_{\mu}^{\nu})\int \mathcal{D}\epsilon\,\mathcal{D}\gamma \,\exp\left(\i\int \d^{d}x \sqrt{g}\,\left(\nabla^{\mu}B_{\mu\nu}+\Box\epsilon_{\nu}+\theta_{\nu}\right)\gamma^{\nu}\right)= 1\,.\nonumber
\end{align}

\section{More details on scalar heat kernel}\label{app:scalarkernel}
This appendix contains some details on the scalar heat kernel which are useful in the main text. We start from \eqref{2.8 K}
\begin{equation}
    K_\text{scal}(x\rvert s)=\frac{2\pi}{\beta} \frac{1}{(4\pi s)^{d/2}}\,e^{-\frac{r^{2}}{2s}}\sum_{l=-\infty}^{\infty}I_{\frac{2\pi |\ell|}{\beta}}\left(\frac{r^{2}}{2s}\right)\,.\label{a.1 def}
\end{equation}
Using the integral representation of the modified Bessel in DLMF 10.32.12, one finds that
\begin{equation}\label{a.3 contour}
    e^{-\frac{r^{2}}{2s}}\sum_{l=-\infty}^{\infty}I_{\frac{2\pi |\ell|}{\beta}}\left(\frac{r^{2}}{2s}\right)=\frac{1}{2\pi\i}e^{-\frac{r^2}{2s}}\int_{\infty-\i\pi}^{\infty+\i \pi}\d t\,\frac{e^{\frac{r^2}{2s}\cosh t}}{\tanh(\frac{\pi t}{\beta})}\,.
\end{equation}
The integrand has poles at $t=\i n\beta$ which are outside of (to the left) of the integration contour because they stem from divergences in the integrand for each Bessel function at $t=-\infty$ prior to summing over $\ell$, which thus also lie to the left of the integration contour. One can deform this contour as shown in figure \ref{fig:contourdeform} below.

\begin{figure}[h]
\begin{center}
\begin{tikzpicture}

\draw[->,black] (-6,-2.3) -- (-6,2.3);
\draw[->,black] (-7,0) -- (-3.5,0);
\draw[black](-4.2,1.8) -- (-3.8,1.8);
\draw[black](-4.2,1.8) -- (-4.2,2.2);
\node at (-4,2) {$t$};

\draw[blue,thick,->] (-3.3,-0.8) -- (-3.3,0.1);
\draw[blue,thick] (-3.3,0.1) -- (-3.3,0.8);

\node at (-3.1,0) {$\color{blue} \gamma$};
\node at (-3.3,-1.15) {$\infty-\i\pi$};
\node at (-3.3,1.15) {$\infty+\i\pi$};
\node[scale=2] at (-6,-1.6) {$\color{red} \vdots$};

\draw[red] (-6.1,-1.1) -- (-5.9,-0.9);
\draw[red] (-6.1,-0.9) -- (-5.9,-1.1);
\node at (-6.6,-1) {$-2\i\beta$};

\draw[red] (-6.1,-0.6) -- (-5.9,0.-0.4);
\draw[red] (-6.1,-0.4) -- (-5.9,-0.6);
\node at (-6.5,-0.5) {$-\i\beta$};

\draw[red] (-6.1,-0.1) -- (-5.9,0.1);
\draw[red] (-6.1,0.1) -- (-5.9,-0.1);

\draw[red] (-6.1,0.4) -- (-5.9,0.6);
\draw[red] (-6.1,0.6) -- (-5.9,0.4);
\node at (-6.4,0.5) {$\i\beta$};

\draw[red] (-6.1,0.9) -- (-5.9,1.1);
\draw[red] (-6.1,1.1) -- (-5.9,0.9);
\node at (-6.45,1) {$2\i\beta$};

\node[scale=2] at (-6,2) {$\color{red} \vdots$};

\draw[->,black,thick] (-1,0) -- (1,0);

\draw[->,black] (3,-2.3) -- (3,2.3);
\draw[->,black] (2,0) -- (5.5,0);
\draw[black](4.8,1.8) -- (5.2,1.8);
\draw[black](4.8,1.8) -- (4.8,2.2);
\node at (5,2) {$t$};

\draw[blue,thick,->] (3,-0.8) -- (3,-0.2);
\draw[blue,thick] (3,-0.2) -- (3,0.8);
\draw[blue,thick,->] (3,0.8) -- (4.6,0.8);
\draw[blue,thick] (4.6,0.8) -- (5.8,0.8);
\draw[blue,thick,->] (5.8,-0.8) -- (4.4,-0.8);
\draw[blue,thick] (4.4,-0.8) -- (3,-0.8);

\node at (4.7,-1.1) {$\color{blue} \gamma_{1}$};
\node at (2.7,0.2) {$\color{blue} \gamma_{2}$};
\node at (4.7,1.1) {$\color{blue} \gamma_{3}$};

\node at (6.5,-0.8) {$\infty-\i\pi$};
\node at (6.5,0.8) {$\infty+\i\pi$};

\node[scale=2] at (3,-1.6) {$\color{red} \vdots$};

\draw[red] (3.1,-1.1) -- (2.9,-0.9);
\draw[red] (3.1,-0.9) -- (2.9,-1.1);

\draw[red] (3.1,-0.6) -- (2.9,0.-0.4);
\draw[red] (3.1,-0.4) -- (2.9,-0.6);

\draw[red] (3.1,-0.1) -- (2.9,0.1);
\draw[red] (3.1,0.1) -- (2.9,-0.1);

\draw[red] (3.1,0.4) -- (2.9,0.6);
\draw[red] (3.1,0.6) -- (2.9,0.4);

\draw[red] (3.1,0.9) -- (2.9,1.1);
\draw[red] (3.1,1.1) -- (2.9,0.9);

\node[scale=2] at (3,2) {$\color{red} \vdots$};

\end{tikzpicture}
\end{center}
\caption{Contour deformation for the integral in \eqref{a.3 contour}. There are poles at $t=\i n\beta$ and the contour (blue) is deformed to pick up (half of the residue of) those poles which lie between $-i\pi$ and $i\pi$.}
\label{fig:contourdeform}
\end{figure}
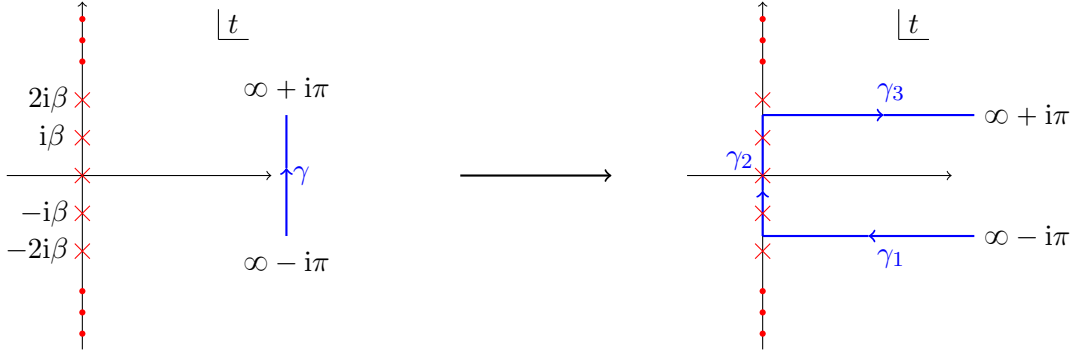

The principal value of this integral along $\gamma_{2}$ vanishes, as the integrand (and boundaries) are odd in $t$. Thus, this piece of the contour integral gives $\i \pi$ times the residues of those poles at $t=\i n\beta$ which lie in between $-\i\pi$ and $\i\pi$. This gives the contribution
\begin{equation}
    e^{-\frac{r^{2}}{2s}}\sum_{l=-\infty}^{\infty}I_{\frac{2\pi |\ell|}{\beta}}\left(\frac{r^{2}}{2s}\right)\bigg|_{\gamma_{2}}= \frac{\beta}{2\pi}\sum_{\abs{n}\beta<\pi}e^{-\frac{r^2}{s}\sin^2(\frac{n\beta}{2})}\,.\label{a.4 half result}
\end{equation}
We then introduce $y=t+\i \pi$ on the top piece $\gamma_{3}$ and $y=t-\i\pi$ on the bottom piece $\gamma_{1}$ in \eqref{a.3 contour} which results in
\begin{align}
    e^{-\frac{r^{2}}{2s}}\sum_{l=-\infty}^{\infty}I_{\frac{2\pi |\ell|}{\beta}}\left(\frac{r^{2}}{2s}\right)\bigg|_{\gamma_{1}+\gamma_{3}} &= -\frac{1}{4\pi}\int_{-\infty}^{\infty}\d y\,e^{-\frac{r^{2}}{s}\cosh^{2}\left(\frac{y}{2}\right)}\left\{\cot\left(\frac{\pi^2}{\beta}+\i \frac{\pi y}{\beta}\right)+\cot\left(\frac{\pi^2}{\beta}-\i \frac{\pi y}{\beta}\right)\right\}\nonumber\\ &=
    \frac{1}{2\pi}\sin(\frac{2\pi^2}{\beta})\int_{-\infty}^{+\infty}\d y\,\frac{e^{-\frac{r^2}{s} \cosh(\frac{y}{2})^2}}{\cosh(\frac{2\pi y}{\beta})-\cos(\frac{2\pi^2}{\beta})}\,.\label{a.5 halfresult}
\end{align}
Combined with \eqref{a.4 half result} this gives the result for the scalar heat kernel \eqref{2.9 Kres} quoted in the main text.

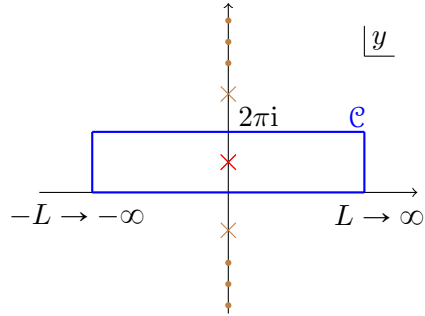
\begin{figure}[h]
\begin{center}
\begin{tikzpicture}

\draw[->,black] (0,-1.6) -- (0,2.5);
\draw[->,black] (-2.5,0) -- (2.5,0);
\draw[black](1.8,1.8) -- (2.2,1.8);
\draw[black](1.8,1.8) -- (1.8,2.2);
\node at (2,2) {$y$};

\draw[blue,thick] (-1.8,0) -- (1.8,0);
\draw[blue,thick] (1.8,0) -- (1.8,0.8);
\draw[blue,thick] (-1.8,0) -- (-1.8,0.8);
\draw[blue,thick] (-1.8,0.8) -- (1.8,0.8);

\node at (1.7,1) {$\color{blue} \mathcal{C}$};
\node at (2,-0.3) {$L \to \infty$};
\node at (-2,-0.3) {$-L \to -\infty$};
\node at (0.4,1) {$2\pi \i$};

\draw[red] (-0.1,0.3) -- (0.1,0.5);
\draw[red] (-0.1,0.5) -- (0.1,0.3);

\draw[brown] (-0.1,-0.4) -- (0.1,-0.6);
\draw[brown] (-0.1,-0.6) -- (0.1,-0.4);

\draw[brown] (-0.1,1.2) -- (0.1,1.4);
\draw[brown] (-0.1,1.4) -- (0.1,1.2);

\node[scale=2] at (0,2.2) {$\color{brown} \vdots$};
\node[scale=2] at (0,-1) {$\color{brown} \vdots$};

\draw[red] (-0.1,0.3) -- (0.1,0.5);
\draw[red] (-0.1,0.5) -- (0.1,0.3);

\end{tikzpicture}
\caption{Rectangular contour for the integral in \eqref{eqn:kernelgamma1+3}. The width $L$ of this contour is taken to infinity.}
\label{fig:yintcontour}
\end{center}
\end{figure}

Next, we consider the trace of the heat kernel. The contribution due to the poles \eqref{a.4 half result} is immediate
\begin{equation}
    \int_\text{cone} \d^d x \sqrt{g}\,K_\text{scal}(x|s)\bigg|_{\gamma_{2}} = \frac{1}{(4\pi s)^{d/2}}\,\int_\text{cone} \d^d x \sqrt{g}+ \frac{A}{(4\pi s)^{d/2}}\frac{s\beta}{2}\underset{\substack{\abs{n}\beta<\pi\,\\n\neq 0}}{\sum}\frac{1}{\sin(\frac{n\beta}{2})^2}\,.\label{a.6 trace}
\end{equation}
The contribution from the integral \eqref{a.5 halfresult} can be rewritten as a contour integral in the complex $y$-plane illustrated in figure \ref{fig:yintcontour} with the result
\begin{equation}\label{eqn:kernelgamma1+3}
    \int_\text{cone} \d^d x \sqrt{g}\,K_\text{scal}(x|s)\bigg|_{\gamma_{1}+\gamma_{3}} = -\frac{A}{(4\pi s)^{d/2}} \frac{s}{4}\oint_\mathcal{C}\d y\frac{\cot\left(\frac{\pi^2}{\beta}+\i \frac{\pi y}{\beta}\right)}{\cosh(\frac{y}{2})^2}\,.\qquad 
\end{equation}

The contributions from the sides of the rectangle vanish due to the exponential in the integrand, while the top and bottom pieces of the rectangular contour give the integral appearing on the first line of \eqref{a.5 halfresult}. We now apply the residue theorem, picking up a third order pole inside the contour at $y=\i \pi$ and simple poles at $y=\i(\pi-n\beta)$ which are inside the contour $\mathcal{C}$ if $\abs{n}\beta<\pi$. This results in
\begin{equation}
     \int_\text{cone} \d^d x \sqrt{g}\,K_\text{scal}(x|s)\bigg|_{\gamma_{1}+\gamma_{3}} = -\frac{A}{(4\pi s)^{d/2}}\frac{s\beta}{2}\underset{\substack{\abs{n}\beta<\pi\,\\n\neq 0}}{\sum}\frac{1}{\sin(\frac{n\beta}{2})^2}+\frac{A}{(4\pi s)^{d/2}}\bigg(\frac{2\pi^2}{3}\frac{s}{\beta}-\frac{s\beta}{6}\bigg)\,.
\end{equation}
The first term cancels with the contributions from the $n\neq0$ poles in \eqref{a.6 trace}, while the other terms reproduce the result \eqref{2.10 traceK} quoted in the main text.

Finally, we consider the Taylor expansion of $K_\text{scal}(x|s)$ near the horizon $r=0$. The constant is clear already from \eqref{a.1 def} because of the identity
\begin{equation}
    I_{\frac{2\pi\abs{\ell}}{\beta}}(0)=\delta_{\ell\,0}\,.
\end{equation}
The first subleading term in the $r$ expansion is computed as follows. Consider for simplicity $\pi<\beta<2\pi$. For convenience, we recall
\begin{equation}
    K_\text{scal}(r|s)\supset -\frac{1}{(4\pi s)^{d/2}}\,\frac{1}{\beta}\sin(\frac{2\pi^2}{\beta})\int_{-\infty}^{+\infty}\d y\,\frac{e^{-\frac{r^2}{s} \cosh(\frac{y}{2})^2}}{\cosh(\frac{2\pi y}{\beta})-\cos(\frac{2\pi^2}{\beta})}\,.
\end{equation}
Therefore,
\begin{align}\label{eqn:STTcontactterm_explicit}
\frac{1}{r}\partial_{r}K_\text{scal}(0|s) &= \frac{1}{(4\pi s)^{d/2}}\,\frac{1}{2s\beta}\sin\left(\frac{2\pi^{2}}{\beta}\right)\int_{-\infty}^{\infty}\d y\,\frac{1+\cosh(y)}{\cosh^{2}\left(\frac{\pi y}{\beta}\right)-\cos^{2}\left(\frac{\pi^{2}}{\beta}\right)}=-\frac{1}{(4\pi s)^{d/2}}\,\frac{2\pi}{\beta}\frac{1}{s}\,,
\end{align}
where we used the following two integrals
\begin{align}
    \int_{-\infty}^{\infty}\d y\,\frac{1}{\cosh^{2}\left(\frac{\pi y}{\beta}\right)-\cos^{2}\left(\frac{\pi^{2}}{\beta}\right)}&= \frac{2\beta-4\pi}{\sin\left(\frac{2\pi^{2}}{\beta}\right)}\,,\quad\int_{-\infty}^{\infty}\d y\,\frac{\cosh y}{\cosh^{2}\left(\frac{\pi y}{\beta}\right)-\cos^{2}\left(\frac{\pi^{2}}{\beta}\right)}&=-\frac{2\beta}{\sin\left(\frac{2\pi^{2}}{\beta}\right)}\,.
\end{align}
The first integral is computed using 2.12.9 and 3.7.11 in \cite{erdelyi1953higher}. The second integral uses 3.7.11, 3.2.3 and 2.8.12 in \cite{erdelyi1953higher}. For $0<\beta<\pi$ one has to be careful with applying these equations, however, contributions from \eqref{a.4 half result} make sure that the final answer in \eqref{eqn:STTcontactterm_explicit} remains valid. One can also plot the first derivative of the full heat kernel \eqref{2.9 Kres} numerically and one finds that the answer matches \eqref{eqn:STTcontactterm_explicit}.\footnote{Note that the integral \eqref{2.9 Kres} vanishes for $\beta=2\pi$ and diverges for $\beta>2\pi$.}

We remark that it is important in this calculation to take the analytic continuation to $\beta=2\pi$ from the region $\beta<2\pi$. The Taylor series around $r=0$ is not an analytic function of $\beta$, as the associated $y$ integral only converges for small enough $\beta$, which is a prerequisite to allow exchanging the integration for the Taylor series. The correct procedure to recover the exact answer is to analytically continue from the region where the Taylor series converges, which is small enough $\beta$. We comment how this is related with a certain amount of debate in the literature concerning which values of $\beta$ to analytically continue from in the computation of the graviton conical entropy in the discussion section \ref{sect:disc}.

The above calculations greatly simplify when we consider orbifolds, i.e.,
\begin{equation}
    \beta=\frac{2\pi}{m}\,, \qquad m = 1,2,3,\ldots\,,
\end{equation}
in which case the cotangents on the first line of \eqref{a.5 halfresult} cancel, leaving only the discrete sum in \eqref{2.9 Kres}. Alternatively, one can directly apply the known summation formula of integer index Bessel functions
\begin{equation}
     K_\text{scal}(x\rvert s)=\frac{m}{(4\pi s)^{d/2}}e^{-\frac{r^2}{2s}}\sum_{\ell=-\infty}^{\infty}I_{m|\ell|}\left(\frac{r^2}{2s}\right)=\frac{1}{(4\pi s)^{d/2}}\sum_{n=0}^{m-1}e^{-\frac{r^2}{s}\sin\left(\frac{\pi n}{m}\right)^2}\,.
\end{equation}
The trace of the scalar heat kernel matches immediately with the generic case \eqref{2.10 traceK}
\begin{align}\label{eqn:scalarheatkerneltrace_orbifold}
   \int_\text{cone} \d^d x \sqrt{g}\,K_\text{scal}(x|s)-\frac{1}{(4\pi s)^{\frac{d}{2}}}\int_\text{cone} \d^d x \sqrt{g}&=\frac{A}{(4\pi s)^{\frac{d}{2}}}\frac{2\pi}{m}\frac{s}{2}\sum_{n=1}^{m-1}\frac{1}{\sin\left(\frac{\pi n}{m}\right)^2}=\frac{A}{(4\pi s)^{\frac{d}{2}}}\frac{2\pi}{m}\frac{s}{6}(m^2-1)\,.
\end{align}
Furthermore, the first term in the Taylor series around $r=0$ matches with \eqref{eqn:STTcontactterm_explicit}
\begin{equation}
    \frac{1}{r}\partial_r K_\text{scal}(0|s)=\frac{1}{(4\pi s)^{\frac{d}{2}}}\frac{1}{s}\sum_{n=0}^{m-1}\bigg(\cos(\frac{2\pi n}{m})-1\bigg)=\frac{1}{(4\pi s)^{\frac{d}{2}}}\frac{1}{s}(-m+\delta_{m\,1})\,.
\end{equation}
As explained, we should analytically continue from $m>1$. For higher spin fields one needs higher order terms in this Taylor series. These would be simpler to obtain via orbifolds instead of working with the generic $\beta$ case.

\section{Mode orthonormality}\label{app:modesortho}
Here we prove orthonormality of the eigenmodes \eqref{2.30 modesgen} of the d'Alembertian $-\Box$ on the Rindler cone. The proof goes as follows. Consider the inner product of two modes
\begin{align}\label{eqn:innerproduct}
    &\int_\text{cone}\d^{d} x \sqrt{G}\,\phi_{\l_1\,\m_1\m_2\dots}^{(\a_1\a_2\dots)}(x)^*\phi_{\l_2\,\n_1\n_2\dots}^{(\b_1\b_2\dots)}(x)\,G^{\m_1\n_1}G^{\m_2\n_2}\dots\nonumber\\&\qquad\qquad\qquad\qquad =\int_\text{cone}\d^{d} x \sqrt{G}\,e^{(\a_1)}_{\m_1} e^{(\a_2)}_{\m_2}\dots \phi_{\lambda_1}(x)^* e^{(\b_1)}_{\n_1} e^{(\b_2)}_{\n_2}\dots \phi_{\lambda_2}(x)\,G^{\m_1\n_1}G^{\m_2\n_2}\dots\nonumber\\&\qquad\qquad\qquad\qquad=\int_\text{cone}\d^{d} x \sqrt{G}\,(-1)^{\#\text{special}}\phi_{\lambda_1}(x)^* e^{\a_1}_{\m_1}e^{\b_1}_{\n_1} G^{\m_1\n_1}\,e^{\a_2}_{\m_2}e^{\b_2}_{\n_2}G^{\m_1\n_1}G^{\m_2\n_2}\dots \phi_{\lambda_2}(x)\,.
\end{align}
Here an index is special if $\a=0,1$ and in those cases one uses integration by parts with the assumption the modes or suitable derivatives vanish on the boundary. We return to this assumption later. Now we have
\begin{equation}
    e^{(\a)}_{\m}e^{(\b)}_{\n} G^{\m\n}\p_\lambda(x) =(-1)^\text{special}\delta^{\a\b}\p_\lambda(x)\,,
\end{equation}
which comes from the fact that for $a$ labels the vectors simply contract, but for special labels they give $\Box^\text{2d}/k^2$ with eigenvalues $-1$. Orthogonality is obvious since the Levi-Cevita contraction of nablas vanishes. Thus, we find
\begin{equation}
    \delta^{\a_1\b_1}\delta^{\a_2\b_2}\dots \int_\text{cone} \d^{d} x\sqrt{G}\,\p_{\lambda_1}(x)^*\p_{\lambda_2}(x)=\delta^{\a_1\b_1}\delta^{\a_2\b_2}\dots\delta_{\lambda_1\lambda_2}\,.
\end{equation}

Regarding the integration by parts, one has to be careful with modes with $0,1$ indices as these have radial derivatives of Bessel functions which may not vanish on the boundary. For photons, one checks that the modes are orthogonal without extra assumptions. For gravitons, the most tricky cases are the $\phi^{(00)}$, $\phi^{(11)}$, and $\phi^{(01)}$ modes with $\abs{\ell}=1$, in which case integration by parts gives a boundary term at $r=0$ that scales as $r^{2(2\pi/\beta-1)}$. Therefore, mode orthonormality requires that we compute the entropy via analytic continuation from the region
\begin{equation}
    \beta<2\pi\,,
\end{equation}
where indeed the boundary terms in the inner product vanish. This is precisely the domain of $\beta$ from which we analytically continued in appendix \ref{app:scalarkernel} to obtain the pieces needed for the graviton heat kernel, as discussed below \eqref{eqn:STTcontactterm_explicit}. See also section \ref{sec:makingsense} and the discussion section \ref{sect:disc}.

\section{More details on graviton heat kernel}
In this appendix we collect various mathematical details required in the graviton calculations of section \ref{sect:kabatgraviton}.

\subsection{Symmetric tensor path integral}\label{app:tensorpathintegral}
The purpose of this appendix is to prove in detail equation \eqref{4.8 tensor determinant} in the main text. The equation states
\begin{align}
    \log \left[\int \mathcal{D}h\,\exp\bigg\{-\frac{1}{4}\int_\text{cone} \d^{d}x\, \sqrt{G}\left(-h_{\m\n}\Box h^{\m\n}+\frac{1}{2}h\Box h\right)\bigg\}\right]=\int_0^\infty \frac{\d s}{2s}\int_\text{cone}\d^d x\sqrt{G}\,K_\text{sym tensor}(x|s)\,,\label{c.1 hPI}
\end{align}
with the symmetric tensor heat kernel given by
\begin{equation}
    K_\text{sym tensor}(x\rvert s)=\sum_{\alpha\leq \beta}\sum_\lambda e^{-s\lambda}\,G^{\mu\sigma}G^{\nu\kappa}\phi^{(\alpha\beta)}_{\lambda\,\mu\nu}(x)^*\phi^{(\alpha\beta)}_{\lambda\,\sigma\kappa}(x)\,.\label{c.2 hPI}
\end{equation}
It is convenient to choose modes as in \eqref{eqn:sym2eigenfns}, except for two diagonal modes for which we flip the signs
\begin{equation}
    \p^{(00)}_{\lambda\,\m\n}=-e^{(0)}_\m e^{(0)}_\n\p_\lambda\,,\quad \p^{(11)}_{\lambda\,\m\n}=-e^{(1)}_\m e^{(1)}_\n\p_\lambda\,.
\end{equation}
This does not affect orthonormality but it removes some minus signs in what follows as it results in the pleasantly simple traces
\begin{equation}
    G^{\mu\nu}\p^{(\a \a)}_{\lambda\,\m\n}=\phi_\l\,.
\end{equation}
The modes are orthonormal with respect to the inner product \eqref{eqn:innerproduct}. The path integral measure is then
\begin{equation}
    \int \mathcal{D}h=\prod_\lambda \prod_{\alpha\leq \beta}\int_{-\infty}^{+\infty}\d c_\lambda^{(\a\b)}\,,\quad h_{\mu\nu}=\sum_\lambda\sum_{\a\leq\b}c_\lambda^{(\a\b)}\,\p_{\lambda\,\mu\nu}^{(\a\b)}\,.
\end{equation}
To evaluate \eqref{c.1 hPI}, it is convenient to do a basis transformation on the diagonal modes as follows. Define
\begin{equation}
    \phi^\text{trace}_{\lambda\,\mu\nu}=\frac{1}{\sqrt{d}}\sum_\alpha \p^{(\a \a)}_{\lambda\,\m\n}\,,\quad \phi^{(i)}_{\lambda\,\mu\nu}=\sum_\alpha u^i_\a\, \p^{(\a \a)}_{\lambda\,\m\n}\,,\quad  \sum_\a u^i_\a=0\,,\quad \sum_\a {u^i_\a}^* u^j_\a=\delta^{i j}\,.\label{C.6 modes}
\end{equation}
An explicit solution for the $u^{j}$ can be found:
\begin{equation}
    u^j = \frac{1}{\sqrt{j(j+1)}}(1^{\otimes j}, -j,0\dots)\,.
\end{equation}
These $d-1$ modes $\phi^{(i)}_{\lambda\,\mu\nu}$ are traceless and together with $\phi^\text{trace}_{\lambda\,\mu\nu}$ they form an orthonormal basis. Thus, the symmetric tensor decomposes as
\begin{equation}
    h_{\mu\nu}=\sum_\lambda\sum_{\a<\b}c_\lambda^{(\a\b)}\,\p_{\lambda\,\mu\nu}^{(\a\b)}+\sum_\lambda c^\text{trace}_\lambda\,\phi^\text{trace}_{\lambda\,\mu\nu}+\sum_i c^{(i)}_\lambda\, \phi^{(i)}_{\lambda\,\mu\nu}\,,
\end{equation}
with the measure invariant under such basis rotation in the diagonal sector
\begin{equation}
    \int \mathcal{D}h=\prod_\lambda \prod_{\alpha<\beta}\int_{-\infty}^{+\infty}\d c_\lambda^{(\a\b)}\prod_{i}\int_{-\infty}^{+\infty}\d c_\lambda^{(i)}\int_{-\i\infty}^{+\i\infty}\d c_\lambda^\text{trace}\,.\label{c.8 meas}
\end{equation}
The action in \eqref{c.1 hPI} evaluates to
\begin{equation}
   e^{-S_{\text{gauge-fixed}}} = \exp\bigg(-\frac{1}{4}\sum_{\a<\b}\sum_\l{c_{\lambda}^{(\a\b)}}^2-\frac{1}{4}\sum_{i}\sum_\l{c_{\lambda}^{(i)}}^2+\frac{d-2}{8}\,\sum_\l{c^\text{trace}_\lambda}^2 \bigg)\,.
\end{equation}

The fact that the prefactor of the last term is positive is the famous conformal factor problem \cite{Gibbons:1978ac}. We fixed this already in the standard manner in \eqref{c.8 meas} by integrating $c_\lambda^\text{trace}$ over an imaginary contour. Since overall multiplicative constants in a determinant do not give contributions to the entropy (see footnote \ref{fn:constant}), the trace-full modes (despite appearances in the action) contribute in the same manner to the path integral \eqref{c.1 hPI} as other modes. Therefore, (ignoring the aforementioned constants, as one always does)
\begin{align}
    &\log \left[\int \mathcal{D}h\,\exp\bigg\{-\frac{1}{4}\int_\text{cone} \d^{d}x\, \sqrt{G}\left(-h_{\m\n}\Box h^{\m\n}+\frac{1}{2}h\Box h\right)\bigg\}\right]\nonumber
    \\&\qquad=\int_0^\infty \frac{\d s}{2s}\int_\text{cone}\d^d x\sqrt{G}\,G^{\mu\sigma}G^{\nu\kappa}\bigg(\sum_{\a<\b}\sum_\lambda e^{-s\l} \phi^{(\alpha\beta)}_{\lambda\,\mu\nu}(x)^*\phi^{(\alpha\beta)}_{\lambda\,\sigma\kappa}(x)\nonumber\\&\qquad\qquad\qquad\qquad\qquad\qquad\qquad+\sum_i\sum_\lambda e^{-s\lambda}\phi^{(i)}_{\lambda\,\mu\nu}(x)^*\phi^{(i)}_{\lambda\,\sigma\kappa}(x)+\sum_\lambda e^{-s\lambda}\phi^{\text{trace}}_{\lambda\,\mu\nu}(x)^*\phi^{\text{trace}}_{\lambda\,\sigma\kappa}(x) \bigg)\,.
\end{align}
The terms on the last line can be rewritten by undoing the basis transformation \eqref{C.6 modes} using completeness
\begin{align}
    &\sum_i\phi^{(i)}_{\lambda\,\mu\nu}(x)^*\phi^{(i)}_{\lambda\,\sigma\kappa}(x)+\phi^{\text{trace}}_{\lambda\,\mu\nu}(x)^*\phi^{\text{trace}}_{\lambda\,\sigma\kappa}(x) \nonumber  \\&\qquad\qquad\qquad\qquad=\sum_{\alpha}\sum_\beta\bigg(\frac{1}{d}+\sum_i {u^i_\a}^*u^i_\b\bigg)\,\phi^{(\a\a)}_{\lambda\,\mu\nu}(x)^*\phi^{(\b\b)}_{\lambda\,\sigma\kappa}(x)=\sum_\alpha \phi^{(\a\a)}_{\lambda\,\mu\nu}(x)^*\phi^{(\a\a)}_{\lambda\,\sigma\kappa}(x)\,.
\end{align}
With this we have arrived at the (unsurprising) expression \eqref{c.2 hPI} for the symmetric tensor heat kernel.

\subsection{Alternative modes}\label{app:differentsymtracelessmodes}
The above evaluation, even though highly symmetric, was a bit subtle due to the fact that our modes \eqref{eqn:sym2eigenfns} are not traceless. As a sanity check on our answer for the graviton heat kernel in section \ref{sect:4.2 gravK}, here we evaulate the graviton heat kernel using the inherently traceless basis of modes in $d=4$ constructed in \cite{Iellici:1996gv}. Their orthonormal modes are given by
\begin{align}\label{eqn:altmodes}
\begin{split}
    (h_{\alpha\beta}^{(1)})_{k,\ell,\mathbf{k}} &= \frac{\sqrt{2}}{k^{2}}\nabla_{\alpha}\nabla_{\beta}\phi_{k,\ell,\mathbf{k}}+\frac{1}{\sqrt{2}}g_{\alpha\beta}\phi_{k,\ell,\mathbf{k}}\,, \qquad (h_{\alpha i}^{(1)})_{k,\ell,\mathbf{k}} = (h_{ij}^{(1)})_{k,\ell,\mathbf{k}} = 0
\\    (h_{\alpha\beta}^{(2)})_{k,\ell,\mathbf{k}} &= \frac{\sqrt{2}}{k^{2}}\epsilon_{(\alpha\gamma}\nabla^{\gamma}\nabla_{\beta)}\phi_{k,\ell,\mathbf{k}}\,, \qquad (h_{\alpha i}^{(2)})_{k,\ell,\mathbf{k}} = (h_{ij}^{(2)})_{k,\ell,\mathbf{k}} = 0
\\    (h_{\alpha i}^{(3)})_{k,\ell,\mathbf{k}} &= \frac{1}{\sqrt{2}kk}\nabla_{\alpha}^{\text{2d}}\nabla_{i}\phi_{k,\ell,\mathbf{k}}\,, \qquad (h_{\alpha \beta}^{(3)})_{k,\ell,\mathbf{k}} = (h_{ij}^{(3)})_{k,\ell,\mathbf{k}} = 0
\\    (h_{\alpha i}^{(4)})_{k,\ell,\mathbf{k}} &= \frac{1}{\sqrt{2}kk}\epsilon_{\alpha\beta}\nabla^{\beta}\nabla_{i}\phi_{k,\ell,\mathbf{k}}\,, \qquad (h_{\alpha \beta}^{(4)})_{k,\ell,\mathbf{k}} = (h_{ij}^{(4)})_{k,\ell,\mathbf{k}} = 0
\\    (h_{\alpha i}^{(5)})_{k,\ell,\mathbf{k}} &= \frac{1}{\sqrt{2}kk}\epsilon_{\alpha\beta}\epsilon_{ij}\nabla^{\beta}\nabla^{j}\phi_{k,\ell,\mathbf{k}}\,, \qquad (h_{\alpha \beta}^{(5)})_{k,\ell,\mathbf{k}} = (h_{ij}^{(5)})_{k,\ell,\mathbf{k}} = 0
\\    (h_{\alpha i}^{(6)})_{k,\ell,\mathbf{k}} &= \frac{1}{\sqrt{2}kk}\epsilon_{ij}\nabla_{\alpha}\nabla^{j}\phi_{k,\ell,\mathbf{k}}\,, \qquad (h_{\alpha \beta}^{(6)})_{k,\ell,\mathbf{k}} = (h_{ij}^{(6)})_{k,\ell,\mathbf{k}} = 0
\\    (h_{ij}^{(7)})_{k,\ell,\mathbf{k}} &= \frac{\sqrt{2}}{k^{2}}\nabla_{i}\nabla_{j}\phi_{k,\ell,\mathbf{k}}+\frac{1}{\sqrt{2}}g_{ij}\phi_{k,\ell,\mathbf{k}}\,, \qquad (h_{\alpha \beta}^{(7)})_{k,\ell,\mathbf{k}} = (h_{\alpha i}^{(7)})_{k,\ell,\mathbf{k}} = 0
\\    (h_{ij}^{(8)})_{k,\ell,\mathbf{k}} &= \frac{\sqrt{2}}{k^{2}}\epsilon_{(ik}\nabla^{k}\nabla_{j)}\phi_{k,\ell,\mathbf{k}}\,, \qquad (h_{\alpha \beta}^{(8)})_{k,\ell,\mathbf{k}} = (h_{\alpha i}^{(2)})_{k,\ell,\mathbf{k}} = 0
\\    (h_{\alpha\beta}^{(9)})_{k,\ell,\mathbf{k}} &= \frac{1}{2}g_{\alpha\beta}\phi_{k,\ell,\mathbf{k}}\,, \qquad (h_{ij}^{(9)})_{k,\ell,\mathbf{k}} = -\frac{1}{2}g_{ij}\phi_{k,\ell,\mathbf{k}}, \qquad (h_{\alpha i}^{(9)})_{k,\ell,\mathbf{k}} = 0
\end{split}
\end{align}
where $\alpha,\beta,\ldots$ label coordinates along the 2d cone and $i,j,\ldots$ label the transverse coordinates. The contribution to the heat kernel of these modes is
\begin{equation}\label{eqn:heatkernelmodes12}
K^{(1)}(r|s) = K^{(2)}(r|s) = K_\text{scal}(r|s)+\frac{2s}{r}\partial_r K_\text{scal}(r|s)+\frac{1}{4\pi s}\frac{s}{r}\partial_{r}K_\text{vec}^{\text{2d}}(r\rvert s)
\end{equation}
and
\begin{equation}\label{eqn:heatkernelmodes3456}
K^{(3)}(r|s) = K^{(4)}(r|s) = K^{(5)}(r|s) = K^{(6)}(r|s) = K_\text{scal}(r|s)+\frac{s}{r}\partial_r K_\text{scal}(r|s)
\end{equation}
and
\begin{equation}\label{eqn:heatkernelmodes789}
K^{(7)}(r|s) = K^{(8)}(r|s) = K^{(9)}(r|s) = K_\text{scal}(r|s)\,. 
\end{equation}
The results are summarized in table \ref{tab:entropycount2} and agree with our final result \eqref{eqn:gravprop} in the main text for $d=4$.

\begin{table}[ht]
\caption{Summary of entropy contributions using alternate modes}
\centering
\begin{tabular}{|c|c|c|c|}
\hline
Modes & Scalar term & Vector contact term & New term \\ \hline
$h^{(1)}$ & $1$ & $1$ & $1$ \\ \hline
$h^{(2)}$ & $1$ & $1$ & $1$ \\ \hline
$h^{(3)}$, $h^{(4)}$, $h^{(5)}$, $h^{(6)}$ & $4$ & $2$ & $0$ \\ \hline
$h^{(7)}$, $h^{(8)}$, $h^{(9)}$ & $3$ & $0$ & $0$ \\ \hline
$h$ & $1$ & $0$ & $0$\\ \hline
vector ghosts & $-8$ & $-2$ & 0 \\ \hline
Total & $2$ & $2$ & $2$\\ \hline
\end{tabular}
\label{tab:entropycount2}
\end{table}

\subsection{Some heat kernel manipulations}\label{app:c2}
In this short appendix we derive equation \eqref{c2 eq}. We drop most of the $\dots^\text{2d}$ superscripts in this appendix for notational comfort. The derivation is
\begin{align}
    &\int_{s}^{\infty}\d p\,\int_0^\infty \d k\, e^{-p k^2}\frac{1}{k^2}\sum_\ell \epsilon^{\nu\alpha}\epsilon_{\rho\beta}\nabla^\rho \nabla_\alpha\phi_{k, l}(x)^*\nabla_\nu \nabla^\beta \phi_{k, l}(x)
    \nonumber\\ \nonumber  &\qquad\qquad\qquad= \int_{s}^{\infty}\d p\,\int_0^\infty \d k\, e^{-p k^2}\sum_\ell \left(\frac{1}{k^2}\nabla_{\beta}\nabla^{\nu}\left(\phi_{k, l}(x)^*\nabla_{\nu}\nabla^{\beta}\phi_{k, l}(x)\right)+\Box\left(\phi_{k, l}(x)^*\phi_{k, l}(x)\right)\right)
    \\ \nonumber &\qquad\qquad\qquad= \frac{1}{2}\int_{s}^{\infty}\d p\,\int \d k\, e^{-p k^2}\sum_\ell \Box\left(\frac{1}{k^2}\nabla_{\nu}\phi_{k, l}(x)^*\nabla^{\nu}\phi_{k, l}(x)+\phi_{k, l}(x)^*\phi_{k, l}(x)\right) 
     \\  &\qquad\qquad\qquad= \frac{1}{2}\int_{s}^{\infty}\d p\,\Box\left(\frac{1}{2}K^{\text{2d}}_\text{vec}(r|p)+K_\text{scal}^{2d}(r|p)\right)\,.\label{eqn:newterm2}
\end{align}
In the first equality, we used that in 2d
\begin{equation}
    \varepsilon_{\mu\nu}\varepsilon^{\alpha\beta}=\delta_{\mu}^\alpha\delta_\nu^\beta-\delta^\alpha_\nu\delta_\mu^\beta\,,
\end{equation}
then by writing out the covariant derivatives on the second line via the Leibniz rule one indeed recovers the first line.
In the second equality, we used the following series of manipulations
\begin{align}
\nonumber &\nabla_{\beta}\nabla^{\nu}\left(\phi_{k, l}(x)^*\nabla_{\nu}\nabla^{\beta}\phi_{k, l}(x)\right) = \nabla_{\beta}\nabla^{\nu}\left(\nabla_{\nu}\left(\phi_{k, l}(x)^*\nabla^{\beta}\phi_{k, l}(x)\right)-\nabla_{\nu}\phi_{k, l}(x)^*\nabla^{\beta}\phi_{k, l}(x)\right)
\\ \nonumber &\qquad\qquad= \frac{1}{2}\Box\Big(\Box\left(\phi_{k, l}(x)^*\phi_{k, l}(x)\right)\Big)-\nabla_{\beta}\left(-k^{2}\phi_{k, l}(x)^*\nabla^{\beta}\phi_{k, l}(x)+\nabla_{\nu}\phi_{k, l}(x)^*\nabla^{\nu}\nabla^{\beta}\phi_{k, l}(x)\right)
\\ \nonumber  &\qquad\qquad= \frac{1}{2}\Box\Big(\Box\left(\phi_{k, l}(x)^*\phi_{k, l}(x)\right)+k^{2}\phi_{k, l}(x)^*\phi_{k, l}(x)-\nabla_{\nu}\phi_{k, l}(x)^*\nabla^{\nu}\phi_{k, l}(x)\Big)
\\  &\qquad\qquad= \frac{1}{2}\Box\Big(\nabla_{\nu}\phi_{k, l}(x)^*\nabla^{\nu}\phi_{k, l}(x)-k^{2}\phi_{k, l}(x)^*\phi_{k, l}(x)\Big).
\end{align}
The final equality in \eqref{eqn:newterm2} follows from recognizing the scalar \eqref{2.8 K} and vector \eqref{eqn:Kvec} heat kernels in 2d.

\bibliographystyle{ourbst}
\bibliography{Refs}

\providecommand{\href}[2]{#2}\begingroup\raggedright\begin{thebibliography}{10}

\bibitem{Donnelly:2011hn}
W.~Donnelly, ``{Decomposition of entanglement entropy in lattice gauge theory},'' \href{http://dx.doi.org/10.1103/PhysRevD.85.085004}{{\em Phys. Rev. D} {\bfseries 85} (2012) 085004}, \href{http://arxiv.org/abs/1109.0036}{{\ttfamily arXiv:1109.0036 [hep-th]}}.

\bibitem{Casini:2013rba}
H.~Casini, M.~Huerta, and J.~A. Rosabal, ``{Remarks on entanglement entropy for gauge fields},'' \href{http://dx.doi.org/10.1103/PhysRevD.89.085012}{{\em Phys. Rev.} {\bfseries D89} no.~8, (2014) 085012},
\href{http://arxiv.org/abs/1312.1183}{{\ttfamily arXiv:1312.1183 [hep-th]}}.

\bibitem{Dong:2008ft}
S.~Dong, E.~Fradkin, R.~G. Leigh, and S.~Nowling, ``{Topological Entanglement Entropy in Chern-Simons Theories and Quantum Hall Fluids},'' \href{http://dx.doi.org/10.1088/1126-6708/2008/05/016}{{\em JHEP} {\bfseries 05} (2008) 016}, \href{http://arxiv.org/abs/0802.3231}{{\ttfamily arXiv:0802.3231 [hep-th]}}.

\bibitem{Qi_2012}
X.-L. Qi, H.~Katsura, and A.~W.~W. Ludwig, ``General relationship between the entanglement spectrum and the edge state spectrum of topological quantum states,'' \href{http://dx.doi.org/10.1103/PhysRevLett.108.196402}{{\em Phys. Rev. Lett.} {\bfseries 108} no.~19, (2012) 196402}, \href{http://arxiv.org/abs/1103.5437}{{\ttfamily arXiv:1103.5437 [cond-mat.mes-hall]}}.

\bibitem{Das:2015oha}
D.~Das and S.~Datta, ``{Universal features of left-right entanglement entropy},'' \href{http://dx.doi.org/10.1103/PhysRevLett.115.131602}{{\em Phys. Rev. Lett.} {\bfseries 115} no.~13, (2015) 131602}, \href{http://arxiv.org/abs/1504.02475}{{\ttfamily arXiv:1504.02475 [hep-th]}}.

\bibitem{Wen:2016snr}
X.~Wen, S.~Matsuura, and S.~Ryu, ``{Edge theory approach to topological entanglement entropy, mutual information and entanglement negativity in Chern-Simons theories},'' \href{http://dx.doi.org/10.1103/PhysRevB.93.245140}{{\em Phys. Rev. B} {\bfseries 93} no.~24, (2016) 245140}, \href{http://arxiv.org/abs/1603.08534}{{\ttfamily arXiv:1603.08534 [cond-mat.mes-hall]}}.

\bibitem{Wong:2017pdm}
G.~Wong, ``{A note on entanglement edge modes in Chern Simons theory},'' \href{http://dx.doi.org/10.1007/JHEP08(2018)020}{{\em JHEP} {\bfseries 08} (2018) 020}, \href{http://arxiv.org/abs/1706.04666}{{\ttfamily arXiv:1706.04666 [hep-th]}}.

\bibitem{Fliss:2020cos}
J.~R. Fliss and R.~G. Leigh, ``{Interfaces and the extended Hilbert space of Chern-Simons theory},'' \href{http://dx.doi.org/10.1007/JHEP07(2020)009}{{\em JHEP} {\bfseries 07} (2020) 009}, \href{http://arxiv.org/abs/2004.05123}{{\ttfamily arXiv:2004.05123 [hep-th]}}.

\bibitem{Geiller:2019bti}
M.~Geiller and P.~Jai-akson, ``{Extended actions, dynamics of edge modes, and entanglement entropy},'' \href{http://dx.doi.org/10.1007/JHEP09(2020)134}{{\em JHEP} {\bfseries 09} (2020) 134}, \href{http://arxiv.org/abs/1912.06025}{{\ttfamily arXiv:1912.06025 [hep-th]}}.

\bibitem{Fliss:2023dze}
J.~R. Fliss and S.~Vitouladitis, ``{Entanglement in BF theory I: Essential topological entanglement},'' \href{http://arxiv.org/abs/2306.06158}{{\ttfamily arXiv:2306.06158 [hep-th]}}.

\bibitem{Fliss:2023uiv}
J.~R. Fliss and S.~Vitouladitis, ``{Entanglement in BF theory II: Edge-modes},'' \href{http://arxiv.org/abs/2310.18391}{{\ttfamily arXiv:2310.18391 [hep-th]}}.

\bibitem{Buividovich:2008gq}
P.~V. Buividovich and M.~I. Polikarpov, ``{Entanglement entropy in gauge theories and the holographic principle for electric strings},'' \href{http://dx.doi.org/10.1016/j.physletb.2008.10.032}{{\em Phys. Lett. B} {\bfseries 670} (2008) 141--145}, \href{http://arxiv.org/abs/0806.3376}{{\ttfamily arXiv:0806.3376 [hep-th]}}.

\bibitem{Radicevic:2014kqa}
D.~Radicevic, ``{Notes on Entanglement in Abelian Gauge Theories},'' \href{http://arxiv.org/abs/1404.1391}{{\ttfamily arXiv:1404.1391 [hep-th]}}.

\bibitem{Donnelly:2014fua}
W.~Donnelly and A.~C. Wall, ``{Entanglement entropy of electromagnetic edge modes},'' \href{http://dx.doi.org/10.1103/PhysRevLett.114.111603}{{\em Phys. Rev. Lett.} {\bfseries 114} no.~11, (2015) 111603}, \href{http://arxiv.org/abs/1412.1895}{{\ttfamily arXiv:1412.1895 [hep-th]}}.

\bibitem{Donnelly:2015hxa}
W.~Donnelly and A.~C. Wall, ``{Geometric entropy and edge modes of the electromagnetic field},'' \href{http://dx.doi.org/10.1103/PhysRevD.94.104053}{{\em Phys. Rev. D} {\bfseries 94} no.~10, (2016) 104053}, \href{http://arxiv.org/abs/1506.05792}{{\ttfamily arXiv:1506.05792 [hep-th]}}.

\bibitem{Huang:2014pfa}
K.-W. Huang, ``{Central Charge and Entangled Gauge Fields},'' \href{http://dx.doi.org/10.1103/PhysRevD.92.025010}{{\em Phys. Rev. D} {\bfseries 92} no.~2, (2015) 025010}, \href{http://arxiv.org/abs/1412.2730}{{\ttfamily arXiv:1412.2730 [hep-th]}}.

\bibitem{Ghosh:2015iwa}
S.~Ghosh, R.~M. Soni, and S.~P. Trivedi, ``{On The Entanglement Entropy For Gauge Theories},'' \href{http://dx.doi.org/10.1007/JHEP09(2015)069}{{\em JHEP} {\bfseries 09} (2015) 069}, \href{http://arxiv.org/abs/1501.02593}{{\ttfamily arXiv:1501.02593 [hep-th]}}.

\bibitem{Soni:2015yga}
R.~M. Soni and S.~P. Trivedi, ``{Aspects of Entanglement Entropy for Gauge Theories},'' \href{http://dx.doi.org/10.1007/JHEP01(2016)136}{{\em JHEP} {\bfseries 01} (2016) 136}, \href{http://arxiv.org/abs/1510.07455}{{\ttfamily arXiv:1510.07455 [hep-th]}}.

\bibitem{Soni:2016ogt}
R.~M. Soni and S.~P. Trivedi, ``{Entanglement entropy in (3 + 1)-d free U(1) gauge theory},'' \href{http://dx.doi.org/10.1007/JHEP02(2017)101}{{\em JHEP} {\bfseries 02} (2017) 101}, \href{http://arxiv.org/abs/1608.00353}{{\ttfamily arXiv:1608.00353 [hep-th]}}.

\bibitem{Blommaert:2018rsf}
A.~Blommaert, T.~G. Mertens, H.~Verschelde, and V.~I. Zakharov, ``{Edge State Quantization: Vector Fields in Rindler},'' \href{http://dx.doi.org/10.1007/JHEP08(2018)196}{{\em JHEP} {\bfseries 08} (2018) 196}, \href{http://arxiv.org/abs/1801.09910}{{\ttfamily arXiv:1801.09910 [hep-th]}}.

\bibitem{Blommaert:2018oue}
A.~Blommaert, T.~G. Mertens, and H.~Verschelde, ``{Edge dynamics from the path integral \textemdash{} Maxwell and Yang-Mills},'' \href{http://dx.doi.org/10.1007/JHEP11(2018)080}{{\em JHEP} {\bfseries 11} (2018) 080}, \href{http://arxiv.org/abs/1804.07585}{{\ttfamily arXiv:1804.07585 [hep-th]}}.

\bibitem{Ball:2024hqe}
A.~Ball, A.~Law, and G.~Wong, ``{Dynamical Edge Modes and Entanglement in Maxwell Theory},'' \href{http://arxiv.org/abs/2403.14542}{{\ttfamily arXiv:2403.14542 [hep-th]}}.

\bibitem{Harlow:2015lma}
D.~Harlow, ``{Wormholes, Emergent Gauge Fields, and the Weak Gravity Conjecture},'' \href{http://dx.doi.org/10.1007/JHEP01(2016)122}{{\em JHEP} {\bfseries 01} (2016) 122}, \href{http://arxiv.org/abs/1510.07911}{{\ttfamily arXiv:1510.07911 [hep-th]}}.

\bibitem{Donnelly:2016auv}
W.~Donnelly and L.~Freidel, ``{Local subsystems in gauge theory and gravity},'' \href{http://dx.doi.org/10.1007/JHEP09(2016)102}{{\em JHEP} {\bfseries 09} (2016) 102}, \href{http://arxiv.org/abs/1601.04744}{{\ttfamily arXiv:1601.04744 [hep-th]}}.

\bibitem{Geiller:2017xad}
M.~Geiller, ``{Edge modes and corner ambiguities in 3d Chern\textendash{}Simons theory and gravity},'' \href{http://dx.doi.org/10.1016/j.nuclphysb.2017.09.010}{{\em Nucl. Phys. B} {\bfseries 924} (2017) 312--365}, \href{http://arxiv.org/abs/1703.04748}{{\ttfamily arXiv:1703.04748 [gr-qc]}}.

\bibitem{Geiller:2017whh}
M.~Geiller, ``{Lorentz-diffeomorphism edge modes in 3d gravity},'' \href{http://dx.doi.org/10.1007/JHEP02(2018)029}{{\em JHEP} {\bfseries 02} (2018) 029}, \href{http://arxiv.org/abs/1712.05269}{{\ttfamily arXiv:1712.05269 [gr-qc]}}.

\bibitem{Freidel:2019ees}
L.~Freidel, E.~R. Livine, and D.~Pranzetti, ``{Gravitational edge modes: from Kac\textendash{}Moody charges to Poincar\'e networks},'' \href{http://dx.doi.org/10.1088/1361-6382/ab40fe}{{\em Class. Quant. Grav.} {\bfseries 36} no.~19, (2019) 195014}, \href{http://arxiv.org/abs/1906.07876}{{\ttfamily arXiv:1906.07876 [hep-th]}}.

\bibitem{Freidel:2020xyx}
L.~Freidel, M.~Geiller, and D.~Pranzetti, ``{Edge modes of gravity. Part I. Corner potentials and charges},'' \href{http://dx.doi.org/10.1007/JHEP11(2020)026}{{\em JHEP} {\bfseries 11} (2020) 026}, \href{http://arxiv.org/abs/2006.12527}{{\ttfamily arXiv:2006.12527 [hep-th]}}.

\bibitem{Donnelly:2020xgu}
W.~Donnelly, L.~Freidel, S.~F. Moosavian, and A.~J. Speranza, ``{Gravitational edge modes, coadjoint orbits, and hydrodynamics},'' \href{http://dx.doi.org/10.1007/JHEP09(2021)008}{{\em JHEP} {\bfseries 09} (2021) 008}, \href{http://arxiv.org/abs/2012.10367}{{\ttfamily arXiv:2012.10367 [hep-th]}}.

\bibitem{Ciambelli:2021vnn}
L.~Ciambelli and R.~G. Leigh, ``{Isolated surfaces and symmetries of gravity},'' \href{http://dx.doi.org/10.1103/PhysRevD.104.046005}{{\em Phys. Rev. D} {\bfseries 104} no.~4, (2021) 046005}, \href{http://arxiv.org/abs/2104.07643}{{\ttfamily arXiv:2104.07643 [hep-th]}}.

\bibitem{Ciambelli:2021nmv}
L.~Ciambelli, R.~G. Leigh, and P.-C. Pai, ``{Embeddings and Integrable Charges for Extended Corner Symmetry},'' \href{http://dx.doi.org/10.1103/PhysRevLett.128.171302}{{\em Phys. Rev. Lett.} {\bfseries 128} (2022) 171302}, \href{http://arxiv.org/abs/2111.13181}{{\ttfamily arXiv:2111.13181 [hep-th]}}.

\bibitem{Ciambelli:2022cfr}
L.~Ciambelli and R.~G. Leigh, ``{Universal corner symmetry and the orbit method for gravity},'' \href{http://dx.doi.org/10.1016/j.nuclphysb.2022.116053}{{\em Nucl. Phys. B} {\bfseries 986} (2023) 116053}, \href{http://arxiv.org/abs/2207.06441}{{\ttfamily arXiv:2207.06441 [hep-th]}}.

\bibitem{Donnelly:2022kfs}
W.~Donnelly, L.~Freidel, S.~F. Moosavian, and A.~J. Speranza, ``{Matrix Quantization of Gravitational Edge Modes},'' \href{http://dx.doi.org/10.1007/JHEP05(2023)163}{{\em JHEP} {\bfseries 05} (2027) 163}, \href{http://arxiv.org/abs/2212.09120}{{\ttfamily arXiv:2212.09120 [hep-th]}}.

\bibitem{David:2022jfd}
J.~R. David and J.~Mukherjee, ``{Entanglement entropy of gravitational edge modes},'' \href{http://dx.doi.org/10.1007/JHEP08(2022)065}{{\em JHEP} {\bfseries 08} (2022) 065}, \href{http://arxiv.org/abs/2201.06043}{{\ttfamily arXiv:2201.06043 [hep-th]}}.

\bibitem{bekenstein1973black}
J.~D. Bekenstein, ``Black holes and entropy,'' {\em Physical Review D} {\bfseries 7} no.~8, (1973) 2333.

\bibitem{Page:1993df}
D.~N. Page, ``{Average entropy of a subsystem},'' \href{http://dx.doi.org/10.1103/PhysRevLett.71.1291}{{\em Phys. Rev. Lett.} {\bfseries 71} (1993) 1291--1294}, \href{http://arxiv.org/abs/gr-qc/9305007}{{\ttfamily arXiv:gr-qc/9305007}}.

\bibitem{Kabat:1995eq}
D.~N. Kabat, ``{Black hole entropy and entropy of entanglement},'' \href{http://dx.doi.org/10.1016/0550-3213(95)00443-V}{{\em Nucl. Phys. B} {\bfseries 453} (1995) 281--299}, \href{http://arxiv.org/abs/hep-th/9503016}{{\ttfamily arXiv:hep-th/9503016}}.

\bibitem{Carlip:1993sa}
S.~Carlip and C.~Teitelboim, ``{The Off-shell black hole},'' \href{http://dx.doi.org/10.1088/0264-9381/12/7/011}{{\em Class. Quant. Grav.} {\bfseries 12} (1995) 1699--1704}, \href{http://arxiv.org/abs/gr-qc/9312002}{{\ttfamily arXiv:gr-qc/9312002}}.

\bibitem{Susskind:1994vu}
L.~Susskind, ``{The World as a hologram},'' \href{http://dx.doi.org/10.1063/1.531249}{{\em J. Math. Phys.} {\bfseries 36} (1995) 6377--6396}, \href{http://arxiv.org/abs/hep-th/9409089}{{\ttfamily arXiv:hep-th/9409089}}.

\bibitem{Blommaert:2024}
A.~Blommaert and S.~Colin-Ellerin, ``{Graviton edge modes},'' {\em in progress} , \href{http://arxiv.org/abs/24xx.xxxxx}{{\ttfamily arXiv:24xx.xxxxx}}.

\bibitem{Susskind:1993ws}
L.~Susskind, ``{Some speculations about black hole entropy in string theory},'' \href{http://arxiv.org/abs/hep-th/9309145}{{\ttfamily arXiv:hep-th/9309145}}.

\bibitem{Susskind:1994sm}
L.~Susskind and J.~Uglum, ``{Black hole entropy in canonical quantum gravity and superstring theory},'' \href{http://dx.doi.org/10.1103/PhysRevD.50.2700}{{\em Phys. Rev. D} {\bfseries 50} (1994) 2700--2711}, \href{http://arxiv.org/abs/hep-th/9401070}{{\ttfamily arXiv:hep-th/9401070}}.

\bibitem{Kabat:1995jq}
D.~N. Kabat, S.~H. Shenker, and M.~J. Strassler, ``{Black hole entropy in the O(N) model},'' \href{http://dx.doi.org/10.1103/PhysRevD.52.7027}{{\em Phys. Rev. D} {\bfseries 52} (1995) 7027--7036}, \href{http://arxiv.org/abs/hep-th/9506182}{{\ttfamily arXiv:hep-th/9506182}}.

\bibitem{Balasubramanian:2018axm}
V.~Balasubramanian and O.~Parrikar, ``{Remarks on entanglement entropy in string theory},'' \href{http://dx.doi.org/10.1103/PhysRevD.97.066025}{{\em Phys. Rev. D} {\bfseries 97} no.~6, (2018) 066025}, \href{http://arxiv.org/abs/1801.03517}{{\ttfamily arXiv:1801.03517 [hep-th]}}.

\bibitem{Ahmadain:2022eso}
A.~Ahmadain and A.~C. Wall, ``{Off-Shell Strings II: Black Hole Entropy},'' \href{http://arxiv.org/abs/2211.16448}{{\ttfamily arXiv:2211.16448 [hep-th]}}.

\bibitem{Donnelly:2020teo}
W.~Donnelly, Y.~Jiang, M.~Kim, and G.~Wong, ``{Entanglement entropy and edge modes in topological string theory. Part I. Generalized entropy for closed strings},'' \href{http://dx.doi.org/10.1007/JHEP10(2021)201}{{\em JHEP} {\bfseries 10} (2021) 201}, \href{http://arxiv.org/abs/2010.15737}{{\ttfamily arXiv:2010.15737 [hep-th]}}.

\bibitem{Witten:2018xfj}
E.~Witten, ``{Open Strings On The Rindler Horizon},'' \href{http://dx.doi.org/10.1007/JHEP01(2019)126}{{\em JHEP} {\bfseries 01} (2019) 126}, \href{http://arxiv.org/abs/1810.11912}{{\ttfamily arXiv:1810.11912 [hep-th]}}.

\bibitem{He:2014gva}
S.~He, T.~Numasawa, T.~Takayanagi, and K.~Watanabe, ``{Notes on Entanglement Entropy in String Theory},'' \href{http://dx.doi.org/10.1007/JHEP05(2015)106}{{\em JHEP} {\bfseries 05} (2015) 106}, \href{http://arxiv.org/abs/1412.5606}{{\ttfamily arXiv:1412.5606 [hep-th]}}.

\bibitem{Dabholkar:1994ai}
A.~Dabholkar, ``{Strings on a cone and black hole entropy},'' \href{http://dx.doi.org/10.1016/0550-3213(95)00050-3}{{\em Nucl. Phys. B} {\bfseries 439} (1995) 650--664}, \href{http://arxiv.org/abs/hep-th/9408098}{{\ttfamily arXiv:hep-th/9408098}}.

\bibitem{Dabholkar:1994gg}
A.~Dabholkar, ``{Quantum corrections to black hole entropy in string theory},'' \href{http://dx.doi.org/10.1016/0370-2693(95)00056-Q}{{\em Phys. Lett. B} {\bfseries 347} (1995) 222--229}, \href{http://arxiv.org/abs/hep-th/9409158}{{\ttfamily arXiv:hep-th/9409158}}.

\bibitem{Mertens:2016tqv}
T.~G. Mertens, H.~Verschelde, and V.~I. Zakharov, ``{String Theory in Polar Coordinates and the Vanishing of the One-Loop Rindler Entropy},'' \href{http://dx.doi.org/10.1007/JHEP08(2016)113}{{\em JHEP} {\bfseries 08} (2016) 113}, \href{http://arxiv.org/abs/1606.06632}{{\ttfamily arXiv:1606.06632 [hep-th]}}.

\bibitem{Mertens:2015adr}
T.~G. Mertens, H.~Verschelde, and V.~I. Zakharov, ``{Revisiting noninteracting string partition functions in Rindler space},'' \href{http://dx.doi.org/10.1103/PhysRevD.93.104028}{{\em Phys. Rev. D} {\bfseries 93} no.~10, (2016) 104028}, \href{http://arxiv.org/abs/1511.00560}{{\ttfamily arXiv:1511.00560 [hep-th]}}.

\bibitem{Fursaev:1996uz}
D.~V. Fursaev and G.~Miele, ``{Cones, spins and heat kernels},'' \href{http://dx.doi.org/10.1016/S0550-3213(96)00631-1}{{\em Nucl. Phys. B} {\bfseries 484} (1997) 697--723}, \href{http://arxiv.org/abs/hep-th/9605153}{{\ttfamily arXiv:hep-th/9605153}}.

\bibitem{Solodukhin:2015hma}
S.~N. Solodukhin, ``{Newton constant, contact terms and entropy},'' \href{http://dx.doi.org/10.1103/PhysRevD.91.084028}{{\em Phys. Rev. D} {\bfseries 91} no.~8, (2015) 084028}, \href{http://arxiv.org/abs/1502.03758}{{\ttfamily arXiv:1502.03758 [hep-th]}}.

\bibitem{Anninos:2020hfj}
D.~Anninos, F.~Denef, Y.~T.~A. Law, and Z.~Sun, ``{Quantum de Sitter horizon entropy from quasicanonical bulk, edge, sphere and topological string partition functions},'' \href{http://dx.doi.org/10.1007/JHEP01(2022)088}{{\em JHEP} {\bfseries 01} (2022) 088}, \href{http://arxiv.org/abs/2009.12464}{{\ttfamily arXiv:2009.12464 [hep-th]}}.

\bibitem{Witten:2023xze}
E.~Witten, ``{A background-independent algebra in quantum gravity},'' \href{http://dx.doi.org/10.1007/JHEP03(2024)077}{{\em JHEP} {\bfseries 03} (2024) 077}, \href{http://arxiv.org/abs/2308.03663}{{\ttfamily arXiv:2308.03663 [hep-th]}}.

\bibitem{Bastianelli:2013tsa}
F.~Bastianelli and R.~Bonezzi, ``{One-loop quantum gravity from a worldline viewpoint},'' \href{http://dx.doi.org/10.1007/JHEP07(2013)016}{{\em JHEP} {\bfseries 07} (2013) 016}, \href{http://arxiv.org/abs/1304.7135}{{\ttfamily arXiv:1304.7135 [hep-th]}}.

\bibitem{Veltman:1975vx}
M.~J.~G. Veltman, ``{Quantum Theory of Gravitation},'' {\em Conf. Proc. C} {\bfseries 7507281} (1975) 265--327.

\bibitem{Grisaru:1975ei}
M.~T. Grisaru, P.~van Nieuwenhuizen, and C.~C. Wu, ``{Background Field Method Versus Normal Field Theory in Explicit Examples: One Loop Divergences in S Matrix and Green's Functions for Yang-Mills and Gravitational Fields},'' \href{http://dx.doi.org/10.1103/PhysRevD.12.3203}{{\em Phys. Rev. D} {\bfseries 12} (1975) 3203}.

\bibitem{Schwartz:2014sze}
M.~D. Schwartz, {\em {Quantum Field Theory and the Standard Model}}.
\newblock Cambridge University Press, 3, 2014.

\bibitem{Colin-Ellerin:2025dgq}
S.~Colin-Ellerin, G.~Lin, and G.~Penington, ``{Generalized entropy of gravitational fluctuations},'' \href{http://arxiv.org/abs/2501.08308}{{\ttfamily arXiv:2501.08308 [hep-th]}}.

\bibitem{t1974one}
G.~t~Hooft and M.~Veltman, ``One-loop divergencies in the theory of gravitation,'' in {\em Annales de l'IHP Physique th{\'e}orique}, vol.~20, pp.~69--94.
\newblock 1974.

\bibitem{Christensen:1979iy}
S.~M. Christensen and M.~J. Duff, ``{Quantizing Gravity with a Cosmological Constant},'' \href{http://dx.doi.org/10.1016/0550-3213(80)90423-X}{{\em Nucl. Phys. B} {\bfseries 170} (1980) 480--506}.

\bibitem{Louko:1995jw}
J.~Louko and R.~D. Sorkin, ``{Complex actions in two-dimensional topology change},'' \href{http://dx.doi.org/10.1088/0264-9381/14/1/018}{{\em Class. Quant. Grav.} {\bfseries 14} (1997) 179--204}, \href{http://arxiv.org/abs/gr-qc/9511023}{{\ttfamily arXiv:gr-qc/9511023}}.

\bibitem{Nelson:1994na}
W.~Nelson, ``{A Comment on black hole entropy in string theory},'' \href{http://dx.doi.org/10.1103/PhysRevD.50.7400}{{\em Phys. Rev. D} {\bfseries 50} (1994) 7400--7402}, \href{http://arxiv.org/abs/hep-th/9406011}{{\ttfamily arXiv:hep-th/9406011}}.

\bibitem{Fursaev:1995ef}
D.~V. Fursaev and S.~N. Solodukhin, ``{On the description of the Riemannian geometry in the presence of conical defects},'' \href{http://dx.doi.org/10.1103/PhysRevD.52.2133}{{\em Phys. Rev. D} {\bfseries 52} (1995) 2133--2143}, \href{http://arxiv.org/abs/hep-th/9501127}{{\ttfamily arXiv:hep-th/9501127}}.

\bibitem{Frolov:1995xe}
V.~P. Frolov, D.~V. Fursaev, and A.~I. Zelnikov, ``{Black hole entropy: Off-shell versus on-shell},'' \href{http://dx.doi.org/10.1103/PhysRevD.54.2711}{{\em Phys. Rev. D} {\bfseries 54} (1996) 2711--2731}, \href{http://arxiv.org/abs/hep-th/9512184}{{\ttfamily arXiv:hep-th/9512184}}.

\bibitem{KayStuder}
B.~S. Kay and U.~M. Studer, ``Boundary conditions for quantum mechanics on cones and fields around cosmic strings,'' \href{https://doi.org/10.1007/BF02102731}{{\em Communications in Mathematical Physics} {\bfseries 139} no.~1, (1991) 103--139}.

\bibitem{Callan:1994py}
C.~G. Callan, Jr. and F.~Wilczek, ``{On geometric entropy},'' \href{http://dx.doi.org/10.1016/0370-2693(94)91007-3}{{\em Phys. Lett. B} {\bfseries 333} (1994) 55--61}, \href{http://arxiv.org/abs/hep-th/9401072}{{\ttfamily arXiv:hep-th/9401072}}.

\bibitem{Larsen:1995ax}
F.~Larsen and F.~Wilczek, ``{Renormalization of black hole entropy and of the gravitational coupling constant},'' \href{http://dx.doi.org/10.1016/0550-3213(95)00548-X}{{\em Nucl. Phys. B} {\bfseries 458} (1996) 249--266}, \href{http://arxiv.org/abs/hep-th/9506066}{{\ttfamily arXiv:hep-th/9506066}}.

\bibitem{Vassilevich:2003xt}
D.~V. Vassilevich, ``{Heat kernel expansion: User's manual},'' \href{http://dx.doi.org/10.1016/j.physrep.2003.09.002}{{\em Phys. Rept.} {\bfseries 388} (2003) 279--360}, \href{http://arxiv.org/abs/hep-th/0306138}{{\ttfamily arXiv:hep-th/0306138}}.

\bibitem{gradshteyn2014table}
I.~S. Gradshteyn and I.~M. Ryzhik, {\em Table of integrals, series, and products}.
\newblock Academic press, 2014.

\bibitem{Donnelly:2013tia}
W.~Donnelly and A.~C. Wall, ``{Unitarity of Maxwell theory on curved spacetimes in the covariant formalism},'' \href{http://dx.doi.org/10.1103/PhysRevD.87.125033}{{\em Phys. Rev. D} {\bfseries 87} no.~12, (2013) 125033}, \href{http://arxiv.org/abs/1303.1885}{{\ttfamily arXiv:1303.1885 [hep-th]}}.

\bibitem{Barvinsky:1995dp}
A.~D. Barvinsky and S.~N. Solodukhin, ``{Nonminimal coupling, boundary terms and renormalization of the Einstein-Hilbert action and black hole entropy},'' \href{http://dx.doi.org/10.1016/0550-3213(96)00438-5}{{\em Nucl. Phys. B} {\bfseries 479} (1996) 305--318}, \href{http://arxiv.org/abs/gr-qc/9512047}{{\ttfamily arXiv:gr-qc/9512047}}.

\bibitem{DeNardo:1996kp}
L.~De~Nardo, D.~V. Fursaev, and G.~Miele, ``{Heat kernel coefficients and spectra of the vector Laplacians on spherical domains with conical singularities},'' \href{http://dx.doi.org/10.1088/0264-9381/14/5/013}{{\em Class. Quant. Grav.} {\bfseries 14} (1997) 1059--1078}, \href{http://arxiv.org/abs/hep-th/9610011}{{\ttfamily arXiv:hep-th/9610011}}.

\bibitem{Iellici:1996jx}
D.~Iellici and V.~Moretti, ``{Kabat's surface terms in the zeta function approach},'' in {\em {12th Italian Conference on General Relativity and Gravitational Physics}}, pp.~317--321.
\newblock 9, 1996.
\newblock \href{http://arxiv.org/abs/hep-th/9703088}{{\ttfamily arXiv:hep-th/9703088}}.

\bibitem{Cognola:1997xp}
G.~Cognola and P.~Lecca, ``{Electromagnetic fields in Schwarzschild and Reissner-Nordstrom geometry. Quantum corrections to the black hole entropy},'' \href{http://dx.doi.org/10.1103/PhysRevD.57.1108}{{\em Phys. Rev. D} {\bfseries 57} (1998) 1108--1111}, \href{http://arxiv.org/abs/hep-th/9706065}{{\ttfamily arXiv:hep-th/9706065}}.

\bibitem{Kabat:2012ns}
D.~Kabat and D.~Sarkar, ``{Cosmic string interactions induced by gauge and scalar fields},'' \href{http://dx.doi.org/10.1103/PhysRevD.86.084021}{{\em Phys. Rev. D} {\bfseries 86} (2012) 084021}, \href{http://arxiv.org/abs/1206.5642}{{\ttfamily arXiv:1206.5642 [hep-th]}}.

\bibitem{Donnelly:2012st}
W.~Donnelly and A.~C. Wall, ``{Do gauge fields really contribute negatively to black hole entropy?},'' \href{http://dx.doi.org/10.1103/PhysRevD.86.064042}{{\em Phys. Rev. D} {\bfseries 86} (2012) 064042}, \href{http://arxiv.org/abs/1206.5831}{{\ttfamily arXiv:1206.5831 [hep-th]}}.

\bibitem{Solodukhin:2012jh}
S.~N. Solodukhin, ``{Remarks on effective action and entanglement entropy of Maxwell field in generic gauge},'' \href{http://dx.doi.org/10.1007/JHEP12(2012)036}{{\em JHEP} {\bfseries 12} (2012) 036}, \href{http://arxiv.org/abs/1209.2677}{{\ttfamily arXiv:1209.2677 [hep-th]}}.

\bibitem{Lin:2018xkj}
J.~Lin, ``{Entanglement entropy in Jackiw-Teitelboim Gravity},'' \href{http://arxiv.org/abs/1807.06575}{{\ttfamily arXiv:1807.06575 [hep-th]}}.

\bibitem{Jafferis:2019wkd}
D.~L. Jafferis and D.~K. Kolchmeyer, ``{Entanglement Entropy in Jackiw-Teitelboim Gravity},'' \href{http://arxiv.org/abs/1911.10663}{{\ttfamily arXiv:1911.10663 [hep-th]}}.

\bibitem{Mertens:2022ujr}
T.~G. Mertens, J.~Sim\'on, and G.~Wong, ``{A proposal for 3d quantum gravity and its bulk factorization},'' \href{http://arxiv.org/abs/2210.14196}{{\ttfamily arXiv:2210.14196 [hep-th]}}.

\bibitem{Chen:2023tvj}
H.~Z. Chen, R.~C. Myers, and A.-M. Raclariu, ``{Entanglement, Soft Modes, and Celestial Holography},'' \href{http://arxiv.org/abs/2308.12341}{{\ttfamily arXiv:2308.12341 [hep-th]}}.

\bibitem{Takayanagi:2019tvn}
T.~Takayanagi and K.~Tamaoka, ``{Gravity Edges Modes and Hayward Term},'' \href{http://dx.doi.org/10.1007/JHEP02(2020)167}{{\em JHEP} {\bfseries 02} (2020) 167}, \href{http://arxiv.org/abs/1912.01636}{{\ttfamily arXiv:1912.01636 [hep-th]}}.

\bibitem{Blommaert:2018iqz}
A.~Blommaert, T.~G. Mertens, and H.~Verschelde, ``{Fine Structure of Jackiw-Teitelboim Quantum Gravity},'' \href{http://dx.doi.org/10.1007/JHEP09(2019)066}{{\em JHEP} {\bfseries 09} (2019) 066}, \href{http://arxiv.org/abs/1812.00918}{{\ttfamily arXiv:1812.00918 [hep-th]}}.

\bibitem{zwiebach2004first}
B.~Zwiebach, {\em A first course in string theory}.
\newblock Cambridge university press, 2004.

\bibitem{Witten:1998qj}
E.~Witten, ``{Anti-de Sitter space and holography},'' \href{http://dx.doi.org/10.4310/ATMP.1998.v2.n2.a2}{{\em Adv. Theor. Math. Phys.} {\bfseries 2} (1998) 253--291}, \href{http://arxiv.org/abs/hep-th/9802150}{{\ttfamily arXiv:hep-th/9802150}}.

\bibitem{Jacobson:2012ek}
T.~Jacobson and A.~Satz, ``{Black hole entanglement entropy and the renormalization group},'' \href{http://dx.doi.org/10.1103/PhysRevD.87.084047}{{\em Phys. Rev. D} {\bfseries 87} no.~8, (2013) 084047}, \href{http://arxiv.org/abs/1212.6824}{{\ttfamily arXiv:1212.6824 [hep-th]}}.

\bibitem{Townsend:1979hd}
P.~K. Townsend, ``{Covariant Quantization of Antisymmetric Tensor Gauge Fields},'' \href{http://dx.doi.org/10.1016/0370-2693(79)90122-9}{{\em Phys. Lett. B} {\bfseries 88} (1979) 97--101}.

\bibitem{Siegel:1980jj}
W.~Siegel, ``{Hidden Ghosts},'' \href{http://dx.doi.org/10.1016/0370-2693(80)90119-7}{{\em Phys. Lett. B} {\bfseries 93} (1980) 170--172}.

\bibitem{Schwarz:1984wk}
A.~S. Schwarz and Y.~S. Tyupkin, ``{Quantization of antisymmetric tensors and ray-singer torsion},'' \href{http://dx.doi.org/10.1016/0550-3213(84)90403-6}{{\em Nucl. Phys. B} {\bfseries 242} (1984) 436--446}.

\bibitem{Buchbinder:1988tj}
I.~L. Buchbinder and S.~M. Kuzenko, ``{Quantization of the classically equivalent theories in the superspace of simple supergravity and equivalence},'' \href{http://dx.doi.org/10.1016/0550-3213(88)90047-8}{{\em Nucl. Phys. B} {\bfseries 308} (1988) 162--190}.

\bibitem{Moitra:2018lxn}
U.~Moitra, R.~M. Soni, and S.~P. Trivedi, ``{Entanglement Entropy, Relative Entropy and Duality},'' \href{http://dx.doi.org/10.1007/JHEP08(2019)059}{{\em JHEP} {\bfseries 08} (2019) 059}, \href{http://arxiv.org/abs/1811.06986}{{\ttfamily arXiv:1811.06986 [hep-th]}}.

\bibitem{Mukherjee:2023ihb}
J.~Mukherjee, ``{Entanglement entropy and the boundary action of edge modes},'' \href{http://arxiv.org/abs/2310.14690}{{\ttfamily arXiv:2310.14690 [hep-th]}}.

\bibitem{erdelyi1953higher}
A.~Erd{\'e}lyi, ``Higher transcendental functions,'' {\em Higher transcendental functions} (1953) 59.

\bibitem{Gibbons:1978ac}
G.~W. Gibbons, S.~W. Hawking, and M.~J. Perry, ``{Path Integrals and the Indefiniteness of the Gravitational Action},'' \href{http://dx.doi.org/10.1016/0550-3213(78)90161-X}{{\em Nucl. Phys. B} {\bfseries 138} (1978) 141--150}.

\bibitem{Iellici:1996gv}
D.~Iellici and V.~Moretti, ``{Thermal partition function of photons and gravitons in a Rindler wedge},'' \href{http://dx.doi.org/10.1103/PhysRevD.54.7459}{{\em Phys. Rev. D} {\bfseries 54} (1996) 7459--7469}, \href{http://arxiv.org/abs/hep-th/9607015}{{\ttfamily arXiv:hep-th/9607015}}.

\end{thebibliography}\endgroup

\end{document}